# A direct microscopic approach to transition strengths in Pre-Equilibrium reactions

F. B. Guimaraes[1] and B. V. Carlson[2]

[1]*Instituto de Estudos Avançados/DCTA,*
*12228-001 São José dos Campos, São Paulo, Brazil*
*e-mail: fbraga@ieav.cta.br*
[2]*Departamento de Física,*
*Instituto Tecnológico de Aeronáutica/DCTA,*
*12228-900 São José dos Campos, São Paulo, Brazil*
*e-mail: brett@ita.br*

**Abstract**

*We present a microscopic formalism that extends the traditional formulation of Williams, Ericson and Bloch and permits to obtain the transition strengths (TS) of pre-equilibrium nuclear reactions directly from their quantum microscopic description. We calculate the TS without resorting to the Laplace transform approach and the use of the saddle point approximation. We also analyze some problems that may appear in connection with these mathematical tools and the Darwin-Fowler approach in this case.*
*We show that, analogously to the nuclear densities, the strengths for transitions that change the exciton number by two or leave it unchanged can be estimated microscopically as convolutions of the functions of simpler states. When using the HO basis for the Model Space we obtained important departure from the results of the exciton model (EXM), which can partially invalidate our previous analysis on the attainment of equilibrium during the PE stage. On the other hand, by using constant grid of energies for the sp-basis we were able to reproduce the results of EXM quite well in a large range of excitation energies.*
*A new model code, TRANSNU, was developed that can be ported to traditional semi-classical codes like TNG for nuclear data evaluation.*

## 1. Introduction

A common description of the preequilibrium stage of nuclear reactions (PE) uses the exciton model (EXM),[1, 2, 3, 4] to analyze nuclear states and particle emissions before the formation of the compound nucleus. In this model the nuclear states are analyzed in terms of their "complexity" defined by the number of excited *single particle states* (sp-states) in comparison with the fundamental state, where the total nuclear excitation is zero and all component nucleons occupy the lowest possible energies levels up to a maximum called Fermi level, $E_F$.

To describe the excited nuclear states, one initially considers the fundamental state and the sp-states that can be excited on it, and call them "holes" when created below $E_F$ and "particles" when created above it. In addition, all initial sp-states above $E_F$ are also defined as holes. If at a given moment of the evolution of the nuclear system there are "$h$" holes "$p$" particles the total number $n$=$p$+$h$ of excited sp-states is called the *exciton number* of the system. For nuclear states with excitation energy greater that zero, one assumes that new particles can be created only in states previously occupied by holes and, vice-versa, holes can only be created in states previously occupied by particles.

When using the Model Space description though, these simple definitions are expanded to consider particles and holes as independent fermion fields, i. e., which can be created or destroyed independently. This is achieved by defining a sequence of independent possible levels for each field and $E_F$ as an independent phenomenological parameter of the model. Another central aspect is the basis adopted for the single particle wave-functions, which in this work is the Hamonic Oscillator (H.O.) one.[5]

The connection with actual physical systems is realized by assuming that the sp-energies cannot be greater than a given phenomenological maximum and that particles and holes can only be created or destroyed simultaneously, i. e., in pairs particle-hole sp-states (ph-pairs). Then the total number of excitons can only vary in steps of $\pm 2$.

From the initial excited state the system is supposed to evolve by increasingly sharing the total excitation among the largest possible number of sp-states. The sharing occurs at the microscopic level as a consequence of the interaction between sp-states, yielding on the average a systematically growing number of excitons up to a maximum, which is the "most probable exciton number" at the PE stage.

On the other hand, the interaction of the excited sp-states is usually defined by a phenomenological "residual term" added to the nuclear Hamiltonian and the basis of sp-states can be defined self-consistently or phenomenologically.

The increase of $n$ is counterbalanced by possible PE emission of particles or annihilation of a ph-pairs, and after the system reaches some level of "complexity" it is supposed to evolve preferrably towards full equilibrium (compound state) instead of emitting more particles at the PE stage.

In the EXM the rates of transition between nuclear states of increasing complexity are usually given in an essentially phenomenological way. In particular, the transition matrix elements can be considered as phenomenological *constants* or simple functions of the nuclear excitation energy, $U$, as exemplified in Ref.[6].

In addition, the EXM relies on approximations of statistical nature that may not be very precisely defined at the microscopic level, which often makes it difficult to evaluate the importance of the details of the microscopic interaction in the description of the PE process. For example, the semi-classical formulations[7] give the density of states for a given $U$ as a convolution of the densities for $p$ and $h$,[1]

$$\omega_{ph}(U) = \int_0^U \omega_p(E)\omega_h(U-E)dE \ , \tag{1.1}$$

which results from the continuum approximation ($CAP$) and does not come directly from the quantum microscopic description.

The traditional approach to the Shell Model defines the moments of the Hamiltonian in terms of Laplace transforms and their inverse to obtain expressions like (1.1) for the nuclear density and in the microscopic description of Ref.[8] a similar approach is followed to obtain the expressions for the transition strengths.

In this work we follow closely the approach of Ref.[8] and present a detailed *direct microscopic formalism* in which one is able to produce the results of the Shell Model without effectively having to resort to the Laplace transform and, therefore, without having to rely on the validity of $CAP$ and other common approximations of the semi-classical models.

In this respect the direct microscopic formalism is a more natural and intuitive description and it reveals itself as a more appropriate tool to describe, for example, the nuclear transition strengths, bringing a complementary view to the usual statistical approach to describe PE states.

One of the central aspects of this formalism is the proportionality of the degeneracy of a given nuclear state (with given total excitation energy, $U$, angular momentum, $M$, and number of excitons, $p$ and $h$), $d(U,M,p,h)$, to the corresponding density of nuclear states

$$d(U, M, p, h) \propto \omega(U, M, p, h), \tag{1.2}$$

which permits to connect to the traditional approach, using $CAP$, and to reinterpret the moments of the nuclear Hamiltonian in this limit in terms of convolutions over the excitation energy of the moments of less excited states in accordance with the general proposal of Ref.[8].

In Sec.2 we analyze the definition of the nuclear density in the direct microscopic formalism in connection with the degeneracy of nuclear levels, the basic definitions of the Shell Model and the Darwin-Fowler statistics.

In Sec.3 we present the formal definitions of the direct microscopic algebra and present the specific ones that apply to the computation of the momenta of the nuclear Hamiltonian. In Sec.4 we use the limit of $CAP$ and the usual description of the grand canonical ensemble of Statistical Mechanics to make the connection with the Laplace transform, in agreement with the traditional approaches.

Sections 5 and 6 describe the specific application of the direct approach to the evaluation of the momenta of the one-body interaction and the residual term of the PE Hamiltonian, respectively, and in Sec.7 we present our results and conclusions.

## 2. The nuclear level density

In the usual "statistical description" of the nucleus inspired by the Shell Model,[3] the grand canonical ensemble can be defined by the following relation for the nuclear mass number,

$$A = \sum_{(i)} n_i \ , \tag{2.1}$$

where $n_i \in \{0, 1\}$, are the occupation numbers of the single particle (sp) states associated with the corresponding set of sp-levels with energies

$$\epsilon_i = \nu_i \epsilon \ . \tag{2.2}$$

The $\nu_i$ are integers and $\epsilon$ is a fixed real number, defining the spacing between any two consecutive sp-levels of the approximate "equidistant spacing model". $\epsilon$ can also be considered as an average spacing of more realistic bases for the sp-states as, e. g., the H.O. basis. The total nuclear energy is then given by

$$E = \mathcal{N}\epsilon = \sum_{(i)} n_i \nu_i \epsilon \ . \tag{2.3}$$

If a given physical quantity has discrete eigenvalues defined as a function of a set of integers, we will call the latter the set of *characteristic integers* associated with that quantity. Then, in the above case, the set $\{\nu_i\}$ are the characteristic integers associated with the *energies* of the sp-levels, and the integers $\{n_i\}$ indicate the *occupations* of the sp-states as either filled with one fermion, $n_i = 1$, or empty, $n_i = 0$. This specific definition can be useful when dealing with the microscopic definition of nuclear states explicitly in terms of the component sp-states. In this case one may have the formation of quasi-continuum sequences of nuclear levels, which nonetheless will still be univocally related to a set of integers because the bound component sp-states have discrete energies.

If more than one element of the set $\{\nu_i, i = 1, .., \infty\}$ are equal, the corresponding elements,

$$\nu_{i1} = \cdots = \nu_{ik} \in \{\nu_i, i = 1, .., \infty\}\,, \tag{2.4}$$

define the set of the degenerate sp-states of the energy level

$$\epsilon_i = \nu_{i1}\epsilon = \cdots = \nu_{ik}\epsilon\,. \tag{2.5}$$

Similarly, the number of elements of the set of all *configurations of sp-states* (nuclear states) associated with a given *nuclear energy* $E$ is, by definition, the degeneracy of the corresponding *nuclear level.* For given $A$ and $E$, the nuclear degeneracy is equal to the number of solutions of the Eqs.(2.1) and (2.3). Therefore, if one considers the microscopic distribution of nuclear states as a function of the nuclear energy, *the cummulative number of states increase in steps equal to the degeneracy of each nuclear level.*

In analogy to sp-states, the degeneracy of a given nuclear level $E$ can be considered as a possible characteristic integer associated with $E$, as the occupation number or "population" of configurations associted with each level.

This number is usually large because there are many possible sets with $A$ non-null elements, $\{n_{i1}, \cdots, n_{iA}\}$, satisfying these equations, with values of $\nu_i$ varying from 0 to $\mathcal{N}$. The degeneracy as a function of $E$ depends on the reference energy used to define the levels of particles and holes. For exemple, when the description of sp-states is made in terms of excitons, the particle and hole states may have energies conveniently defined with respect to $\epsilon_F$ and the nuclear energy will be equal to the *excitation energy.* In this case, the *ground state* has zero energy and all characteristic integers associated with the energies of the sp-levels, $\{\nu_i\}$, are zero.

Notice that, in the microscopic description the degeneracy of the sp-levels is defined by the specific features of the system hamiltonian, more precisely the symmetries of the nuclear system, while the nuclear degeneracy has a more "combinatorial" meaning, in terms of the distribution of nucleons into a "pre-defined" structure of sp-states.

In this context, the nuclear level density at the energy $E$, $\rho(E)$, can be intuitively *defined* as the ratio between the nuclear degeneracy and the sp-level spacing $\epsilon$. This is reasonable because when the nuclear energy varies from a given value $E$ to its next $E_{next}$, corresponding to the variation from the corresponding characteristic integer to its next, the cumulative number of nuclear states varies by an amount equal to the degeneracy of the nuclear level at $E_{next}$ and, by hypothesis, the corresponding variation in the nuclear energy, $E_{next} - E$, should be proportional to $\epsilon$,

$$E_{next} - E \propto \epsilon \ . \tag{2.6}$$

Now, if one considers the *continuum approximation* (*CAP*) and assumes that the nuclear energies vary very little from $E$ to $E_{next}$, then the assumption of these variations to be proportional to $\epsilon$ forces it logically to be even smaller. Then in the limit of a set of nuclear states with "very large" density Eq.(2.6) becomes,

$$E_{next} - E \ \approx \ dE \ \approx \ \epsilon \tag{2.7}$$

and the corresponding variation of the cumulative number of states would be a function of the energy given by

$$d\mathcal{N} \approx \mathcal{N}_{next} - \mathcal{N} = \mathcal{D}_k(E)\,, \tag{2.8}$$

where $\mathcal{D}_k(E)$ is the nuclear degeneracy at the excitation $E$.

Then, the nuclear density $\rho(E)$ becomes approximately,

$$\rho(E) = \frac{d\mathcal{N}}{dE} \approx \frac{\mathcal{D}_k(E)}{\epsilon} \tag{2.9}$$

The same result can be obtained using a more formal argument, by considering $\epsilon$ constant in (2.3) and applying the corresponding definition of the statistical Shell Model formalism using the Darwin-Fowler method.[3] The generating function for the corresponding grand canonical ensemble is given by the expression

$$f(x,y) = \prod_i (1 + xy^{\nu_i}) = \prod_i (1 + x_i)\ , \tag{2.10}$$

where the last simple form takes into account the fact that $x$ is indiferentiated for the various component sp-states as it accounts for the *number* of "relevant" sets of sp-states independently of their specific characteristics, i. e., $x$ is the same for all sp-states. Both $x$ and $y$ have physical meaning under the statistical description of the grand canonical ensemble, for many-body system with variable number of "bodies" and variable energy, but $x$ has a more strictly combinatorial meaning while $y$ is related with the probability distribution associated with the various microscopic systems of the ensemble.

Then, the nuclear level density can be directly defined as an adequate *pole* of the generating function divided by $\epsilon$,[3]

$$\rho(A,E) = \frac{1}{(2\pi i)^2 \epsilon} \oint \oint \frac{f(x,y)dxdy}{x^{A+1} y^{\mathcal{N}+1}} \ , \tag{2.11}$$

while the generating function can be rewritten as

$$f(x,y) = 1 + x \sum_{(j)} y^{\nu_J} + x^2 \sum_{(j1,j2)} y^{(\nu_{J1}+\nu_{J2})} + \cdots + x^A \sum_{(j1,\cdots,jA)} y^{(\nu_{J1}+\cdots+\nu_{JA})} + \cdots \ , \tag{2.12}$$

which, therefore, describes all nuclear systems with all possible “mass numbers” (relevant sp-states in each microstate) and energies (total energy of the microstate).

In other words, for nuclear systems each configuration of sp-states is also a microstate of the canonical ensemble with fixed mass number and temperature,[9] and the term proportional to $x^A$ is the sum over all possible configurations with fixed nuclear mass $A$ and variable energy.

For each nuclear level, $E_k$, corresponds usually many different configurations of sp-states and to each nuclear mass $A$ a term, $\mathcal{Y}_A$, is defined in Eq. (2.12) as follows,

$$\mathcal{Y}_A = \sum_{(j1,\cdots,jA)} y^{\left(\nu_{j1}+\cdots+\nu_{jA}\right)} \ . \tag{2.13}$$

which can be rewritten in terms of the degeneracies for the various nuclear levels, $\mathcal{D}_k$, as follows

$$\mathcal{Y}_A = \sum_{(k)} \mathcal{D}_k y^{\mathcal{N}_k} \tag{2.14}$$

where

$$\mathcal{N}_k = \sum_{i=1}^{A} \nu_{ki} = \frac{E_k}{\epsilon} \tag{2.15}$$

and $\mathcal{D}_k = \mathcal{D}_k(E_k, A)$ is the degeneracy of the nuclear level $E_k$, for a given nuclear mass number $A$.

Taking Eq. (2.14) into Eq. (2.12) and using the definition (2.11) for the nuclear level density yields

$$\rho(E_k, A) = \frac{1}{\epsilon} \mathcal{D}_k(E_k, A) \ , \tag{2.16}$$

where usually the dependence in $A$ is not explicitly written. Therefore, the definition in Eq.(2.11) is coherent with the “intuitive” description given at the beginning of this section.

Equation (2.12) can be rewritten as

$$f(x,y) = 1 + x\sum_{(k1)}\mathcal{D}_{k1}(E_{k1},1)y^{\mathcal{N}_{k1}(1)} +$$

$$x^2\sum_{(k2)}\mathcal{D}_{k2}(E_{k2},2)y^{\mathcal{N}_{k2}(2)} + \cdots + x^A\sum_{(k)}\mathcal{D}_k(E_k,A)y^{\mathcal{N}_k(A)} + \cdots + \tag{2.17}$$

where different indices have been used for each term to reinforce the fact that the corresponding nuclear levels may not be the same. In these expressions the sum over $k$ is equivalent to the sum over $E_k$, then $f(x,y)$ can be rewritten as

$$f(x,y) = \sum_{(A,U)}\mathcal{D}(A,U)x^A y^{U/\epsilon} , \tag{2.18}$$

and also as a sum over individual configurations, with all degneracies are equal one,

$$f(x,y) = \sum_{(conf)} x^A y^{U/\epsilon} . \tag{2.19}$$

## 3. Formal definitions of the direct microscopic approach

To describe a given nuclear excited state, with variable mass number $A$ and energy $E$, one may assume that it belongs to the corresponding grand canonical distribution defined over the nuclear configurations obtained as the solutions of the Eqs.(2.1) and (2.3).

In this case one may always assume that the elements of the set of characteristic integers of the energies of sp-states, $\{\nu_i\}$ in (2.3), are ordered according to increasing values as a function of $i$ and, by definition, there are only $A$ nonzero elements in the set $\{n_i\}$.

For a given nuclear mass $A$ the nuclear state, $|\psi_A\rangle$, can be represented by a product of sp-states, which are usually approximated by self-consistent quasi-particle states, and can be written as linear combinations of the corresponding complete set of eigenvectors of the sp-hamiltonian,

$$|\psi_A\rangle = \prod_{i=1}^{A}|\psi_i(t)\rangle = \prod_{i=1}^{A}\sum_{(\mathtt{k}_i)}c_{i\mathtt{k}_i}\mathsf{e}^{iE_{\mathtt{k}_i}t/\hbar}|u_{\mathtt{k}_i}\rangle$$

$$= \sum_{(\mathtt{k}_1)}\cdots\sum_{(\mathtt{k}_A)}c_{1\mathtt{k}_1}\cdots c_{A\mathtt{k}_A}\mathsf{e}^{i\left(E_{\mathtt{k}_1}+\cdots+E_{\mathtt{k}_A}\right)t/\hbar}|u_{\mathtt{k}_1}\cdots u_{\mathtt{k}_A}\rangle . \tag{3.1}$$

which can be rewritten as

$$|\psi_A\rangle = \sum_{(\mathtt{j})}\mathcal{C}_j\mathsf{e}^{iE_\mathtt{j}t/\hbar}|\mathcal{W}_j\rangle \tag{3.2}$$

where $\mathtt{j}$ is a characteristic integer for the nuclear energy, corresponding to a specific sequence of the enumerable set of components of $\psi_A$. In this case the notation $(\mathtt{j})$ is used to designate the corresponding degenerate set of nuclear configurations with energy $E_{\mathtt{j}}$.

The grand canonical ensemble vector corresponding to the various nuclear states $|\psi_A\rangle$ can also be written[10] directly in terms of the occupation numbers of all sp-states defined in (2.3) as

$$|\psi\rangle = \sum_{(n_1,\cdots,n_\infty)} |n_1\cdots n_\infty\rangle \ , \tag{3.3}$$

and if one defines the symbol "$(\mathtt{k}_1\cdots\mathtt{k}_A)$" for the set of all configurations containing $A$ and only $A$ sp-states with non null occupations as

$$(\mathtt{k}_1\cdots\mathtt{k}_A) = \Big\{\{\mathtt{k}_1,\cdots,\mathtt{k}_A\}/\!: n_{\mathtt{k}_1} = \cdots = n_{\mathtt{k}_A} = 1\Big\}, \quad \text{for given } A, \tag{3.4}$$

where $\mathtt{k}_1,\cdots,\mathtt{k}_A$ are also supposed to be ordered by increasing values, then (3.3) can be written as

$$|\psi\rangle = \sum_{(\cdots,\mathtt{k}_1,\cdots,\mathtt{k}_A,\cdots)} |n_1,\cdots,n_{k1},\cdots,n_{kA},\cdots,n_\infty\rangle = \sum_{(A)}\sum_{(\mathtt{k}_1,\cdots,\mathtt{k}_A)} |(\mathtt{k}_1\cdots\mathtt{k}_A)\rangle = \sum_{(A)}|\psi_A\rangle \tag{3.5}$$

The Fock space operator of the excitons, "particles" $(p)$ and "holes" $(h)$, of an arbitrary nuclear system, with arbitrary $A$, that describes the corresponding grand canonical ensemble of nuclear states, with expected values on the states $|\psi\rangle$ given by (2.10), can be defined in the case of two fermion fields, "$p$" and "$h$", as[8]

$$F_0 = \prod_{\mu\nu}\Big(a_\nu {a_\nu}^\dagger + x_{p\nu}{a_\nu}^\dagger a_\nu\Big)\Big(b_\mu {b_\mu}^\dagger + x_{h\mu}{b_\mu}^\dagger b_\mu\Big) = \prod_{\mu\nu} F_{p\nu}F_{h\mu} \tag{3.6}$$

If one considers initially the simpler one fermion expression

$$F_0 = \prod_l \Big(a_l a_l^\dagger + x_{pl} a_l^\dagger a_l\Big), \tag{3.7}$$

then it can be rewritten as

$$F_0 = \prod_{l=0}^{\infty}\Big[\Big(a_l a_l^\dagger\Big)\cdots\Big] + \sum_{s=1}^{\infty}\left[\prod_{\binom{l_1,l_2}{\neq s}} \Big(a_{l1}a_{l1}^\dagger\Big)\cdots\Big(a_s^\dagger a_s\Big)\cdots\Big(a_{l2}a_{l2}^\dagger\Big)\cdots\right] x_{ps} \tag{3.8}$$

$$+ \sum_{(s1<s2)}\left[\prod_{\binom{l_1,l_2,l_3}{\neq(s1,s2)}} \Big(a_{l1}a_{l1}^\dagger\Big)\cdots\Big(a_{s1}^\dagger a_{s1}\Big)\cdots\Big(a_{l2}a_{l2}^\dagger\Big)\cdots\Big(a_{s2}^\dagger a_{s2}\Big)\cdots\Big(a_{l3}a_{l3}^\dagger\Big)\cdots\right] x_{ps1}x_{ps2} \tag{3.9}$$

$$+\sum_{(s1<s2<s3)}(\cdots)x_{ps1}x_{ps2}x_{ps3}+\cdots \tag{3.10}$$

$$=\sum_{(N=0)}^{\infty}\sum_{(s_N)}\left(\prod_{\mathtt{j}=1}^{N}x_{ps_j}\right)\Pi(s_N) \tag{3.11}$$

where, for fixed $N$ and configuration $(s_N)$, $\Pi(s_N)$ is given by

$$\Pi(s_N)=\prod_{\binom{\mathtt{l}_1,\cdots,\mathtt{l}_N,\mathtt{l}_{N+1}}{\neq(s1,\cdots,sN)}}\left(a_{l1}a_{l1}^{\dagger}\right)\cdots\left(a_{s1}^{\dagger}a_{s1}\right)\cdots\left(a_{lN}a_{lN}^{\dagger}\right)\cdots\left(a_{sN}^{\dagger}a_{sN}\right)\cdots\left(a_{l(N+1)}a_{l(N+1)}^{\dagger}\right)\cdots \tag{3.12}$$

where there are $N$ terms of the type $\left(a_{s\mathtt{j}}^{\dagger}a_{s\mathtt{j}}\right)$ and infinite terms of the type $\left(a_{l\mathtt{j}}a_{l\mathtt{j}}^{\dagger}\right)$.

Then it is clear that

$$\Pi(s_N)|(\mathtt{k}_1\cdots\mathtt{k}_A)\rangle=\delta_{N,A}\delta(s_1\cdots s_N|\mathtt{k}_1\cdots\mathtt{k}_A)|(\mathtt{k}_1\cdots\mathtt{k}_A)\rangle \tag{3.13}$$

and therefore

$$F_0|\psi\rangle=F_0\sum_{(A)}|\psi_A\rangle= \tag{3.14}$$

$$\sum_{(N)}\sum_{(s_N)}\left(\prod_{\mathtt{j}=1}^{N}x_{ps_j}\right)\Pi(s_N)\sum_{(A)}\sum_{(\mathtt{k}_A)}|(\mathtt{k}_1\cdots\mathtt{k}_A)\rangle= \tag{3.15}$$

$$\sum_{(N,A)}\sum_{(s_N)(\mathtt{k}_A)}\left(\prod_{\mathtt{j}=1}^{N}x_{ps_j}\right)\delta_{N,A}\delta(s_1\cdots s_N|\mathtt{k}_1\cdots\mathtt{k}_A)|(\mathtt{k}_1\cdots\mathtt{k}_A)\rangle= \tag{3.16}$$

$$\sum_{(N)}\sum_{(s_1\cdots s_N)}\left(\prod_{\mathtt{j}=1}^{N}x_{ps_j}\right)|(s_1\cdots s_N)\rangle=\sum_{(N)}\left|\widetilde{\psi_N}\right\rangle \tag{3.17}$$

which, by comparison with (3.5) shows that $F_0$ projects the grand canonical ensemble vector $|\psi\rangle$ of Eq.(3.3) into another linear combination of its various components, where the coefficients have changed from 1 to

$$\left(\prod_{\mathtt{j}=1}^{N}x_{ps_j}\right), \tag{3.18}$$

which are proportional (for given $N$) to the canonical ensemble probabilities of the component configuration $|s_1\cdots s_N\rangle$.[9]

Eq. (3.17) also implies that for two given independent grand canonical vectors

$$|1\rangle = \sum_{(N_1)}\sum_{(\mathtt{k}_{N_1})} |(\mathtt{k}_1\cdots\mathtt{k}_{N_1})\rangle \quad \text{and} \quad |2\rangle = \sum_{(N_2)}\sum_{(\mathtt{l}_{N_2})} |(\mathtt{l}_1\cdots\mathtt{l}_{N_2})\rangle \tag{3.19}$$

results,

$$\langle 1\,|F_0|\,2\rangle = \langle 1|\sum_{(N)}\sum_{(s_N)}\left(\prod_{\mathtt{j}=1}^{N} x_{ps_j}\right)|(s_1\cdots s_N)\rangle \tag{3.20}$$

$$= \sum_{(N)}\sum_{(s_N)}\left(\prod_{\mathtt{j}=1}^{N} x_{ps_j}\right)\delta(s_N|1)\delta(s_N|2) \tag{3.21}$$

Let $O$ be an operator on the Fock space of the sp-states, which can modify the nuclear configuration. Then one can write the corresponding *transition strength*, defined as the square of the transition moment $\langle 1\,|O|\,2\rangle$ summed over all possible transitions, as[11]

$$S_O = \sum_{(trans.)} |O|^2 = \sum_{(12)} |\langle 1\,|O|\,2\rangle|^2 = \sum_{(12)} \langle 1\,|O|\,2\rangle\langle 2\,|O^\dagger|\,1\rangle$$

due to possible variation in the number of excitons and nuclear excitation $S_O$ must be redefined in the framework of the grand canonical ensemble using the grand canonical distribution, which can be written schematically as

$$|O|^2 = \left\langle F_0 O F_0' O^\dagger \right\rangle = \sum_{(1)} \langle 1\,|F_0 O F_0' O^\dagger|\,1\rangle = \sum_{\binom{12}{34}} \langle 1\,|F_0|\,2\rangle\langle 2\,|O|\,3\rangle\langle 3\,|F_0'|\,4\rangle\langle 4\,|O^\dagger|\,1\rangle \tag{3.22}$$

$$= \sum_{(N)}\sum_{(s_N)}\left(\prod_{\mathtt{j}}^{N} x_{ps_\mathtt{j}}\right)\langle s_N\,|O|\,r_M\rangle \sum_{(M)}\sum_{(r_M)}\left(\prod_{k}^{M} x'_{pr_k}\right)\langle r_M\,|O^\dagger|\,s_N\rangle \tag{3.23}$$

In the case of two Fermion fields (e. g., "particles" and "holes") $F_0$ is given by Eq. (3.6)

$$F_0 = F_p F_h = \prod_{\mu}\Big(a_\mu a_\mu^\dagger + x_{p\mu} a_\mu^\dagger a_\mu\Big)\prod_{\nu}\Big(b_\mu b_\mu^\dagger + x_{h\mu} b_\mu^\dagger b_\mu\Big), \tag{3.24}$$

which can be rewritten as

$$F_p = \sum_{p=0}^{\infty}\sum_{(s_p)}\left(\prod_{k=1}^{p} x_{ps_k}\right)\Pi(s_p) \tag{3.25}$$

and

$$F_h = \sum_{h=0}^{\infty}\sum_{(r_h)}\left(\prod_{\mathtt{j}=1}^{h} x_{hr_\mathtt{j}}\right)\Pi(r_h) \tag{3.26}$$

and, as we saw in (3.13), due to the properties of the single particle fermion operators one can also identify $\Pi(s_p)$ and $\Pi(r_h)$ with the components of the respective projection operator,

$$\Pi(s_p) = |s_p\rangle\langle s_p| \qquad \text{and} \qquad \Pi(r_h) = |r_h\rangle\langle r_h| \tag{3.27}$$

and

$$\Pi(s_p)\Pi(r_h) = |s_p r_h\rangle\langle s_p r_h|\,. \tag{3.28}$$

Then the analogous of (3.20) becomes,

$$\langle 1\,|\,F_pF_h|\,2\rangle = \langle 1|\sum_{(ph)}\sum_{(s_pr_h)}\left(\prod_{k,\mathrm{j}=1}^{ph} x_{ps_k}x_{hr_\mathrm{j}}\right)\delta(s_pr_h|2)|(s_pr_h)\rangle \tag{3.29}$$

$$= \sum_{(ph)}\sum_{(s_pr_h)}\left(\prod_{k,\mathrm{j}=1}^{ph} x_{ps_k}x_{hr_\mathrm{j}}\right)\delta(s_pr_h|1)\delta(s_pr_h|2) \tag{3.30}$$

and

$$\langle 1\,|\,F'_pF'_h|\,2\rangle = \langle 1|\sum_{(p'h')}\sum_{(s_{p'}r_{h'})}\left(\prod_{l,i=1}^{p'h'} x'_{ps_l}x'_{hr_i}\right)\delta(s_{p'}r_{h'}|2)\big|\big(s_{p'}r_{h'}\big)\big\rangle \tag{3.31}$$

$$= \sum_{(p'h')}\sum_{(s_{p'}r_{h'})}\left(\prod_{l,i=1}^{p'h'} x'_{ps_l}x'_{hr_i}\right)\delta(s_{p'}r_{h'}|1)\delta(s_{p'}r_{h'}|2) \tag{3.32}$$

and the analogous of (3.22) and (3.23) are

$$\left\langle F_pF_h\Omega F'_pF'_hO^\dagger\right\rangle = \sum_{(1)}\langle 1\,|\,F_pF_h\Omega F'_pF'_hO^\dagger|\,1\rangle \tag{3.33}$$

$$= \sum_{\binom{12}{34}}\langle 1\,|\,F_pF_h|\,2\rangle\langle 2\,|\,\Omega|\,3\rangle\langle 3\,|\,F'_pF'_h|\,4\rangle\langle 4\,|\,O^\dagger|\,1\rangle \tag{3.34}$$

$$= \sum_{(ph)}\sum_{(s_pr_h)}\left(\prod_{k,\mathrm{j}=1}^{ph} x_{ps_k}x_{hr_\mathrm{j}}\right)\langle s_pr_h\,|\,\Omega|\,s_{p'}r_{h'}\rangle\sum_{(p'h')}\sum_{(s_{p'}r_{h'})}\left(\prod_{l,i=1}^{p'h'} x'_{ps_l}x'_{hr_i}\right)\langle s_{p'}r_{h'}\,|\,O^\dagger|\,s_pr_h\rangle \tag{3.35}$$

which can be rewritten as

$$= \sum_{\binom{ph}{p'h'}} \sum_{\binom{s_p r_h}{s_{p'} r_{h'}}} \left( \prod_{k,\mathtt{j}=1}^{ph} \prod_{l,i=1}^{p'h'} x_{ps_k} x_{hr_\mathtt{j}} x'_{ps_l} x'_{hr_i} \right) \langle s_p r_h | \Omega | s_{p'} r_{h'} \rangle \langle s_{p'} r_{h'} | O^\dagger | s_p r_h \rangle \tag{3.36}$$

Now we make the usual change of variables that defines the explicit connection with the microscopic statistical parameters of the sp-states and also brings the possibility of the Laplace transform interpretation

$$x_{ps_k} = x \mathsf{e}^{(-\beta \epsilon_{s_k} - \gamma \mathtt{m}_{\mathtt{s_k}})}$$

$$x'_{ps_l} = x' \mathsf{e}^{(-\beta' \epsilon_{s_l} - \gamma' \mathtt{m}_{\mathtt{s_l}})}$$

$$x_{hr_\mathtt{j}} = y \mathsf{e}^{(-\beta \epsilon_{r_\mathtt{j}} - \gamma \mathtt{m}_{\mathtt{r_j}})}$$

$$x'_{hr_i} = y' \mathsf{e}^{(-\beta' \epsilon_{r_i} - \gamma' \mathtt{m}_{\mathtt{r_i}})} \tag{3.37}$$

then (3.36) can be rewritten as

$$= \sum_{\binom{ph}{p'h'}} x^p x'^{p'} y^h y'^{h'} \sum_{\binom{s_p r_h}{s_{p'} r_{h'}}} \left( \prod_{k,\mathtt{j}=1}^{ph} \prod_{l,i=1}^{p'h'} \mathsf{e}^{(-\beta \epsilon_{s_k} - \gamma \mathtt{m}_{\mathtt{s_k}})} \mathsf{e}^{(-\beta \epsilon_{r_\mathtt{j}} - \gamma \mathtt{m}_{\mathtt{r_j}})} \mathsf{e}^{(-\beta' \epsilon_{s_l} - \gamma' \mathtt{m}_{\mathtt{s_l}})} \mathsf{e}^{(-\beta' \epsilon_{r_i} - \gamma' \mathtt{m}_{\mathtt{r_i}})} \right)$$

$$\langle s_p r_h | \Omega | s_{p'} r_{h'} \rangle \langle s_{p'} r_{h'} | O^\dagger | s_p r_h \rangle \tag{3.38}$$

or

$$= \sum_{\binom{ph}{p'h'}} x^p x'^{p'} y^h y'^{h'} \sum_{\binom{s_p r_h}{s_{p'} r_{h'}}} \mathsf{e}^{-\beta U - \gamma M - \beta' U' - \gamma' M'} \langle s_p r_h | \Omega | s_{p'} r_{h'} \rangle \langle s_{p'} r_{h'} | O^\dagger | s_p r_h \rangle \tag{3.39}$$

where

$$U = \sum_{(k,\mathtt{j}=1)}^{ph} \epsilon_{s_k} + \epsilon_{r_\mathtt{j}} \quad \text{and} \quad M = \sum_{(k,\mathtt{j}=1)}^{ph} \mathtt{m}_{\mathtt{s_k}} + \mathtt{m}_{\mathtt{r_j}} \tag{3.40}$$

$$U' = \sum_{(l,i=1)}^{p'h'} \epsilon_{s_l} + \epsilon_{r_i} \quad \text{and} \quad M' = \sum_{(l,i=1)}^{p'h'} \mathtt{m}_{\mathtt{s_l}} + \mathtt{m}_{\mathtt{r_i}} \tag{3.41}$$

As a shorthand practical notation that includes the essential features of the above expressions one may define

$$\left\langle F_p F_h \Omega F'_p F'_h O^\dagger \right\rangle = (\Omega|O^\dagger) = \sum_{(12)(UM)} \langle s_p r_h \,|\Omega|\, s_{p'} r_{h'} \rangle\langle s_{p'} r_{h'} \,|O^\dagger|\, s_p r_h \rangle \tag{3.42}$$

$$= \sum_{(1)} \sum_{(2)} \mathsf{e}^{[UM]} \langle s_p r_h \,|\Omega|\, s_{p'} r_{h'} \rangle\langle s_{p'} r_{h'} \,|O^\dagger|\, s_p r_h \rangle \tag{3.43}$$

where

$$\sum_{(1)} = \sum_{\binom{ph}{p'h'}} x^p x'^{p'} y^h y'^{h'} \tag{3.44}$$

represents the sum over all possible numbers of excitons and

$$\sum_{(2)} = \sum_{\binom{s_p r_h}{s_{p'} r_{h'}}} \tag{3.45}$$

represents the sum over all configurations for a given exciton number and

$$\mathsf{e}^{[UM]} = \mathsf{e}^{-\beta U - \gamma M - \beta' U' - \gamma' M'} \ , \tag{3.46}$$

which is the non normalized gand canonical distribution function.

At last one may just drop the $s$'s and $r$'s and write

$$(\Omega|O^\dagger) = \sum_{(1)} \sum_{(2)} \mathsf{e}^{[UM]} \langle ph \,|\Omega|\, p'h' \rangle\langle p'h' \,|O^\dagger|\, ph \rangle, \tag{3.47}$$

which now has a precise meaning, where $|p,h\rangle$, $|p',h'\rangle$ represent *the possible configurations for given exciton numbers* $p$,$h$ and $p'$,$h'$.

In the case of the simple expected values of $\Omega$ the expressions are totally analogous,

$$\langle F_p F_h \Omega \rangle = \langle \Omega \rangle = \sum_{(12)(UM)} \langle s_p r_h \,|\Omega|\, s_p r_h \rangle = \sum_{(1)} \sum_{(2)} \mathsf{e}^{[UM]} \langle s_p r_h \,|\Omega|\, s_p r_h \rangle \tag{3.48}$$

with the two sums given by

$$\sum_{(1)} = \sum_{(ph)} x^p y^h \quad \text{and} \quad \sum_{(2)} = \sum_{(s_p r_h)} \tag{3.49}$$

and

$$\mathsf{e}^{[UM]} = \mathsf{e}^{-\beta U - \gamma M} \tag{3.50}$$

Again one may drop the $s$'s and $r$'s to obtain the simplified expression

$$(\Omega) = \sum_{(1)}\sum_{(ph)} \mathsf{e}^{[UM]} \langle ph|\Omega|ph\rangle \tag{3.51}$$

Now, it is clear that the application of $CAP$ on Eq. (3.44), for example, will produce a very large number of levels per unit energy on $\sum_2$ and permit the approximate replacement of the sum by an integral. The details of this procedure, its interpretation and consequences are analyzed in the next section.

# 4. The connection with the Laplace transform

From the definitions of Sec.2 it results that the nuclear excitation, $U$, is a parameter that varies between two finite extremes

$$E_{min} \leq U \leq E_{max} \tag{4.1}$$

where it takes a sequence of discrete values with degeneracy $\mathcal{D}(A,U,M)$, as defined by Eqs.(2.1) and (2.3), with $A$ interpreted as the total number of excitons,

$$A = n = p + h. \tag{4.2}$$

Similarly the total nuclear momentum also varies in a stepwise manner between $M_{min}$ and $M_{max}$ and any additive quantum number of the total system can be treated similarly.

As it is shown in Eq.(2.17), in expressions like (3.43) or (3.47) for given numbers ($p$,$h$), there is in general a subset of the configurations ($s_p r_h$) for which the quantum numbers ($U$,$M$) take the same values and, by definition, the number of elements of this subset is equal to the degeneracy of the corresponding nuclear state,

$$\left\{(s_p)_i, (r_h)_i\right\} = \left\{\{s_1, \cdots, s_p, r_1, \cdots, r_h\}_i,\ i \in \{1, \cdots, \mathcal{D}(A, U, M)\}\right\}. \tag{4.3}$$

Then, e. g., one can rewrite Eq.(3.48) as

$$\langle F_p F_h \Omega \rangle = \sum_{(12)} \mathsf{e}^{[UM]} \langle s_p r_h \,|\Omega|\, s_p r_h \rangle$$

$$= \sum_{(1)} \sum_{M=M_{min}}^{M_{max}} \sum_{U=E_{min}}^{E_{max}} \sum_{i=1}^{\mathcal{D}(A,U,M)} \mathsf{e}^{[UM]} \big\langle (s_p)_i (r_h)_i \,|\Omega|\, (s_p)_i (r_h)_i \big\rangle \tag{4.4}$$

From the phenomenological calculations with nuclear level densities[12] one knows that the cumulative number of nuclear states can be very high even for not very high excitations. Then, it is clear that the use of $CAP$ to replace the sum over $U$ by an integral in (4.4) is a reasonable procedure, although approximate, and one can explore this possibility using an ad hoc definition of nuclear density, inspired by the analysis of Sec.2.

Note that for each $U$ in the sum in the RHS of (4.4) there are $\mathcal{D}(A,U,M)$ states with the same energy and angular momentum, for which $\mathsf{e}^{[UM]}$ has the same value. Then, if $\delta U = (U - U_{prev})$ is the variation of the nuclear excitation between its present and "previous" value, the corresponding approximate nuclear density for each index $i$ on the last sum of the RHS of (4.4) will be

$$\omega(A,U,M) \approx \mathcal{D}(A,U,M)/\delta U = constant,$$

and

$$\mathcal{D}(A,U,M) = \sum_{(i=1)}^{\mathcal{D}(A,U,M)} (1) \approx \omega(A,U,M)\delta U = \int_{U_{prev}}^{U} \omega(A,U,M)(1)dU. \tag{4.5}$$

Then, replacing "(1)" by an arbitrary integrable function "(...)" and summing over all $U$ gives,

$$\sum_{(U=E_{min})}^{E_{max}} \sum_{(i=1)}^{\mathcal{D}(A,U,M)} (...) \approx \int_{U=E_{min}}^{E_{max}} \omega(A,U,M)(...)dU. \tag{4.6}$$

and the sum over configurations with given a number of excitons in (4.4) becomes

$$\sum_{M=M_{min}}^{M_{max}} \sum_{U=E_{min}}^{E_{max}} \sum_{i=1}^{\mathcal{D}(A,U,M)} \mathsf{e}^{[UM]} \approx \int_{U=E_{min}}^{E_{max}} \mathsf{e}^{-\beta U} \left( \sum_{(M=M_{min})}^{M_{max}} \omega(A,U,M) \mathsf{e}^{-\gamma M} \right) dU \tag{4.7}$$

where the definition (3.50) for $\mathsf{e}^{[UM]}$ was used.

Now for the nuclear excitation energies the minimum is the ground state corresponding to $E_{min} = 0$ and the maximum is unbounded and one can take,[1]

$$E_{max} \approx \infty, \text{ with good approximation,} \tag{4.8}$$

then (4.7) becomes the Laplace transform of the part of the integrand inside the parenthesis.

More generally, one can write (4.4) as

$$\langle F_p F_h \Omega\rangle = \sum_{(1)} \sum_{(M=M_{min})}^{M_{max}} \sum_{(U=0)}^{\infty} \sum_{(\alpha)} \mathsf{e}^{[UM]} d_\alpha(p,h,U,M) \langle\Omega\rangle_\alpha(p,h,U,M) \tag{4.9}$$

where $\alpha$ indicates all configurations for which "$\langle phUM|\Omega|phUM\rangle$" has the same value, i. e., the configurations degenerated with respect to the action of $\Omega$ or the observation of the physical quantity represented by $\Omega$, and "$d_\alpha(p,h,U,M)$" is the corresponding degeneracy of states, and

$$\sum_{(\alpha)} d_\alpha(A,U,M) = \mathcal{D}(A,U,M)\ . \tag{4.10}$$

Then, the expression corresponding to (4.7) is

$$\langle F_p F_h \Omega\rangle \approx \sum_{(1)} \sum_{(M=M_{min})}^{M_{max}} \int_0^\infty dU \mathsf{e}^{-\beta U-\gamma M} \sum_{(\alpha)} \omega_\alpha(p,h,U,M) \langle\Omega\rangle_\alpha(p,h,U,M) \tag{4.11}$$

where the nuclear density is approximately defined as

$$\omega_\alpha(A,U,M) \approx d_\alpha(p,h,U,M)/\delta U = constant,$$

for each $\alpha$, and one can rewrite

$$d_\alpha(p,h,U,M) \approx \omega_\alpha(A,U,M)\times(U-U_{prev}) \approx \omega_\alpha(A,U,M)\delta U\ , \tag{4.12}$$

which corresponds, for given number of excitons $(p,h)$, to

$$\sum_{(M=M_{min})}^{M_{max}} \int_0^\infty dU \mathsf{e}^{-\beta U-\gamma M} \sum_{(\alpha)} \omega_\alpha(p,h,U,M) \langle\Omega\rangle_\alpha(p,h,U,M)\ . \tag{4.13}$$

Therefore, by definition,

$$\langle F_p F_h \Omega\rangle \approx \mathcal{L}\left\{ \sum_{(M=M_{min})}^{M_{max}} \mathsf{e}^{-\gamma M} \sum_{(\alpha)} \omega_\alpha(p,h,U,M) \langle p\ h\ UM|\Omega|p\ h\ UM\rangle_\alpha \right\}, \tag{4.14}$$

where the symbol $\mathcal{L}\{X\}$ indicates the Laplace transform of $(X)$.

In particular, in the case of the unitary operator, $\Omega = \mathbf{1}$ , all expected values are equal "1" and $\langle F_p F_h \rangle$ becomes essentially the Laplace transform of the state density, then

$$\sum_{(M=M_{min})}^{M_{max}} \mathsf{e}^{-\gamma M} \omega(p,h,U,M) \approx \mathcal{L}^{-1}(\langle F_p F_h \rangle) = \mathcal{L}^{-1}\left( \prod_{\mu}(1+x_{p\mu}) \prod_{\nu}(1+x_{h\nu}) \right) . \qquad (4.15)$$

In Appendix I it is shown that this result is equivalent to the traditional one of (2.11) because the RHS of Eq.(2.11) can be interpreted, using *CAP*, as the *inverse* Laplace transform of the grand canonical generating function.[1, 3]

Therefore, the microscopic formalism yields the expression of the inverse Laplace transform of the expected values of the interacting operators *without having to actually evaluate it.*

## 4.1 Convolutions and discrete functions of the configurations

Regarding the idea of nuclear degeneracy, one notices that in (4.4) all the expected values "$\left\langle (s_p)_i (r_h)_i | \Omega | (s_p)_i (r_h)_i \right\rangle$" correspond to configurations with energy $U$ and total degeneracy $\mathcal{D}(A,U,M)$, which we will designate by $(u)$

$$(u) = \{(s_{pi}, r_{ri}); i = 1, \mathcal{D}(A,U,M)\} , \qquad (4.16)$$

with $A$=$n$=$p$+$h$.

If the operator $\Omega$ describes the measurement of the *spin* of the nuclear state, for example, it will have in general a sequence of different *discrete values* for the different elements of $(u)$ and the expected values, $\langle \Omega \rangle$, will also be degenerated, i. e., in general there will be more than one configuration for each value of the total spin. Then, one may write

$$\sum_{\substack{(u)\\(conf)}} \langle \Omega \rangle_u = \sum_{(i=1)}^{\mathcal{D}(A,U,M)} \left\langle (s_p)_i (r_h)_i | \Omega | (s_p)_i (r_h)_i \right\rangle = \sum_{(i=1)}^{\mathcal{D}(A,U,M)} \langle \Omega \rangle_i = \sum_{\substack{\beta=1\\(spins)}}^{\beta_{max}} \langle \Omega \rangle_\beta d_\beta(A,U,M) \qquad (4.17)$$

where $\{\beta = 1, \beta_{max}\}$ is a sequence of *integers* in biunivocal (one to one) correspondence with the discrete set of "spin values" $\left\{ \langle \Omega \rangle_\beta \right\}_\beta$ and $d_\beta(A,U,M$ is the "spin degeneracy" satisfying,

$$\sum_{(\beta)}^{\beta_{max}} d_\beta(A,U,M) = \mathcal{D}(A,U,M) \qquad (4.18)$$

This idea can be straightforwardly generalized to the "measurement" of a quantity that changes the total number of excitons, $A$=$n$=($p$+$h$), for example: $\Omega = \sum_{(\alpha)} O_\alpha a_\alpha^\dagger a_\alpha$, where "$\alpha$" indicates all possible exciton states of "particle" type. Then, the expected values in (4.4) would select all configurations in ($u$), Eq. (4.16), that have one sp-state "$\alpha$" in them.

As we have seen, the number of such configurations define their "degeneracy", for which the destruction of an sp-state "$\alpha$" has non null expected value. If this number is designated by $d_\alpha(A,U,M)$, then the resulting configuration after the "observation" of $a_\alpha$, the *intermediary state*, would have *the same degeneracy*, but the corresponding grid of nuclear energies and angular momenta would be displaced by $\epsilon_\alpha$ and $\mathtt{m}_\alpha$ respctively,

$$d_\alpha(A-1, U-\epsilon_\alpha, M-\mathtt{m}_\alpha) = d_\alpha(A, U, M) \ . \tag{4.19}$$

In this context the introduction of *CAP* naturally brings the idea of convolution between states of different levels of complexity as it transforms the sums of discrete sets of values times the respective degeneracies, Eq. (4.17), into integrals involving the corresponding nuclear densities.

The sums and integrals have their ranges defined by the idea of *available states*, i. e. the states that give non zero expected value for the measured quantity. For example, in (4.13) the density $\omega_\alpha(A,U,M)$ corresponds to available states associated with the operation or measurement of $\Omega$, in the subset of the grand canonical ensemble with given $A$, $U$ and $M$.

On the other hand, the set of configurations for each $\alpha$ is not directly connected with the nuclear excitation energy, but if the density of states is high all configurations with energies between two given energies, $U$ and $U_{prev}$, can be considered as having energy $U$ and the variation over $\alpha$ can be associated with densities corresponding to $U$.

In particular, when one sums over an sp-state "$\alpha$" the previous discussion shows that it could have some degeneracy and, therefore, it is not equivalent to the sum over $\epsilon_\alpha$. This idea is further developed in Appendix II where the sum over $\epsilon_\alpha$ is then reduced to a convolution of nuclear densities, i. e., the sp-energies have an *associated nuclear density* that, in the *CAP* limit, naturally defines a convolution with the nuclear density present in the expected values.

Notice that the sum over the total nuclear spin projection $M$, and the corresponding sums over single particle angular momenta, $\mathtt{m}_\alpha$, *cannot* be transformed in the same way as $U$ and $\epsilon_\alpha$, using *CAP*, because the interval between sucessive values of $M$ or $\mathtt{m}_\alpha$ is never smaller than 0.5 and therefore it cannot be considered as an infinitesimal even for a large number of configurations.

Next we present a brief application of the above formalism for the calculation of the momenta of the operator of one body transitions to illustrate the main techniques to evaluate the transition strengths (TS) of the PE nuclear Hamiltonian.

## 5. Brief discussion about the one body operator

This section presents a detailed calculation of the transition strengths of a generic one body operator using the direct microscopic formalism. The results are given either as discrete sums of the microscopic approach or, using $CAP$, as approximate convolutions involving the nuclear density of the intermediary state and the densities of sp-states.

The one body operator

$$O = \sum_{\binom{\alpha \in p}{\beta \in p}} O_{\alpha\beta} a^\dagger_\alpha a_\beta \ , \tag{5.1}$$

could describe a physical process involving a single particle transition like a $\gamma$ transition or the particle emission into the continuum in a direct statistical process.[8]

When the "particles" are redefined as excitons and the Model Space also includes "holes" it becomes necessary to take the contribution of the latter into account. In the EXM the particles and holes of the Model Space are considered as independent and complementary fields, in which the creation of one exciton can also be described as the anihilation of the complementary one. Then, the description of this pair of independent and complementary fields is achieved by replacing

$$a_\beta \ \text{ by } \ (a_\beta + b^\dagger_\beta) \qquad \text{and} \qquad a^\dagger_\alpha \ \text{ by } \ (a^\dagger_\alpha + b_\alpha) \ , \tag{5.2}$$

then (5.1) becomes,

$$O = \sum_{\binom{\alpha \in p}{\beta \in p}} O_{\alpha\beta} a^\dagger_\alpha a_\beta + \sum_{\binom{\alpha \in p}{\beta \in h}} O_{\alpha\beta} a^\dagger_\alpha b^\dagger_\beta + \sum_{\binom{\alpha \in h}{\beta \in p}} O_{\alpha\beta} b_\alpha a_\beta + \sum_{\binom{\alpha \in h}{\beta \in h}} O_{\alpha\beta} b_\alpha b^\dagger_\beta \ . \tag{5.3}$$

This expression is a little more complicated than it looks and some epistemological clarification is presented in Appendix III, where it is shown that (5.3) can be rewritten as

$$O = \underset{\text{(a)}}{\sum_{\binom{\alpha \in -p}{\beta \in \texttt{filled}-p}} O_{\alpha\beta} a^\dagger_\alpha a_\beta} + \underset{\text{(b)}}{\sum_{(\alpha \in \texttt{filled}-p)} O_\alpha a^\dagger_\alpha a_\alpha} + \underset{\text{(c)}}{\sum_{\binom{\alpha \in -p}{\beta \in -h}} O_{\alpha\beta} a^\dagger_\alpha b^\dagger_\beta} + \underset{\text{(d)}}{\sum_{\binom{\alpha \in \texttt{filled}-h}{\beta \in \texttt{filled}-p}} O_{\alpha\beta} b_\alpha a_\beta}$$

$$- \sum_{\binom{\alpha \in -h}{\beta \in -h}} O_{\alpha\beta} b^\dagger_\beta b_\alpha - \sum_{(\alpha \in \texttt{filled}-h)} b^\dagger_\alpha b_\alpha O_{\alpha\alpha} + O_0 \tag{5.4}$$

(e) (f) (g)

where,

$$O_0 = \sum_{(\alpha \in \mathtt{all}-h)} O_{\alpha\alpha} = \sum_{(\alpha,\beta \in \mathtt{all}-h)} \langle \alpha | O_{\alpha\beta} \delta_{\alpha,\beta} | \beta \rangle \ .$$

Then, if one considers, for example, the *number of particles operator*, $N$, in (5.1), all $O_{\alpha\beta}$ will be one and Eq. (5.4) gives

$$\langle N \rangle = \sum_{(\alpha \in \mathtt{filled}-p)} a^\dagger_\alpha a_\alpha - \sum_{(\alpha \in \mathtt{filled}-h)} b^\dagger_\alpha b_\alpha + \sum_{(\alpha \in \mathtt{all}-h)} (1) \ . \tag{5.5}$$

Then, using the letter $\mathsf{g}$ for the maximum number of "$p$" excitons and $b$ for the maximum number of "$h$" excitons in the Model Space, results

$$\langle ph | N | ph \rangle = p - h + b \ , \tag{5.6}$$

which is equal to the total number of already excited "$p$" sp-states ("filled" $p$ states) plus the total number of "$h$" sp-states that have not yet been excited ("empty" $h$ states).

In Appendix III it is shown that although the above result is true in the strictly combinatorial point of view, the correct physical interpretation of (5.5) is $\langle N \rangle = \mathsf{g}$, which is in agreement with the original idea expressed in (5.1) for $O_{\alpha\beta}$ equal to $\delta_{(\alpha\beta)}$.

## 5.1 Transitions that change the number of excitons

The terms of (5.4) that change the number of particles or holes of the initial state, $|ph\rangle$, in comparison with the intermediary states, $|p'h'\rangle$, are ($c$) and ($d$), corresponding to the description of pair creation and destruction in the Model Space.

The corresponding transition strengths can be calculated using the general algebra developed in Sec.3. For example, for the term ($c$) one has

$$(c|c^+) = \sum_{(12)(UM)} \sum_{(\alpha\beta)} |O_{\alpha\beta}|^2 \langle ph | a^+_\alpha b^+_\beta | p'h' \rangle\langle p'h' | b_\beta a_\alpha | ph \rangle \tag{5.7}$$

and initially one defines all the constraints for ($ph$) and ($p'h'$) determined by the sp-operators

$$= \sum_{(1,2,UM)} \sum_{(\alpha\beta)} |O_{\alpha\beta}|^2 \delta(p=\alpha)\delta(h=\beta)\delta(p' \neq \alpha)\delta(h' \neq \beta)\delta(p|p'+1)\delta(h|h'+1) \tag{5.8}$$

factorize the terms of $\mathtt{e}^{[U,M]}$ selected by the destruction operators,

$$= \sum_{(1)} \sum_{(\alpha\beta)} |O_{\alpha\beta}|^2 \left( \frac{x_{p\alpha} x_{h\beta}}{xy} \right) \sum_{\binom{p=\alpha,h=\beta}{p'\neq\alpha,h'\neq\beta}} \mathtt{e}^{[St]} \delta(p-1|p')\delta(h-1|h') \tag{5.9}$$

and use $CAP$ to obtain,

$$\stackrel{CAP}{\approx} \sum_{(ph)} \delta_{(p-1,p')} \delta_{(h-1,h')} \sum_{\binom{S,t}{S',t'}} \mathtt{e}^{[-\beta U - \beta U' - \gamma M - \gamma M']} \delta(U-S')\delta(M-t') \times \sum_{(\alpha\beta)} |O_{\alpha\beta}|^2 \sum_{(j,k)}^{dd'} (1) \ , \tag{5.10}$$

where

$$S = U - \epsilon_\alpha - \epsilon_\beta \ ; \quad t = M - \mathtt{m}_\alpha - \mathtt{m}_\beta \ ; \quad S' = U' + \epsilon_\alpha + \epsilon_\beta \quad \text{and} \quad t' = M' + \mathtt{m}_\alpha + \mathtt{m}_\beta \ . \tag{5.11}$$

Now one notices that the sum over the intermediary states reduces to *only one term* corresponding to the conditions between configurations $\delta(p-1|p')$ and $\delta(h-1|h')$, then

$$(c|c^\dagger) \approx \sum_{(1)} \sum_{(S,t)} \mathtt{e}^{[UM]} \sum_{(\alpha\beta)} O_{\alpha\beta} d(p-1\neq\alpha, h-1\neq\beta, S, t) \ , \tag{5.12}$$

or more explicitly

$$(c|c^\dagger) \approx \sum_{(1)} \sum_{(S,t)} \mathtt{e}^{[UM]} \sum_{(\alpha\beta)} O_{\alpha\beta} d(p-\alpha, h-\beta, S, t) \ . \tag{5.13}$$

In Appendix III it is shown that (5.13) can be rewritten as convolutions between nuclear densities and the densities of sp-states

$$(c|c^\dagger) \stackrel{CAP}{\approx} \sum_{(1)} \sum_{(S,t)} \mathtt{e}^{[UM]} \int d\epsilon \sum_{(\mathtt{m})} \omega(p-1, h-1, U-\epsilon, M-\mathtt{m}) O_\epsilon(1,1,\epsilon,\mathtt{m}) \ , \tag{5.14}$$

where $\epsilon{=}(\epsilon_1{+}\epsilon_2)$, $\mathtt{m}{=}(\mathtt{m}_\alpha{+}\mathtt{m}_\beta)$ and

$$O_\epsilon(1,1,\epsilon,\mathtt{m}) = \int_0^\epsilon d\epsilon_2 \sum_{(\mathtt{m}_\beta)} O_{\alpha\beta} \omega(1,0,\epsilon-\epsilon_2, \mathtt{m}-\mathtt{m}_\beta) \omega(0,1,\epsilon_2,\mathtt{m}_\beta) \ , \tag{5.15}$$

and the approximate definition of nuclear density of Sec.4 was used.

Similar expressions are valid for the other expected values and transition strengths. For term $(d)$ in (5.4) the procedure is analogous,

$$(d|d^+) = \sum_{(1,2,UM)} \sum_{(\alpha\beta)} \langle ph\,|\, b_\alpha a_\beta \,|\, p'h' \rangle\!\langle p'h' \,|\, a_\beta^+ b_\alpha^+ \,|\, ph \rangle \tag{5.16}$$

$$= \sum_{(12)(UM)} \sum_{(\alpha\beta)} \delta(p{\neq}\beta)\delta(h{\neq}\alpha)\delta\big(p' = \beta\big)\delta\big(h' = \alpha\big)\delta(p|p'-1)\delta(h|h'-1) \tag{5.17}$$

$$= \sum_{(1)} \sum_{(\alpha\beta)} \left(\frac{x'_{p\beta}x'_{h\alpha}}{xy}\right) \sum_{\binom{p'=\beta,h'=\alpha}{p\neq\beta,h\neq\alpha}} \mathsf{e}^{[St]}\delta(p|p'-1)\delta(h|h'-1) \tag{5.18}$$

and using the continuum approximation results,

$$(d|d^{+}) \overset{CAP}{\approx} \sum_{(1)} \sum_{(St)} e^{[UM]}\delta_{(p,p'-1)}\delta_{(h,h'-1)}\delta(U-S')\delta(M-t') \sum_{(\alpha\beta)} \sum_{(j,k)}^{dd'} (1) \ , \tag{5.19}$$

where

$$S = U + \epsilon_\alpha + \epsilon_\beta \qquad \text{and} \qquad t = M + \mathtt{m}_\alpha + \mathtt{m}_\beta \ , \tag{5.20}$$

$$S' = U' - \epsilon_\alpha - \epsilon_\beta \qquad \text{and} \qquad t' = M' - \mathtt{m}_\alpha - \mathtt{m}_\beta \ , \tag{5.21}$$

then

$$(d|d^{\dagger}) \approx \sum_{(1)} \sum_{(S,t)} \mathsf{e}^{[UM]} \sum_{(\alpha\beta)} O_{\alpha\beta} d(p+1{=}\ \beta, h+1{=}\ \alpha, S, t) \ , \tag{5.22}$$

or more explicitly

$$(d|d^{\dagger}) \approx \sum_{(1)} \sum_{(S,t)} \mathsf{e}^{[UM]} \sum_{(\alpha\beta)} O_{\alpha\beta} d(p+\beta, h+\alpha, U+\epsilon_\alpha+\epsilon_\beta, M+\mathtt{m}_\alpha+\mathtt{m}_\beta) \ , \tag{5.23}$$

where the sum over $(\alpha,\beta)$ can be approximated as

$$\sum_{(\alpha\beta)} O_{\alpha\beta} d(p+\beta, h+\alpha, U_{(\neq\alpha\neq\beta)}+\epsilon_\alpha+\epsilon_\beta, M_{(\neq\alpha\neq\beta)}+\mathtt{m}_\alpha+\mathtt{m}_\beta) \tag{5.24}$$

$$\overset{CAP}{\approx} O_{00}\omega(p,h,U,M) - \int d\epsilon \sum_{(\mathtt{m})} \omega(p-1,h-1,U-\epsilon,M-\mathtt{m}) O_\epsilon(1,1,\epsilon,\mathtt{m}) \ , \tag{5.25}$$

with

$$O_{00} = \int d\epsilon_1 d\epsilon_2 \sum_{(\mathtt{m}_\alpha \mathtt{m}_\beta)} O_{\alpha\beta}\omega(1,0,\epsilon_1,\mathtt{m}_\alpha)\omega(0,1,\epsilon_2,\mathtt{m}_\beta) \ , \tag{5.26}$$

and

$$O_\epsilon(1,1,\epsilon,\mathtt{m}) = \int d\epsilon_2 \sum_{(\mathtt{m}_\beta)} O_{\alpha\beta}\omega(1,0,\epsilon-\epsilon_2,\mathtt{m}-\mathtt{m}_\beta)\omega(0,1,\epsilon_2,\mathtt{m}_\beta) \ , \tag{5.27}$$

The transition strength for the term $(a)$ in (5.4) is given by

$$(a|a^+) = \sum_{(12)(UM)} \sum_{(\alpha\beta)} O_{\alpha\beta} \langle ph\,|a_\alpha^+ a_\beta|\,p'h'\rangle\langle p'h'\,|a_\beta^+ a_\alpha|\,ph\rangle \tag{5.28}$$

$$= \sum_{(12)(UM)} \sum_{(\alpha\beta)(\alpha\neq\beta)} O_{\alpha\beta}\delta(p{=}\alpha{\neq}\beta)\delta(p'{=}\beta{\neq}\alpha)\delta(p|p')\delta(h|h') \tag{5.29}$$

$$= \sum_{(1)} \sum_{(\alpha\beta)} O_{\alpha\beta} \left(\frac{x'_{p\beta}x_{p\alpha}}{x^2}\right) \sum_{\binom{p-1\neq\alpha\beta,h}{p'-1\neq\alpha\beta,h'}} \mathsf{e}^{[St]}\delta(p|p')\delta(h|h') \tag{5.30}$$

then

$$(a|a^+) \overset{CAP}{\approx} \sum_{(1)} \sum_{(St)} e^{[UM]}\delta_{(p,p')}\delta_{(h,h')}\delta(U-S')\delta(M-t') \sum_{(\alpha\beta)} O_{\alpha\beta} \sum_{(j,k)}^{dd'} (1) \ , \tag{5.31}$$

where

$$S = U-\epsilon_\alpha+\epsilon_\beta \ , \ \ t = M-\mathtt{m}_\alpha+\mathtt{m}_\beta \ , \ \ S' = U'+\epsilon_\alpha-\epsilon_\beta \text{ and } t' = M'+\mathtt{m}_\alpha-\mathtt{m}_\beta \ , \tag{5.32}$$

yielding

$$(a|a^\dagger) \approx \sum_{(1)} \sum_{(S,t)} \mathsf{e}^{[UM]} \sum_{(\alpha\beta)} O_{\alpha\beta} d(p-1+1{\neq}\alpha{=}\beta,h,S,t) \ , \tag{5.33}$$

or more explicitly

$$(a|a^\dagger) \approx \sum_{(1)} \sum_{(S,t)} \mathsf{e}^{[UM]} \sum_{(\alpha\beta)} O_{\alpha\beta} d(p-\alpha+\beta,h,U-\epsilon_\alpha+\epsilon_\beta,M-\mathtt{m}_\alpha+\mathtt{m}_\beta) \ , \tag{5.34}$$

which can be rewritten as (see Appendix II)

$$(a|a^\dagger) \approx \int_0^U d\epsilon_1 \sum_{(\mathtt{m}_\alpha)} O_\alpha \omega(p-1,h,U-\epsilon_1,M-\mathtt{m}_\alpha)\omega(1,0,\epsilon_1,\mathtt{m}_\alpha) \tag{5.35}$$

$$-\int_0^U d\epsilon \sum_{(\mathtt{m})} O_\epsilon(2,0,\epsilon,\mathtt{m})\omega(p-2,h,U-\epsilon,M-\mathtt{m}) \ , \tag{5.36}$$

where

$$O_\alpha = \left( \int d\epsilon_2 \sum_{(\mathtt{m}_\beta)} \omega(1,0,\epsilon_2,\mathtt{m}_\beta) O_{\alpha\beta} \right) , \tag{5.37}$$

$\epsilon{=}(\epsilon_1{+}\epsilon_2)$, $\mathtt{m}{=}(\mathtt{m}_\alpha{+}\mathtt{m}_\beta)$ and

$$O_\epsilon(2,0,\epsilon,\mathtt{m}) = \int_0^\epsilon d\epsilon_2 \sum_{(\mathtt{m}_\beta)} O_{\alpha\beta}\omega(1,0,\epsilon-\epsilon_2,\mathtt{m}-\mathtt{m}_\beta)\omega(1,0,\epsilon_2,\mathtt{m}_\beta) , \tag{5.38}$$

and the transition strength for the term ($\mathtt{e}$) in (5.4) is given by

$$(\mathtt{e}|\mathtt{e}^+) = \langle ph|b_\beta^+ b_\alpha|p'h'\rangle\langle p'h'|b_\alpha^+ b_\beta|ph\rangle , \tag{5.39}$$

$$\approx \sum_{(1)}\sum_{(S,t)} \mathtt{e}^{[UM]} \sum_{(\alpha\beta)} O_{\alpha\beta} d(p,h-1+1{\neq}\beta{=}\alpha,S,t) , \tag{5.40}$$

or

$$(\mathtt{e}|\mathtt{e}^\dagger) \approx \sum_{(1)}\sum_{(S,t)} \mathtt{e}^{[UM]} \sum_{(\alpha\beta)} O_{\alpha\beta} d(p,h+\alpha-\beta,U+\epsilon_\alpha-\epsilon_\beta,M+\mathtt{m}_\alpha-\mathtt{m}_\beta) . \tag{5.41}$$

which can be rewritten as

$$(\mathtt{e}|\mathtt{e}^\dagger) \approx \int_0^U d\epsilon_1 \sum_{(\mathtt{m}_\alpha)} O_\beta d(p,h-1,U-\epsilon_2,M-\mathtt{m}_\beta)\omega(0,1,\epsilon_2,\mathtt{m}_\beta) \tag{5.42}$$

$$-\int_0^U d\epsilon \sum_{(\mathtt{m})} O_\epsilon(0,2,\epsilon,\mathtt{m}) d(p,h-2,U-\epsilon,M-\mathtt{m}) , \tag{5.43}$$

where

$$O_\beta = \left( \int d\epsilon_1 \sum_{(\mathtt{m}_\alpha)} \omega(1,0,\epsilon_1,\mathtt{m}_\alpha) O_{\alpha\beta} \right) , \tag{5.44}$$

$\epsilon{=}(\epsilon_1{+}\epsilon_2)$, $\mathtt{m}{=}(\mathtt{m}_\alpha{+}\mathtt{m}_\beta)$ and

$$O_\epsilon(0,2,\epsilon,\mathtt{m}) = \int_0^\epsilon d\epsilon_1 \sum_{(\mathtt{m}_\alpha)} O_{\alpha\beta}\omega(0,1,\epsilon-\epsilon_1,\mathtt{m}-\mathtt{m}_\alpha)\omega(0,1,\epsilon_1,\mathtt{m}_\alpha) , \tag{5.45}$$

Similar results are obtained for the expressions involving the crossed products of the terms $(b)$, $(f)$ and $(g)$, For example the product $(b|g^+)$ yields,

$$(b|g^\dagger) = \sum_{\binom{12}{UM}}\sum_{(\alpha)} O_{\alpha\alpha} O_0^\star \langle ph|a_\alpha^+ a_\alpha|p'h'\rangle <p'h'|ph> , \tag{5.46}$$

$$= \sum_{(12)(UM)} \sum_{(\alpha)} O_{\alpha\alpha} O_0^\star \delta(p = \alpha) \delta\big(p' = \alpha\big) \delta(p|p') \delta(h|h') \tag{5.47}$$

$$= \sum_{(1)} \sum_{(\alpha)} O_{\alpha\alpha} O_0^\star \left( \frac{x_{p\alpha} x'_{p'\alpha}}{xy} \right) \sum_{\binom{p'=\beta, h'=\alpha}{p\neq\beta, h\neq\alpha}} \mathtt{e}^{[St]} \delta(p|p') \delta(h|h') \ , \tag{5.48}$$

then

$$(b|g^\dagger) \overset{CAP}{\approx} \sum_{(1)} \sum_{(St)} e^{[UM]} \delta_{(p,p')} \delta_{(h,h')} \delta(U - S') \delta(M - t') \sum_{(\alpha)} \sum_{(j,k)}^{dd'} (1) \ , \tag{5.49}$$

where

$$S = U \ , \qquad t = M \ , \qquad S' = U' \qquad \text{and} \qquad t' = M' \ , \tag{5.50}$$

giving

$$(b|g^\dagger) \approx \sum_{(1)} \sum_{(S,t)} \mathtt{e}^{[UM]} \sum_{(\alpha)} O_{\alpha\alpha} O_0^\star d(p{=}\ \alpha, h, U_\alpha, M_\alpha) \ , \tag{5.51}$$

or more explicitly

$$(b|g^\dagger) \approx \sum_{(1)} \sum_{(S,t)} \mathtt{e}^{[UM]} \sum_{(\alpha)} O_{\alpha\alpha} O_0^\star d(p - \alpha + \alpha, h, U - \epsilon_\alpha + \epsilon_\alpha, M - \mathtt{m}_\alpha + \mathtt{m}_\alpha) \ . \tag{5.52}$$

The total of the nonvanishing terms of the sum over $(\alpha)$ is equal to the number of excited "particle" sp-states in $|p'h'\rangle$ or $|ph\rangle$, i. e. "$p$", then if $O_{\alpha\alpha}$ is approximately the same for all $(\alpha)$, Eq. (5.51) reduces to

$$(b|g^\dagger) \approx \sum_{(1)} \sum_{(S,t)} \mathtt{e}^{[UM]} O_{\alpha\alpha} O_0^\star p d(p{=}\ \alpha, h, U_\alpha, M_\alpha) \tag{5.53}$$

# 6. The pre-equilibrium nuclear Hamiltonian

This section presents the analysis of the momenta of two-body operators, in which the transition strengths may describe the pre-equilibrium nuclear transitions such as those in the multi-step formalisms and semi-classical models.[8]

To distinguish between the "mean-field" and "residual" components of the total interaction one may add a one-body kinetic energy term to the total interacting potential and analyze the resulting moments of the Hamiltonian. Those invariant with respect to one body transitions correspond to the Hartree-Fock mean-field description of the nucleus, without particle emission and with fixed nuclear excitation, the remaining ones define the residual interaction responsible for the pre-equilibrium transitions.[13]

The full Hamiltonian that describes the interacting field of sp-states that compounds the nuclear system during the pre-equilibrium stage can be defined as[8]

$$H = \sum_{\binom{\alpha \in h}{\beta \in p}} t_{\alpha\beta} a^{\dagger}_{\alpha} a_{\beta} + \frac{1}{4} \sum_{\binom{\alpha\beta \in h}{\gamma\delta \in p}} V_{\alpha\beta\gamma\delta} a^{\dagger}_{\alpha} a^{\dagger}_{\beta} a_{\delta} a_{\gamma}, \tag{6.1}$$

which, after normal ordering the single particle operators for particles and holes, see Eq.(5.3), becomes

$$H = E_0 + H_{hf} + V_{res} \tag{6.2}$$

where the ground state energy is

$$E_0 = \sum_{(\alpha \in p)} t_{\alpha\alpha} + \frac{1}{2} \sum_{(\alpha\beta \in p)} V_{\alpha\beta\alpha\beta}, \tag{6.3}$$

and the Hartree-Fock Hamiltonian is

$$H_{hf} = \sum_{(\alpha\beta \in p)} h_{\alpha\beta} a^{+}_{\alpha} a_{\beta} + \sum_{\binom{\alpha \in p}{\beta \in h}} h_{\alpha\beta} b_{\alpha} a_{\beta} + \sum_{\binom{\alpha \in p}{\beta \in h}} h_{\alpha\beta} a^{+}_{\alpha} b^{+}_{\beta} - \sum_{(\alpha\beta \in h)} h_{\alpha\beta} b^{+}_{\beta} b_{\alpha}. \tag{6.4}$$

The condition of diagonalization of $H_{hf}$ (invariable nuclear excitation) is then equivalent to the diagonalization of $h_{\alpha\beta}$ which is expressed by

$$h_{\alpha\beta} = \delta_{\alpha\beta} \epsilon_{\alpha}, \tag{6.5}$$

where $\epsilon_{\alpha}$ is the single particle energy of the sp-state $|\alpha\rangle$ and the terms involving $b_{\alpha} a_{\beta}$ and $a^{+}_{\alpha} b^{+}_{\beta}$ have vanished because particles and holes are supposed to be independent fields and, therefore, cannot be associated with the same sp-state (see Appendix III for detailed comparison of the Model Space and physical interpretations). Equation (6.5) corresponds to the idea that in the Hartree-Fock description of many-body systems composed by "particles" and "holes" there is no creation or destruction of particle-hole pairs (ph-pairs), but only propagation of sp-states.[13]

Therefore, the consideration of possible change in the number of excitons is restricted to the action of the residual interaction, $V_{res}$ in (6.2), which is introduced as a perturbation over the mean-field description and it is not necessarily self-consistent as the Hartree-Fock part usually is.

Using (6.5) $H_{hf}$ becomes

$$H_{hf} = \sum_{(\alpha\in p)} \epsilon_\alpha a^+_\alpha a_\alpha - \sum_{(\beta\in h)} \epsilon_\beta b^+_\beta b_\beta \; . \tag{6.6}$$

and if the energies of "particles" are defined relative to $\epsilon_F$ as

$$\epsilon_p = \epsilon_\alpha - \epsilon_F \tag{6.7}$$

and for "holes"

$$\epsilon_h = \epsilon_F - \epsilon_\beta, \tag{6.8}$$

then the nuclear excitation energy becomes the simple sum of the $p$ and $h$ energies,

$$U = \sum_{(p,h)} (\epsilon_p + \epsilon_h). \tag{6.9}$$

Strictly speaking this definition of the nuclear excitation only makes sense if the numbers of particles and holes are the same, otherwise it will add an arbitrary phenomenological term proportional to $\epsilon_F$. This term can be interpreted as the total excitation at the origin of time, but it is not important in the present discussion.

The last term in the hamiltonian (6.2) is given by

$$\begin{aligned} V_{res} = \sum_{(\alpha\beta\gamma\delta)} \Big( & \underset{\text{(a)}}{\frac{1}{4} V_{\alpha\beta\gamma\delta} a^+_\alpha a^+_\beta b^+_\delta b^+_\gamma} + \underset{\text{(b)}}{\frac{1}{2} V_{\alpha\beta\gamma\delta} a^+_\alpha a^+_\beta b^+_\delta a_\gamma} + \underset{\text{(c)}}{\frac{1}{2} V_{\alpha\beta\gamma\delta} a^+_\alpha b^+_\delta b^+_\gamma b_\beta} + \\ & \underset{\text{(d)}}{\frac{1}{4} V_{\alpha\beta\gamma\delta} a^+_\alpha a^+_\beta a_\delta a_\gamma} - \underset{\text{(e)}}{V_{\alpha\beta\gamma\delta} a^+_\alpha b^+_\delta b_\beta a_\gamma} + \underset{\text{(f)}}{\frac{1}{4} V_{\alpha\beta\gamma\delta} b^+_\delta b^+_\gamma b_\alpha b_\beta} + \\ & \underset{\text{(g)}}{\frac{1}{2} V_{\alpha\beta\gamma\delta} a^+_\alpha b_\beta a_\delta a_\gamma} + \underset{\text{(h)}}{\frac{1}{2} V_{\alpha\beta\gamma\delta} b^+_\delta b_\alpha b_\beta a_\gamma} + \underset{\text{(i)}}{\frac{1}{4} V_{\alpha\beta\gamma\delta} b_\alpha b_\beta a_\delta a_\gamma} \Big), \end{aligned} \tag{6.10}$$

where the various letters below each term will be used for specific reference in the following calculations.

## 6.1 Evaluation of the first moments of $H$

This section shows a detailed calculation of the expected values of the various terms of $H$ in (6.2).

Initially one notices that from (3.51) and (4.15), with $\Omega{=}F_pF_h$, comes

$$\langle F_pF_h\rangle = \sum_{(1)}\sum_{(2)}\mathtt{e}^{[UM]} = \prod_{\mu}(1+x_{p\mu})\prod_{\nu}(1{+}x_{h\nu}) = f_p(x,y)f_h(x,y) = f(x,y)\ , \tag{6.11}$$

then, for the first term of $H_{hf}$ in (6.6) one has

$$\langle a\rangle = \sum_{(12)(UM)}\sum_{(\alpha)}\epsilon_\alpha\langle ph\,|\,a^+_\alpha a_\alpha|\,ph\rangle = \sum_{(12)(UM)}\sum_{(\alpha)}\epsilon_\alpha\delta(p=\alpha) = \sum_{(1)}\sum_{(\alpha)}\sum_{(p=\alpha,h)}\mathtt{e}^{[U_\alpha M_\alpha]}\epsilon_\alpha \tag{6.12}$$

$$= \sum_{(1)}\sum_{(\alpha)}\Big(\frac{x_{p\alpha}}{x}\Big)\sum_{(p-1\neq\alpha,h)}\mathtt{e}^{[St]}\epsilon_\alpha = \sum_{(p,h)}y^hx^p\sum_{(\alpha)}\Big(\frac{x_{p\alpha}}{x}\Big)\sum_{(p-1\neq\alpha,h)}\mathtt{e}^{[St]}\epsilon_\alpha \tag{6.13}$$

where $(p{=}\alpha)$ represents all configurations with $p$ particles in which the sp-state $(\alpha)$ is present,

$$S = U_\alpha - \epsilon_\alpha \qquad \text{and} \qquad t = M_\alpha - \mathtt{m}_\alpha \tag{6.14}$$

and the domains (grids) of $U_\alpha$ and $M_\alpha$ are obtained from the domains of $U$ and $M$ by selecting only the energies and angular momenta of the $(p{=}\alpha)$ configurations. When writing the first expression of (6.13) the following important relation was used

$$\sum_{(p-1\neq\alpha)}(\cdots) = \sum_{(p=\alpha)}(\cdots)\ , \tag{6.15}$$

which is correct in the sense that the sets of configurations in both sums are in bijective correspondence with each other (i. e., for each configuration in the sum of the LHS corresponds one and only one configuration on the RHS) with the same degeneracies.

Therefore, the following relation between the degeneracies of $(p\text{-}1{\neq}\alpha)$ and $(p{=}\alpha)$ is also valid,

$$d(p-1{\neq}\alpha,h,S,t) = d(p-1{\neq}\alpha,h,U_\alpha-\epsilon_\alpha,M_\alpha-\mathtt{m}_\alpha) = d(p=\alpha,h,U_\alpha,M_\alpha)\ , \tag{6.16}$$

which express the invariability of a configuration in which a given sp-state $(\alpha)$ is destroyed and immediately re-created aftwerwards. The number of configurations and the corresponding degeneracies *decrease* by the destruction of $(\alpha)$ but they are kept invariable by its subsequent re-creation. It is usual to write simply $(U\text{-}\epsilon_\alpha)$ for the energy argument of the degeneracy on the LHS instead of ($U_\alpha$-$\epsilon_\alpha$), but this may lead to inconsistencies when comparing the detailed microscopic grid of the various energies.

Then, using (6.11) results

$$\langle a\rangle = \sum_{(\alpha)} x_{p\alpha} \sum_{(p-1,h)} x^{p-1} y^h \sum_{(p-1\neq\alpha,h)} \mathsf{e}^{[St]}\epsilon_\alpha = \sum_{(\alpha)} x_{p\alpha} \frac{\prod_\mu(1+x_{p\mu})\prod_\nu(1+x_{h\nu})}{1+x_{p\alpha}}\epsilon_\alpha \ . \tag{6.17}$$

From (6.15) one can also write

$$\langle a\rangle = \sum_{(1)}\sum_{(\alpha)}\sum_{(St)} \mathsf{e}^{[U_\alpha M_\alpha]}\epsilon_\alpha \sum_{(\mathrm{j}=1)}^{d} (1) \overset{CAP}{\approx} \sum_{(1)}\sum_{(St)} \mathsf{e}^{[St]} \sum_{(\alpha)} \mathsf{e}^{-\beta\epsilon_\alpha - \mathtt{m}_\alpha}\epsilon_\alpha d(p-1\neq\alpha, h, S, t) \ , \tag{6.18}$$

$$= \sum_{(1)}\sum_{(U_\alpha M_\alpha)} \mathsf{e}^{[U_\alpha M_\alpha]} \sum_{(\alpha)} \epsilon_\alpha d(p=\alpha, h, U_\alpha M_\alpha) \ , \tag{6.19}$$

where the approximation used in the last expression is that, under *CAP*, the difference in the domains of the various $S$ and $t$ as a function of ($\alpha$) can be neglected, then

$$\sum_{(\alpha)}\sum_{(St)} \approx \sum_{(St)}\sum_{(\alpha)} \tag{6.20}$$

and the identity

$$\sum_{(St)}(\cdots) = \sum_{(U_\alpha M_\alpha)}(\cdots) \tag{6.21}$$

has been used, wihch is a direct consequence of (6.16), again due to the bijective correspondence between the domains of ($S$,$t$) and ($U_\alpha$,$M_\alpha$). This correspondence is not changed by *CAP*. Expression (6.20) is the *CAP* correspondent of (6.13) and (6.19) is the correspondent of (6.14).

In Appendix IV we analyze some additional aspects of these results and some problems that may appear if the Laplace transform approach is used in this case.

For the second term of (6.6) the calculation is analogous and gives

$$\langle b\rangle = \sum_{(\beta)} x_{h\beta} \sum_{(p,h)} y^{h-1} x^p \sum_{(p,h\neq\beta)} \mathsf{e}^{[St]}\epsilon_\beta = \sum_{(\beta)} x_{h\beta} \frac{\prod_\mu(1+x_{p\mu})\prod_\nu(1+x_{h\nu})}{1+x_{h\beta}}\epsilon_\beta \ . \tag{6.22}$$

Therefore,

$$\langle b\rangle = \sum_{(1)}\sum_{(\beta)}\sum_{(St)} \mathsf{e}^{[UM]}\epsilon_\beta \sum_{(\mathrm{j}=1)}^{d} (1) \approx \sum_{(1)}\sum_{(St)} \mathsf{e}^{[St]} \sum_{(\beta)} \mathsf{e}^{-\beta\epsilon_\beta - \mathtt{m}_\beta}\epsilon_\beta d(p, h-1\neq\beta, S, t) \ , \tag{6.23}$$

where

$$S = U - \epsilon_\beta \qquad \text{and} \qquad t = M - \mathtt{m}_\beta \tag{6.24}$$

Notice that, under $CAP$ and according to Appendix IV, the sp-state exponential terms in $\epsilon_\alpha$ and $\epsilon_\beta$ in (6.18) and (6.23), respectively, should be "absorbed" into the integral of the Laplace transform if the remaining terms are to be interpreted as the inverse Laplace transforms of the corresponding expected values.

In the residual interaction of Eq.(6.10) the only terms that have non null expected values are those indicated by $(d)$, $(e)$ and $(f)$. The calculation is again straightforward, for example for $(d)$ one obtains

$$\langle d\rangle = \sum_{\binom{12}{UM}} \sum_{(\alpha\beta\delta\gamma)} V_{\alpha\beta\gamma\delta} \langle ph\,|\,a^+_\alpha a^+_\beta a_\delta a_\gamma\,|\,ph\rangle \tag{6.25}$$

$$= \sum_{\binom{12}{UM}} \sum_{(\alpha\beta\delta\gamma)} V_{\alpha\beta\gamma\delta}\delta(p=\delta\gamma{\neq}\alpha\beta)(\delta_{\alpha,\delta}\delta_{\beta,\gamma}+\delta_{\alpha,\gamma}\delta_{\beta,\delta}) \tag{6.26}$$

$$= \sum_{(1)} \sum_{(\alpha\beta\delta\gamma)} V_{\alpha\beta\gamma\delta}\Big(\frac{x_{p\delta}x_{p\gamma}}{x^2}\Big) \sum_{(p-2\neq\alpha\beta\delta\gamma,h)} \mathsf{e}^{[St]}(\delta_{\alpha,\delta}\delta_{\beta,\gamma}+\delta_{\alpha,\gamma}\delta_{\beta,\delta}) \tag{6.27}$$

$$= \sum_{(p,h)} y^h x^p \sum_{(\alpha\beta\delta\gamma)} V_{\alpha\beta\gamma\delta}\Big(\frac{x_{p\delta}x_{p\gamma}}{x^2}\Big)(\delta_{\alpha,\delta}\delta_{\beta,\gamma}+\delta_{\alpha,\gamma}\delta_{\beta,\delta}) \sum_{(p-2\neq\alpha\beta\delta\gamma,h)} \mathsf{e}^{[St]} \tag{6.28}$$

$$= \sum_{(p,h)} y^h x^p \sum_{(\alpha\beta)} V_{\alpha\beta\alpha\beta}\Big(\Big(\frac{x_{p\alpha}x_{p\beta}}{x^2}\Big)+\Big(\frac{x_{p\beta}x_{p\alpha}}{x^2}\Big)\Big) \sum_{(p-2\neq\alpha\beta,h)} \mathsf{e}^{[St]} \tag{6.29}$$

$$= 2\sum_{(\alpha\beta)} V_{\alpha\beta\alpha\beta}(x_{p\alpha}x_{p\beta}) \sum_{(p-2,h)} y^h x^{p-2} \sum_{(p-2\neq\alpha\beta,h)} \mathsf{e}^{[St]} \tag{6.30}$$

$$= 2\sum_{(\alpha\beta)} V_{\alpha\beta\alpha\beta}(x_{p\alpha}x_{p\beta}) \frac{\prod_\mu(1+x_{p\mu})\prod_\nu(1{+}x_{h\nu})}{(1+x_{p\alpha})(1+x_{p\beta})}\ . \tag{6.31}$$

The term corresponding to $\langle e\rangle$ is

$$\langle e\rangle = \sum_{\binom{12}{UM}} \sum_{(\alpha\beta\delta\gamma)} V_{\alpha\beta\gamma\delta} \langle ph\,|\,a^+_\alpha b^+_\delta b_\beta a_\gamma\,|\,ph\rangle \tag{6.32}$$

$$= \sum_{\binom{12}{UM}} \sum_{(\alpha\beta\delta\gamma)} V_{\alpha\beta\gamma\delta}\delta(p = \gamma \neq \alpha, h = \beta \neq \delta)(\delta_{\alpha,\gamma}\delta_{\beta,\delta}) \tag{6.33}$$

$$= \sum_{(p,h)} y^h x^p \sum_{(\alpha\beta\delta\gamma)} V_{\alpha\beta\gamma\delta}\left(\frac{x_{p\gamma}x_{h\beta}}{xy}\right)(\delta_{\alpha,\gamma}\delta_{\beta,\delta}) \sum_{(p\neq\alpha\gamma,h\neq\beta\delta)} \mathsf{e}^{[St]} \tag{6.34}$$

$$= \sum_{(\alpha\beta)} V_{\alpha\beta\alpha\beta}(x_{p\alpha}x_{h\beta}) \sum_{(p,h)} y^{h-1}x^{p-1} \sum_{(p\neq\alpha,h\neq\beta)} \mathsf{e}^{[St]} \tag{6.35}$$

$$= \sum_{(\alpha\beta)} V_{\alpha\beta\alpha\beta}(x_{p\alpha}x_{h\beta}) \frac{\prod_\mu(1 + x_{p\mu})\prod_\nu(1+x_{h\nu})}{(1 + x_{p\alpha})(1 + x_{h\beta})}, \tag{6.36}$$

and the last non null term in (6.10) is

$$\langle f \rangle = \sum_{\binom{12}{UM}} \sum_{(\alpha\beta\delta\gamma)} V_{\alpha\beta\gamma\delta}\langle ph\,|b^+_\delta b^+_\gamma b_\alpha b_\beta|\,ph\rangle \tag{6.37}$$

$$= \sum_{\binom{12}{UM}} \sum_{(\alpha\beta\delta\gamma)} V_{\alpha\beta\gamma\delta}\delta(p, h = \alpha\beta \neq \delta\gamma)(\delta_{\alpha,\delta}\delta_{\beta,\gamma} + \delta_{\alpha,\gamma}\delta_{\beta,\delta}) \tag{6.38}$$

$$= \sum_{(p,h)} y^h x^p \sum_{(\alpha\beta\delta\gamma)} V_{\alpha\beta\gamma\delta}\left(\frac{x_{h\alpha}x_{h\beta}}{y^2}\right)(\delta_{\alpha,\delta}\delta_{\beta,\gamma} + \delta_{\alpha,\gamma}\delta_{\beta,\delta}) \sum_{(p,h\neq\alpha\beta\delta\gamma)} \mathsf{e}^{[St]} \tag{6.39}$$

$$= 2\sum_{(\alpha\beta)} V_{\alpha\beta\alpha\beta}(x_{h\alpha}x_{h\beta}) \sum_{(p,h)} y^{h-2}x^p \sum_{(p,h\neq\alpha\beta)} \mathsf{e}^{[St]} \tag{6.40}$$

$$= 2\sum_{(\alpha\beta)} V_{\alpha\beta\alpha\beta}(x_{h\alpha}x_{h\beta}) \frac{\prod_\mu(1 + x_{p\mu})\prod_\nu(1+x_{h\nu})}{(1 + x_{h\alpha})(1 + x_{h\beta})}, \tag{6.41}$$

If one invokes $CAP$ over the expressions of the expected values and follows a reasoning similar to Sec.4 they become the Laplace transforms of the operators' expectated values multiplied by the corresponding degeneracies. Consequently, these linear combinations of nuclear degeneracies can be identified with the inverse Laplace transform of the above expressions. This can be achieved by simply rearranging the sums intead of doing the expansion of $\sum_{(1)}$.

For example, from (6.27) one has

$$\langle d\rangle = \sum_{(1)}\sum_{(\alpha\beta)} V_{\alpha\beta\alpha\beta}\left(\frac{2x_{p\alpha}x_{p\beta}}{x^2}\right)\sum_{(p-2\neq\alpha\beta,h)} \mathtt{e}^{[St]} \tag{6.42}$$

and instead of expanding $\sum_{(1)}$ one may write

$$\langle d\rangle = 2\sum_{(1)}\sum_{(\alpha\beta)} V_{\alpha\beta\alpha\beta}\left(\frac{x_{p\alpha}x_{p\beta}}{x^2}\right)\sum_{(p-2\neq\alpha\beta,h)} \mathtt{e}^{[St]} \approx 2\sum_{(1)}\sum_{(St)} \mathtt{e}^{\left[U_{\alpha\beta}M_{\alpha\beta}\right]}\sum_{(\alpha\beta)} V_{\alpha\beta\alpha\beta}\sum_{(j)}^{d}(1)\ , \tag{6.43}$$

where $d$ is the degeneracy associated with the energy $S$ and angular momentum $t$ of the configurations selected by the destruction of sp-states, given by

$$S = U_{\alpha\beta} - \epsilon_\alpha - \epsilon_\beta \qquad \text{and} \qquad t = M_{\alpha\beta} - \mathtt{m}_\alpha - \mathtt{m}_\beta \tag{6.44}$$

and $U_{\alpha\beta}$ and $M_{\alpha\beta}$ are the energy and angular momentum before the destruction of $(\alpha)$ and $(\beta)$ of the selected configurations in which $(\alpha)$ and $(\beta)$ are simultaneously present. Then,

$$\langle d\rangle \approx 2\sum_{(1)}\sum_{(U_{\alpha\beta}M_{\alpha\beta})} \mathtt{e}^{\left[U_{\alpha\beta}M_{\alpha\beta}\right]}\sum_{(\alpha\beta)} V_{\alpha\beta\alpha\beta}\, d(p-2\neq\alpha\beta,h,S,t)\ . \tag{6.45}$$

$$= 2\sum_{(1)}\sum_{(U_{\alpha\beta}M_{\alpha\beta})} \mathtt{e}^{\left[U_{\alpha\beta}M_{\alpha\beta}\right]}\sum_{(\alpha\beta)} V_{\alpha\beta\alpha\beta}\, d(p=\alpha\beta,h,U_{\alpha\beta},M_{\alpha\beta})\ . \tag{6.46}$$

where the identity

$$d(p-2\neq\alpha\beta,h,S,t) = d(p=\alpha\beta,h,U_{\alpha\beta},M_{\alpha\beta}) \tag{6.47}$$

was used.

Note that (6.45) can be rewritten as

$$\langle d\rangle \approx \sum_{(p,h,U)} y^h x^p \mathtt{e}^{[-\beta U]} 2\sum_{(\alpha\beta)} V_{\alpha\beta\alpha\beta}\sum_{(M_{\alpha\beta})} \mathtt{e}^{-\gamma M_{\alpha\beta}} d(p-2\neq\alpha\beta,h,S,t)\ , \tag{6.48}$$

where the sum over $U$ can be approximated as the Laplace transform of the last two sums in the *CAP* limit.

Analogously, the calculation of $\langle f\rangle$ yields

$$\langle f\rangle \approx 2\sum_{(1)}\sum_{(UM)}\mathtt{e}^{[UM]}\sum_{(\alpha\beta)}V_{\alpha\beta\alpha\beta}d(p,h-2{\neq}\alpha\beta,S,t)\ ,$$

$$= 2\sum_{(1)}\sum_{(UM)}\mathtt{e}^{[UM]}\sum_{(\alpha\beta)}V_{\alpha\beta\alpha\beta}d(p,h-2{\neq}\alpha\beta,U-\epsilon_\alpha-\epsilon_\beta,M-\mathtt{m}_\alpha-\mathtt{m}_\beta)\ , \tag{6.49}$$

with $S$ and $t$ given by (6.45) and ($U$,$M$) were used as a simplified notation for ($U_{\alpha\beta}$,$M_{\alpha\beta}$).

Due to the use of $CAP$ in all approximations, the last sums in (6.46), (6.48) and (6.49) can be interpreted as the inverse Laplace transform of the corresponding sums over all sp-states of the expected values of the residual interaction weighted by the degeneracies for given energy and angular momentum. These sums can be used as the definition of the first moments of the corresponding operators.[11]

## 6.2 Evaluation of the second moments of $H$

The second momenta of the Hamiltonian are also called transition strengths and are connected with the transition rates in pre-equilibrium models.[6, 7]

Some of these terms do not have a dependence on the excitation energy involving all processes with sp-states and therefore are not "on-shell" processes. These "virtual" processes are already taken into account by Brueckner's theory and one should be careful with the possible double counting of these contributions when calculating the transition strengths.[8, 14]

Due to the greater complexity involved in these calculations in comparison with the first momenta, the detailed calculation of $(a|a^+)$ will be presented next in a subsection of its own.

### 6.2.1 Terms that increases the number of $p$ and $h$ by 2

In Eq.(6.10) the term that increases the number of particles and holes by 2 is ($a$). The corresponding expansion is similar to what was done in the previous section,

$$(a|a^+) = \sum_{\binom{12}{UM}}\sum_{(\alpha\beta\delta\gamma)}\langle ph\,|\,a^+_\alpha a^+_\beta b^+_\delta b^+_\gamma|p'h'\rangle\langle p'h'\,|\,b_\gamma b_\delta a_\beta a_\alpha|ph\rangle \tag{6.50}$$

$$= \sum_{\binom{12}{UM}}\sum_{(\alpha\beta\delta\gamma)}|V_{\alpha\beta\gamma\delta}|^2\delta(p=\alpha\beta)\delta(h=\delta\gamma)\delta\big(p'{\neq}\alpha\beta\big)\delta\big(h'{\neq}\delta\gamma\big)\delta(p|p'+2)\delta(h|h'+2)\ . \tag{6.51}$$

Now one factorates the terms in $\mathtt{e}^{[UM]}$ that have been selected by the *destruction* of sp-states, to obtain a grand canonical distribution in terms of the energy and angular momentum of the intermediary state,

$$(a|a^{+}) = \sum_{(1)} \sum_{(\alpha\beta\delta\gamma)} |V_{\alpha\beta\gamma\delta}|^2 \left( \frac{x_{p\alpha}x_{p\beta}x_{h\delta}x_{h\gamma}}{x^2y^2} \right) \sum_{\binom{p=\alpha\beta, h=\delta\gamma}{p'\neq\alpha\beta, h'\neq\delta\gamma}} \mathtt{e}^{[St]}\delta(p|p'+2)\delta(h|h'+2) \; , \tag{6.52}$$

where

$$S = U - \epsilon_\alpha - \epsilon_\beta - \epsilon_\delta - \epsilon_\gamma \qquad \text{and} \qquad t = M - \mathtt{m}_\alpha - \mathtt{m}_\beta - \mathtt{m}_\delta - \mathtt{m}_\gamma \; , \tag{6.53}$$

expands $\sum_{(1)}$

$$(a|a^{+}) = \sum_{\binom{p,h}{p',h'}} y^h x^p y'^{h'} x'^{p'} \sum_{(\alpha\beta\delta\gamma)} |V_{\alpha\beta\gamma\delta}|^2 \left( \frac{x_{p\alpha}x_{p\beta}x_{h\delta}x_{h\gamma}}{x^2y^2} \right) \sum_{\binom{p=\alpha\beta, h=\delta\gamma}{p'\neq\alpha\beta, h'\neq\delta\gamma}} \mathtt{e}^{[St]}\delta(p|p'+2)\delta(h|h'+2) \tag{6.54}$$

and, similarly to (6.18), changes the sums over ($p$=$\alpha\beta$) and ($h$=$\delta\gamma$) into sums over ($p$-2$\neq\alpha\beta$) and ($h$-2$\neq\delta\gamma$), as suggested by the new distribution in terms of $S$ and $t$,

$$= \sum_{(\alpha\beta\delta\gamma)} |V_{\alpha\beta\gamma\delta}|^2 (x_{p\alpha}x_{p\beta}x_{h\delta}x_{h\gamma}) \sum_{\binom{p,h}{p',h'}} y^{h-2} y'^{h'} x^{p-2} x'^{p'} \sum_{\binom{p-2\neq\alpha\beta, h-2\neq\delta\gamma}{p'\neq\alpha\beta, h'\neq\delta\gamma}} \mathtt{e}^{[St]}\delta(p-2|p')\delta(h-2|h') \tag{6.55}$$

$$= \sum_{(\alpha\beta\delta\gamma)} |V_{\alpha\beta\gamma\delta}|^2 (x_{p\alpha}x_{p\beta}x_{h\delta}x_{h\gamma}) \sum_{\binom{p-2}{h-2}} (yy')^{(h-2)} (xx')^{(p-2)} \sum_{\binom{p-2\neq\alpha\beta}{h-2\neq\delta\gamma}} \mathtt{e}^{[St]} \tag{6.56}$$

$$= \sum_{(\alpha\beta\delta\gamma)} |V_{\alpha\beta\gamma\delta}|^2 (x_{p\alpha}x_{p\beta}x_{h\delta}x_{h\gamma}) \frac{\prod_\mu \left(1 + x_{p\mu}x'_{p'\mu}\right) \prod_\nu (1 + x_{h\nu}x'_{h'\nu})}{(1 + x_{p\alpha}x'_{p'\alpha})(1 + x_{p\beta}x'_{p'\beta})(1 + x_{h\delta}x'_{h'\delta})(1 + x_{h\gamma}x'_{h'\gamma})} \; . \tag{6.57}$$

As we have seen before, this expression can be considered, under *CAP*, as the Laplace transform of the expected value of the residual interaction times the corresponding degeneracy, for each ($\alpha$,$\beta$,$\gamma$,$\delta$).

To obtain the expression of the corresponding inverse Laplace transform one may procede as in (6.43), then from (6.52) comes

$$(a^{+}_\alpha a^{+}_\beta b^{+}_\delta b^{+}_\gamma | b_\gamma b_\delta a_\beta a_\alpha) = \sum_{(1)} \sum_{(\alpha\beta\delta\gamma)} |V_{\alpha\beta\gamma\delta}|^2 \left( \frac{x_{p\alpha}x_{p\beta}x_{h\delta}x_{h\gamma}}{x^2y^2} \right) \sum_{\binom{p-2\neq\alpha\beta, h-2\neq\delta\gamma}{p'\neq\alpha\beta, h'\neq\delta\gamma}} \mathtt{e}^{[St]}\delta(p|p'+2)\delta(h|h'+2) \tag{6.58}$$

$$= \sum_{(1)} \delta_{(p,p'+2)}\delta_{(h,h'+2)} \sum_{\binom{St}{S't'}} \mathtt{e}^{[UM]} \sum_{(\alpha\beta\delta\gamma)} |V_{\alpha\beta\gamma\delta}|^2 \delta(U-S')\delta(M-t') \sum_{(j,k)}^{dd'}(1) \ , \tag{6.59}$$

where $S$=$U'$=$U$-($\epsilon_\alpha$+$\epsilon_\beta$+$\epsilon_\delta$+$\epsilon_\gamma$) , $t$=$M'$=$M$-($\mathtt{m}_\alpha$+$\mathtt{m}_\beta$+$\mathtt{m}_\delta$+$\mathtt{m}_\gamma$) and

$$S' = U' + \epsilon_\alpha + \epsilon_\beta + \epsilon_\delta + \epsilon_\gamma \qquad \text{and} \qquad t' = M' + \mathtt{m}_\alpha + \mathtt{m}_\beta + \mathtt{m}_\delta + \mathtt{m}_\gamma \ . \tag{6.60}$$

Here ($U$,$M$) are the energy and momentum of the selected initial configuration in which the sp-states ($\alpha$,$\beta$,$\gamma$,$\delta$) are present and ($U'$,$M'$) are the energy and momentum corresponding to the selected intermediary configuration in which the same sp-states are supposed to not be present. Therefore, the domains of ($U$,$M$) and ($U'$,$M'$) have the "same number of points"[1] and correspond to a simple displacement of each other.

The identity between the energies, ($U$,$U'$) and ($S'$,$S$), and the angular momenta, ($M$,$M'$) and ($t'$,$t$), is a consequence of the definition of ($a|a^\dagger$) if one is not considering *off-shell* transitions.

Due to the supposed orthonormality of the single particle basis, for each configuration $|p-2,h-2\rangle$ associated with the energy $S$ and momentum $t$, one cannot have more than one $|p'h'\rangle$ satisfying the condition $<p-2,h-2|p'h'>=1$. Therefore, the last sum over $(j=1,d')$ is equal to 1 and the entire sum over the intermediary states $|p'h'\rangle$ reduces to only one term.

On the other hand, the sum over $(k=1,d)$ yields the nuclear degeneracy associated with ($p$,$h$,$S$,$t$) and the expression of ($a|a^+$) can be rewritten as

$$(a|a^+) = \sum_{(1)}\sum_{(St)} \mathtt{e}^{[UM]} \sum_{(\alpha\beta\delta\gamma)} |V_{\alpha\beta\gamma\delta}|^2 d(p-2{\neq}\alpha\beta, h-2{\neq}\delta\gamma, S, t) \ , \tag{6.61}$$

where now

$$\mathtt{e}^{[U,M]} = \mathtt{e}^{[-\beta U-\gamma M-\beta U'-\gamma M']} \overset{CAP}{\approx} \mathtt{e}^{2[-\beta U-\gamma M]} \tag{6.62}$$

due to energy conservation, i. e., the probability distribution after the transition, in which the final nuclear state equal to the initial one, is the square of the grand canonical probability of existence of the initial state. The factor of 2 is non important in the $CAP$ limit and (6.61) becomes the usual Laplace transform expression as we saw in Sec.4 (equations (4.11) and (4.14)).

Equation (6.61) is analogous to the first moment expressions in (6.47) and (6.50). There are no essential differences in both cases because the present analysis reduces to on-shell processes in which the intermediary states do not play an important role on the calculated expected values.

[1]This is correct whether the domains are continuous or discrete grids.

In Appendix V one shows that (6.61) can be written as the following convolution involving densities of the nuclear states associated with the given degeneracy,

$$(a|a^{+}) \approx \sum_{(1)}\sum_{(St)} \mathtt{e}^{[UM]} \sum_{(\mathtt{m})} \int d\epsilon V(2,2,\epsilon,\mathtt{m})\, \omega(p-2,h-2,U-\epsilon,M-\mathtt{m}) \tag{6.63}$$

where $\epsilon{=}\epsilon_\alpha{+}\epsilon_\beta{+}\epsilon_\gamma{+}\epsilon_\delta$, $\mathtt{m}{=}\mathtt{m}_\alpha{+}\mathtt{m}_\beta{+}\mathtt{m}_\gamma{+}\mathtt{m}_\delta$ and

$$V(2,2,\epsilon,\mathtt{m}) = \sum_{(\mathtt{m_x},\mathtt{m_y},\mathtt{m}_\delta)} \int dx\, dy\, d\epsilon_4\, |V_{\alpha\beta\gamma\delta}|^2\, \omega(1,0,\epsilon-x,\mathtt{m}-\mathtt{m_x})$$

$$\times\, \omega(1,0,x-y,\mathtt{m_x}-\mathtt{m_y})\, \omega(0,1,y-\epsilon_4,\mathtt{m_y}-\mathtt{m}_\delta)\, \omega(0,1,\epsilon_4,\mathtt{m}_\delta) \tag{6.64}$$

with $y{=}\epsilon_3{+}\epsilon_4$, $x{=}\epsilon_2{+}y$, $\mathtt{m_x}{=}\mathtt{m}_\beta{+}\mathtt{m_y}$ and $\mathtt{m_y}{=}\mathtt{m}_\gamma{+}\mathtt{m}_\alpha$.

### 6.2.2 Terms that increases the number of $p$ and $h$ by one.

The strengths of the transitions that increase the number of both $p$ and $h$ by one correspond to the terms $(b)$ and $(c)$ in Eq.(6.10) and the corresponding terms that *decrease* $p$ and $h$ by one are (g) and $(h)$, respectively.

The transition strength corresponding to term $(b)$ can be calculated in a totally analogous fashion as for $(a|a^\dagger)$ in the previous section,

$$(b|b^\dagger) = \sum_{\binom{12}{UM}} \sum_{(\alpha\beta\delta\gamma)} |V_{\alpha\beta\gamma\delta}|^2 \langle ph\,|\,a^+_\alpha a^+_\beta b^+_\delta a_\gamma\,|\,p'h'\rangle\langle p'h'\,|\,a^+_\gamma b_\delta a_\beta a_\alpha\,|\,ph\rangle \tag{6.65}$$

$$= \sum_{\binom{12}{UM}} \sum_{(\alpha\beta\delta\gamma)} |V_{\alpha\beta\gamma\delta}|^2 \delta(p=\alpha\beta{\neq}\gamma)\delta(h=\delta)\delta\big(p'=\gamma{\neq}\alpha\beta\big)\delta\big(h'{\neq}\delta\big)\delta(p|p'+1)\delta(h|h'+1)\ . \tag{6.66}$$

Now one factorates the terms in $\mathtt{e}^{[UM]}$ that have been selected by the destruction of sp-states *both in* $(ph)$ *and* $(p'h')$, redefining the grand canonical distribution in terms of the energy and angular momentum of the intermediary state,

$$U' = S = (U-\epsilon_\alpha-\epsilon_\beta-\epsilon_\delta+\epsilon_\gamma) \qquad \text{and} \qquad M' = t = (M-\mathtt{m}_\alpha-\mathtt{m}_\beta-\mathtt{m}_\delta+\mathtt{m}_\gamma)\ , \tag{6.67}$$

then,

$$(b|b^\dagger) = \sum_{(1)} \sum_{(\alpha\beta\delta\gamma)} |V_{\alpha\beta\gamma\delta}|^2 \left(\frac{x_{p\alpha}x_{p\beta}x_{h\delta}x'_{p'\gamma}}{x^2 y x'}\right) \sum_{\binom{p=\alpha\beta\neq\gamma,h=\delta}{p'=\gamma\neq\alpha\beta,h'\neq\delta}} \mathtt{e}^{[St]}\delta(p-2|p'-1)\delta(h-1|h')\ . \tag{6.68}$$

Therefore, $(U,M)$ correspond to the selected initial configuration in which the sp-states $(\alpha,\beta,\delta)$ are present and $(\gamma)$ is not, and $(U',M')$ to the selected intermediary state in which the sp-state $(\gamma)$ is present and $(\alpha,\beta,\delta)$ are not.

Notice that the term $(x'_{p'\gamma}/x')=\mathsf{e}^{-\beta\epsilon_\gamma-\gamma\mathtt{m}_\gamma}$ is *not* present in $\mathsf{e}^{[UM]}$ because $(\gamma)$ is not supposed to belong to the initial configuration $(ph)$, but the conservation laws imply that $\epsilon_\gamma$ and $\mathtt{m}_\gamma$ must be added to $(U,M)$ to produce $(U',M')$. Then, to write the grand canonical distribution in terms of $(U',M')$ one must multiply the entire expression by $(x'_{p'\gamma}/x')$.

Now one expands $\sum_{(1)}$ obtaining

$$(b|b^\dagger)=\sum_{\binom{p,h}{p',h'}} y^h x^p y'^{h'} x'^{p'} \sum_{(\alpha\beta\delta\gamma)} |V_{\alpha\beta\gamma\delta}|^2 \left(\frac{x_{p\alpha}x_{p\beta}x_{h\delta}x'_{p'\gamma}}{x^2yx'}\right) \sum_{\binom{p=\alpha\beta\neq\gamma,h=\delta}{p'=\gamma\neq\alpha\beta,h'\neq\delta}} \mathsf{e}^{[St]}\delta(p-2|p'-1)\delta(h-1|h') \tag{6.69}$$

$$=\sum_{(\alpha\beta\delta\gamma)} |V_{\alpha\beta\gamma\delta}|^2 \left(x_{p\alpha}x_{p\beta}x_{h\delta}x'_{p'\gamma}\right) \sum_{\binom{p,h}{p',h'}} y^{h-1}x^{p-2}y'^{h'}x'^{p'-1} \sum_{\binom{p-2\neq\alpha\beta\gamma,h-1\neq\delta}{p'-1\neq\alpha\beta\gamma,h'\neq\delta}} \mathsf{e}^{[St]}\delta(p-2|p'-1)\delta(h-1|h') \tag{6.70}$$

and rewrite some terms a using the deltas

$$= \sum_{(\alpha\beta\delta\gamma)} |V_{\alpha\beta\gamma\delta}|^2 \left(x_{p\alpha}x_{p\beta}x_{h\delta}x'_{p'\gamma}\right) \sum_{\binom{p-2}{h-1}} (yy')^{(h-1)}(xx')^{(p-2)} \sum_{\binom{p-2\neq\alpha\beta\gamma}{h-1\neq\delta}} \mathsf{e}^{[St]}$$

$$= \sum_{(\alpha\beta\delta\gamma)} |V_{\alpha\beta\gamma\delta}|^2 \left(x_{p\alpha}x_{p\beta}x_{h\delta}x'_{p'\gamma}\right) \frac{\prod_\mu \left(1+x_{p\mu}x'_{p'\mu}\right)\prod_\nu \left(1+x_{h\nu}x'_{h'\nu}\right)}{\left(1+x_{p\alpha}x'_{p\alpha}\right)\left(1+x_{p\beta}x'_{p\beta}\right)\left(1+x_{h\delta}x'_{h\delta}\right)\left(1+x_{p\gamma}x'_{p\gamma}\right)} \tag{6.71}$$

To obtain the expression corresponding to the inverse Laplace-transform one procedes as in Eq. (6.58),

$$(a^+_\alpha a^+_\beta b^+_\delta a_\gamma|a^+_\gamma b_\delta a_\beta a_\alpha)=\sum_{(1)}\sum_{(\alpha\beta\delta\gamma)} |V_{\alpha\beta\gamma\delta}|^2 \left(\frac{x_{p\alpha}x_{p\beta}x_{h\delta}x'_{p'\gamma}}{x^2yx'}\right) \sum_{\binom{p=\alpha\beta\neq\gamma,h=\delta}{p'=\gamma\neq\alpha\beta,h'\neq\delta}} \mathsf{e}^{[St]}\delta(p|p'+1)\delta(h|h'+1) \tag{6.72}$$

$$\approx \sum_{(1)} \delta_{(p,p'+1)}\delta_{(h,h'+1)} \sum_{\binom{St}{S't'}} \mathsf{e}^{[UM]} \sum_{(\alpha\beta\delta\gamma)} |V_{\alpha\beta\gamma\delta}|^2 \delta(U-S')\delta(M-t') \sum_{(j,k)}^{dd'} (1)\ . \tag{6.73}$$

where

$$S' = \big(U' + \epsilon_\alpha + \epsilon_\beta + \epsilon_\delta - \epsilon_\gamma\big) \qquad \text{and} \qquad t' = \big(M' + \mathtt{m}_\alpha + \mathtt{m}_\beta + \mathtt{m}_\delta - \mathtt{m}_\gamma\big)\,. \tag{6.74}$$

As happened for $(a|a^\dagger)$, the identification of the energies, $(U,U')=(S',S)$, and the angular momenta, $(M,M')=(t',t)$, is a consequence of the definition of $(b|b^\dagger)$ where only on-shell processes occur, therefore, the sum over intermediary states $(p'h')$ reduces to a single factor of *one* due to the various constraints relating them to the initial and final states. Then,

$$(b|b^+) = \sum_{(1)}\sum_{(S,t)} \mathsf{e}^{[UM]} \sum_{(\alpha\beta\delta\gamma)} |V_{\alpha\beta\gamma\delta}|^2 d(p-2+1{\neq}\alpha\beta\gamma, h-1{\neq}\delta, S, t)\ . \tag{6.75}$$

Here $d(p-2+1{\neq}\alpha\beta\gamma, h-1{\neq}\delta, S, t)$ is the degeneracy associated with the configurations obtained by the destruction of 2 "p" sp-states followed by the creation of 1 "p" sp-state, *plus* the destruction of 1 "h" sp-state and (6.75) could also be written, more explicitly, as

$$(b|b^+) = \sum_{(1)}\sum_{(S,t)} \mathsf{e}^{[UM]} \sum_{(\alpha\beta\delta\gamma)} |V_{\alpha\beta\gamma\delta}|^2 d(p-\alpha-\beta+\gamma, h-\delta, S, t)\ . \tag{6.76}$$

One notices that "$p$"s and "$h$"s are created and destroyed independently and, although $d(p-2+1, h-1, S, t)$ is in general different of $d(p+1-2, h-1, S, t)$ as functions of $S$ and $t$, the sums of the degneracies over all configurations are the same,

$$NOC(p-2+1, h-1) = \sum_{(S,t)} d(p-2+1, h-1, S, t) \tag{6.77}$$

$$= \sum_{(S,t)} d(p+1-2, h-1, S, t) = NOC(p+1-2, h-1)\ , \tag{6.78}$$

where $NOC$ stands for "number of configurations".

Therefore, under $CAP$ it becomes reasonable to drop the explicit dependence on the sp-states in (6.75) and (6.76) and rewrite them as

$$(b|b^+) \overset{CAP}{\approx} \sum_{(1)}\sum_{(S,t)} \mathsf{e}^{[UM]} \sum_{(\alpha\beta\delta\gamma)} |V_{\alpha\beta\gamma\delta}|^2 d(p-2+1, h-1, S, t)\ , \tag{6.79}$$

where the contributions of terms with a given $(S,t)$ are the same even if they correspond to different sets of sp-states.

In Appendix V one shows that (6.79) can be rewritten as a convolution

$$(b|b^+) \approx \sum_{\binom{1}{S,t}} \mathsf{e}^{[UM]} \sum_{(\alpha\beta\delta\gamma)} |V_{\alpha\beta\gamma\delta}|^2 \big(d(p-\alpha-\beta, h-\delta, S, t) - d(p-\alpha-\beta-\gamma, h-\delta, S, t)\big) \tag{6.80}$$

$$= \sum_{\binom{1}{S,t}} \mathtt{e}^{[UM]}[\sum_{(\mathtt{m_x})}\int d\ eV_x(2,1,x,\mathtt{m_x})\,\omega(p-2,h-1,U-x,M-\mathtt{m_x})\ ,$$

$$-\sum_{(\mathtt{m})}\int d\epsilon V(2,2,\epsilon,\mathtt{m})\,\omega(p-2,h-2,U-\epsilon,M-\mathtt{m})]\ , \tag{6.81}$$

where $x{=}(\epsilon_1{+}\epsilon_2{+}\epsilon_4)$, $y{=}(\epsilon_2{+}\epsilon_4)$, $\mathtt{m_x}{=}(\mathtt{m}_\alpha{+}\mathtt{m}_\beta{+}\mathtt{m}_\delta)$ and $\mathtt{m_y}{=}(\mathtt{m}_\beta{+}\mathtt{m}_\delta)$,

$$V_x(2,1,x,\mathtt{m_x}) = \sum_{(\mathtt{m_y},\mathtt{m}_\delta)}\int dy\, d\epsilon_4\, V_{(\alpha\beta\delta)}\,\omega(1,0,x-y,\mathtt{m_x}-\mathtt{m_y})\,\omega(0,1,y-\epsilon_4,\mathtt{m_y}-\mathtt{m}_\delta)\,\omega(0,1,\epsilon_4,\mathtt{m}_\delta), \tag{6.82}$$

$$V_{(\alpha\beta\delta)} = \int d\epsilon_3 \sum_{(\mathtt{m}_\gamma)} |V_{\alpha\beta\gamma\delta}|^2\,\omega(1,0,\epsilon_3,\mathtt{m}_\gamma)\ , \tag{6.83}$$

and

$$V(2,2,\epsilon,\mathtt{m}) = \sum_{(\mathtt{m_x},\mathtt{m_y},\mathtt{m}_\delta)}\int dx\, dy\, d\epsilon_4\, |V_{\alpha\beta\gamma\delta}|^2\,\omega(1,0,\epsilon-x,\mathtt{m}-\mathtt{m_x})$$

$$\times\,\omega(1,0,x-y,\mathtt{m_x}-\mathtt{m_y})\,\omega(0,1,y-\epsilon_4,\mathtt{m_y}-\mathtt{m}_\delta)\,\omega(0,1,\epsilon_4,\mathtt{m}_\delta) \tag{6.84}$$

with $\epsilon{=}(\epsilon_1{+}\epsilon_2{+}\epsilon_3{+}\epsilon_4)$, $\mathtt{m}{=}(\mathtt{m}_\alpha{+}\mathtt{m}_\beta{+}\mathtt{m}_\gamma{+}\mathtt{m}_\delta)$, $x{=}(\epsilon_2{+}\epsilon_3{+}\epsilon_4)$, $\mathtt{m_x}{=}(\mathtt{m}_\beta{+}\mathtt{m}_\gamma{+}\mathtt{m}_\alpha)$, $y{=}(\epsilon_3{+}\epsilon_4)$ and $\mathtt{m_y}{=}(\mathtt{m}_\gamma{+}\mathtt{m}_\alpha)$.

## 6.3 Terms with off-shell contributions

The various terms in (6.30) are defined as sums over sp-states and in some cases both Hermitean conjugates of a given sp-operator, $a_{p\alpha}$ and $a^\dagger_{p\alpha}$, act on the same configuration and the corresponding sp-energies become absent from the final accounting of the energy conservation for the total transition. In these cases the process of creation and destruction of these excitons is off-shell with respect to the entire transition.

For example, in the case of the expression of $(b|b^+)$ in (6.69) with $\gamma$ replaced by $\alpha$ in $(b)$ and $(b+)$ one has

$$(b_{\gamma\to\alpha}|b^+_{\gamma\to\alpha}) = \sum_{\binom{12}{UM}}\sum_{(\alpha\beta\delta)} |V_{\alpha\beta\delta}|^2 \langle ph\,|\,a^+_\alpha a^+_\beta b^+_\delta a_\alpha|\,p'h'\rangle\!\langle p'h'\,|\,a^+_\alpha b_\delta a_\beta a_\alpha|\,ph\rangle \tag{6.85}$$

which gives the following constraints over configurations

$$= \sum_{\binom{12}{UM}} \sum_{(\alpha\beta\delta)} |V_{\alpha\beta\delta}|^2 \delta(p=\alpha\beta)\delta(h=\delta)\delta\big(p'=\alpha{\neq}\beta\big)\delta\big(h'{\neq}\delta\big)\delta(p|p'+1)\delta(h|h'+1)\ . \tag{6.86}$$

Now one factorates the terms in $\mathtt{e}^{[UM]}$ selected by the destruction of sp-states,

$$(b_{\gamma\to\alpha}|b^+_{\gamma\to\alpha}) = \sum_{(1)} \sum_{(\alpha\beta\delta)} |V_{\alpha\beta\delta}|^2 \left(\frac{x_{p\alpha}x_{p\beta}x_{h\delta}x'_{p'\alpha}}{x^2yx'}\right) \sum_{\binom{p=\alpha\beta,h=\delta}{p'=\alpha\neq\beta,h'\neq\delta}} \mathtt{e}^{[St]}\delta(p-2|p'-1)\delta(h-1|h')\ , \tag{6.87}$$

where

$$S = (U - \epsilon_\beta - \epsilon_\delta) \qquad \text{and} \qquad t = (M - \mathtt{m}_\beta - \mathtt{m}_\delta)\ , \tag{6.88}$$

and, therefore, the sp-energy and momentum ($\epsilon_\alpha$,$\mathtt{m}_\alpha$) *are not present* in the formal conditions of energy and momentum conservations.

Now one expands $\sum_{(1)}$ and rewrite the terms using the deltas,

$$= \sum_{\binom{p,h}{p',h'}} y^h x^p y'^{h'} x'^{p'} \sum_{(\alpha\beta\delta)} |V_{\alpha\beta\delta}|^2 \left(\frac{x_{p\alpha}x_{p\beta}x_{h\delta}x'_{p'\alpha}}{x^2yx'}\right) \sum_{\binom{p=\alpha\beta,h=\delta}{p'=\alpha\neq\beta,h'\neq\delta}} \mathtt{e}^{[St]}\delta(p-2|p'-1)\delta(h-1|h') \tag{6.89}$$

$$= \sum_{(\alpha\beta\delta)} |V_{\alpha\beta\delta}|^2 \big(x_{p\alpha}x_{p\beta}x_{h\delta}x'_{p'\alpha}\big) \sum_{\binom{p,h}{p',h'}} y^{h-1}x^{p-2}y'^{h'}x'^{p'-1} \sum_{\binom{p-2\neq\alpha\beta,h-1\neq\delta}{p'-1\neq\alpha\beta,h'\neq\delta}} \mathtt{e}^{[St]}\delta(p-2|p'-1)\delta(h-1|h'), \tag{6.90}$$

to obtain the Laplace transform of the expected values of these operators,

$$= \sum_{(\alpha\beta\delta)} |V_{\alpha\beta\delta}|^2 \big(x_{p\alpha}x_{p\beta}x_{h\delta}x'_{p'\alpha}\big) \sum_{\binom{p-2}{h-1}} (xx')^{(p-2)}(yy')^{(h-1)} \sum_{\binom{p-2\neq\alpha\beta}{h-1\neq\delta}} \mathtt{e}^{[St]}$$

$$= \sum_{(\alpha\beta\delta)} |V_{\alpha\beta\delta}|^2 \big(x_{p\alpha}x_{p\beta}x_{h\delta}x'_{p'\alpha}\big) \frac{\prod_\mu \Big(1+x_{p\mu}x'_{p'\mu}\Big)\prod_\nu \big(1+x_{h\nu}x'_{h'\nu}\big)}{\big(1+x_{p\alpha}x'_{p\alpha}\big)\Big(1+x_{p\beta}x'_{p\beta}\Big)\big(1+x_{h\delta}x'_{h\delta}\big)} \tag{6.91}$$

As before, the expression for the inverse Laplace-transform can be obtained by rearranging the terms and using *CAP*,

$$(a^+_\alpha a^+_\beta b^+_\delta a_\alpha|a^+_\alpha b_\delta a_\beta a_\alpha) = \sum_{(1)} \sum_{(\alpha\beta\delta)} |V_{\alpha\beta\delta}|^2 \left(\frac{x_{p\alpha}x_{p\beta}x_{h\delta}x'_{p'\alpha}}{x^2yx'}\right) \sum_{\binom{p-2\neq\alpha\beta,h-1\neq\delta}{p'-1\neq\alpha\beta,h'\neq\delta}} \mathtt{e}^{[St]}\delta(p|p'+1)\delta(h|h'+1) \tag{6.92}$$

$$\stackrel{CAP}{\approx}\sum_{(1)}\delta_{(p,p'+1)}\delta_{(h,h'+1)}\sum_{\binom{St}{S't'}}\mathtt{e}^{[UM]}\sum_{(\alpha\beta\delta)}|V_{\alpha\beta\delta}|^2\delta(U-S')\delta(M-t')\sum_{(j,k)}^{dd'}(1)\ , \tag{6.93}$$

where

$$S'=\big(U'+\epsilon_\beta+\epsilon_\delta\big)\qquad\text{and}\qquad t'=\big(M'+\mathtt{m}_\beta+\mathtt{m}_\delta\big)\,, \tag{6.94}$$

which reduces to

$$(b_{\gamma\to\alpha}|b^+_{\gamma\to\alpha})\approx\sum_{(1)}\sum_{(S,t)}\mathtt{e}^{[UM]}\sum_{(\alpha\beta\delta)}|V_{\alpha\beta\delta}|^2d(p-2+1{\neq}\beta,h-1{\neq}\delta,S,t)\ , \tag{6.95}$$

and can be written more explicitly as

$$(b_{\gamma\to\alpha}|b^+_{\gamma\to\alpha})\approx\sum_{(1)}\sum_{(S,t)}\mathtt{e}^{[UM]}\sum_{(\alpha\beta\delta)}|V_{\alpha\beta\delta}|^2d(p-\alpha-\beta+\alpha,h-\delta,S,t)\ , \tag{6.96}$$

In Appendix V one shows that (6.96) can be rewritten as a convolution

$$(b_{\gamma\to\alpha}|b^+_{\gamma\to\alpha})\approx\sum_{(1)}\sum_{(S,t)}\mathtt{e}^{[UM]}\int dx\sum_{(\mathtt{m_x})}V_x(2,1,x,\mathtt{m_x})\omega(p-2,h-1,U-x,M-\mathtt{m_x}) \tag{6.97}$$

where

$$V_x(2,1,x,\mathtt{m_x})=\int dyd\epsilon_3\sum_{(\mathtt{m_y}\mathtt{m}_\delta)}|V_{\alpha\beta\gamma\delta}|^2\omega(1,0,x-y,\mathtt{m_x}-\mathtt{m_y})$$

$$\times\omega(1,0,y-\epsilon_3,\mathtt{m_y}-\mathtt{m}_\delta)\,\omega(0,1,\epsilon_3,\mathtt{m}_\delta)\ , \tag{6.98}$$

and

$$x=\epsilon_\alpha+\epsilon_\beta+\epsilon_\delta\qquad\text{and}\qquad\mathtt{m_x}=\mathtt{m}_\alpha+\mathtt{m}_\beta+\mathtt{m}_\delta\ . \tag{6.99}$$

When acting over their corresponding Hermitean conjugates the sequences of sp operators do not need to be the exact conjugates of each other as it has happened so far, for example, the term obtained from the previous one with $\alpha$ replaced by $\gamma{\neq}\alpha$ in $(b+)$ is defined as

$$(b_{\gamma\to\alpha}|b^+_{\alpha\to\gamma})_{\gamma\neq\alpha}=\sum_{\binom{12}{UM}}\sum_{(\alpha\beta\delta\gamma)}|V_{\alpha\beta\gamma\delta}|^2\langle ph\,|a^\dagger_\alpha a^\dagger_\beta b^\dagger_\delta a_\alpha|p'h'\rangle\!\langle p'h'\,|a^\dagger_\gamma b_\delta a_\beta a_\gamma|ph\rangle\ , \tag{6.100}$$

and it can be treated similarly.

Notice that the selection of sp-states must correspond to the action of the Fock operators acting both on the nuclear "ket" and "bra" states, for example, in the present case the operation of $(a^\dagger_\alpha a^\dagger_\beta b^\dagger_\delta a_\alpha)$ on $\langle ph|$ produces the selection of the sp-state $\alpha$ which is not present due to the operation of $(a^\dagger_\gamma b_\delta a_\beta a_\gamma)$ on $|ph\rangle$, etc.. This is a simple rule that results from the conservation laws, because the intermediary state is unique.

Then, the selections are represented by the following constraints over configurations,

$$= \sum_{\binom{12}{UM}} \sum_{(\alpha\beta\delta\gamma)} |V_{\alpha\beta\gamma\delta}|^2 \delta(p = \gamma\beta\alpha)\delta(h = \delta)\delta\big(p' = \alpha\gamma{\neq}\beta\big)\delta\big(h'{\neq}\delta\big)\delta(p-2|p'-1)\delta(h-1|h') \; . \quad (6.101)$$

Now one factorates the terms in $\mathsf{e}^{[UM]}$ selected by the destruction of sp-states

$$= \sum_{(1)} \sum_{(\alpha\beta\delta\gamma)} |V_{\alpha\beta\gamma\delta}|^2 \left( \frac{x_{p\gamma} x_{p\beta} x_{p\alpha} x_{h\delta} x'_{p'\alpha} x'_{p'\gamma}}{x^3 y x'^2} \right) \sum_{\binom{p=\gamma\beta\alpha h=\delta}{p'=\alpha\gamma\neq\beta h'\neq\delta}} \mathsf{e}^{[St]} \delta(p-3|p'-2)\delta(h-1|h') \quad (6.102)$$

expand $\sum_{(1)}$ and rewrite some terms a using the deltas,

$$= \sum_{(\alpha\beta\delta\gamma)} |V_{\alpha\beta\gamma\delta}|^2 \big(x_{p\gamma} x_{p\beta} x_{p\alpha} x_{h\delta} x'_{p'\alpha} x'_{p'\gamma}\big) \sum_{\binom{p-3}{h-1}} (xx')^{(p-3)} (yy')^{(h-1)} \sum_{\binom{p-3\neq\alpha\beta\gamma}{h-1\neq\delta}} \mathsf{e}^{[St]}$$

$$= \sum_{(\alpha\beta\delta\gamma)} |V_{\alpha\beta\gamma\delta}|^2 \frac{(x_{p\gamma} x_{p\beta} x_{p\alpha} x_{h\delta} x'_{p'\alpha} x'_{p'\gamma}) \prod_\mu \Big(1 + x_{p\mu} x'_{p'\mu}\Big) \prod_\nu \big(1 + x_{h\nu} x'_{h'\nu}\big)}{\big(1 + x_{p\alpha} x'_{p\alpha}\big)\Big(1 + x_{p\beta} x'_{p\beta}\Big)\big(1 + x_{h\delta} x'_{h\delta}\big)\big(1 + x_{p\gamma} x'_{p\gamma}\big)} \; . \quad (6.103)$$

The expression corresponding to the inverse Laplace-transform is obtained as before,

$$(b|b^+)_{\gamma\neq\alpha} = \sum_{(1)} \sum_{(\alpha\beta\delta\gamma)} \left( \frac{x_{p\alpha} x'_{p'\alpha} x_{p\beta} x_{p\gamma} x'_{p'\gamma} x_{h\delta}}{x^3 y x'^2} \right) \sum_{\binom{p-3\neq\gamma\beta\alpha,\, h-1\neq\delta}{p'-2\neq\alpha\gamma\beta,\, h'\neq\delta}} \mathsf{e}^{[St]} \delta(p-3|p'-2)\delta(h-1|h') \quad (6.104)$$

$$= \sum_{(1)} \delta_{(p,p'+1)} \delta_{(h,h'+1)} \sum_{\binom{St}{S't'}} \mathsf{e}^{[UM]} \sum_{(\alpha\beta\delta\gamma)} \delta(U - S')\delta(M - t') \sum_{(j,k)}^{dd'} (1) \; . \quad (6.105)$$

which reduces to

$$(b|b^+)_{\gamma\neq\alpha} \approx \sum_{(1)} \sum_{(S,t)} \mathsf{e}^{[UM]} \sum_{(\alpha\beta\delta\gamma)} d(p - 3 + 2{\neq}\beta, h - 1{\neq}\delta, S, t) \; , \quad (6.106)$$

and can also be written as

$$(b|b^{+})_{\gamma\neq\alpha} \approx \sum_{(1)}\sum_{(S,t)} \mathsf{e}^{[UM]} \sum_{(\alpha\beta\delta\gamma)} d(p-\alpha-\gamma-\beta+\alpha+\gamma, h-\delta, S, t) \;, \tag{6.107}$$

where now both $\alpha$ and $\gamma$ are canceled out in the expressions of energy and angular momentum conservation,

$$S = (U - \epsilon_\beta - \epsilon_\delta) \qquad \text{and} \qquad t = (M - \mathtt{m}_\beta - \mathtt{m}_\delta) \tag{6.108}$$

and

$$S' = \big(U' + \epsilon_\beta + \epsilon_\delta\big) \qquad \text{and} \qquad t' = \big(M' + \mathtt{m}_\beta + \mathtt{m}_\delta\big)\,. \tag{6.109}$$

Therefore, the transitions to and from sp-states $\alpha$ and $\gamma$ are described as off-shell.

The degneracies in 6.75 and 6.95 should have similar magnitudes as functions of $U$ and $M$, but they sually would be considerably larger than (6.106), despite the fact that all them correspond to configurations with ($p$-1,$h$-1) excitons, which should have a direct reflex on the magnitudes of the corresponding transition strengths.

In Appendix V one shows that (6.106) can be rewritten as the following convolution,

$$(b_{\gamma\to\alpha}|b^{+}_{\alpha\to\gamma})_{\gamma\neq\alpha} \approx \int d\epsilon \sum_{(\mathtt{m})} V_\epsilon(3,1,\epsilon,\mathtt{m})\omega(p-3,h-1,U-\epsilon,M-\mathtt{m}) \tag{6.110}$$

where

$$V_\epsilon(3,1,\epsilon,\mathtt{m}) = \int dx dy d\epsilon_4 \sum_{\binom{\mathtt{m_x m_y}}{\mathtt{m}_\delta}} |V_{\alpha\beta\gamma\delta}|^2 \omega(1,0,\epsilon-x,\mathtt{m}-\mathtt{m_x})$$

$$\times\omega(1,0,x-y,\mathtt{m_x}-\mathtt{m_y})\,\omega(1,0,y-\epsilon_4,\mathtt{m_y}-\mathtt{m}_\delta)\,\omega(0,1,\epsilon_4,\mathtt{m}_\delta)\;, \tag{6.111}$$

$$\epsilon = \epsilon_\alpha + \epsilon_\beta + \epsilon_\gamma + \epsilon_\delta \qquad \text{and} \qquad \mathtt{m} = \mathtt{m}_\alpha + \mathtt{m}_\beta + \mathtt{m}_\gamma + \mathtt{m}_\delta\;, \tag{6.112}$$

$$x = \epsilon_\beta + \epsilon_\gamma + \epsilon_\delta \qquad \text{and} \qquad \mathtt{m_x} = \mathtt{m}_\beta + \mathtt{m}_\gamma + \mathtt{m}_\delta \tag{6.113}$$

and

$$y = \epsilon_\gamma + \epsilon_\delta \qquad \text{and} \qquad \mathtt{m_y} = \mathtt{m}_\gamma + \mathtt{m}_\delta\;, \tag{6.114}$$

## 6.4 Other terms with on-shell transitions

At this point it is clear that all transitions can be treated in essentially the same way to obtain the final expressions in term of the degneracies, with or without $CAP$.

The basic steps are the fatorization of the exponential terms associated with the sp-states destroyed in the transition and the rearrangement of the sums using the deltas to express them in terms of the quantities corresponding to the intermediary states. We will use this rule in the following analysis.

The term corresponding to $(c|c+)$ in (6.10) is

$$(c|c^{+}) = \sum_{\binom{12}{UM}} \sum_{(\alpha\beta\delta\gamma)} \langle ph\,|\,a^{\dagger}_{\alpha}b^{\dagger}_{\delta}b^{\dagger}_{\gamma}b_{\beta}\,|\,p'h'\rangle\langle p'h'\,|\,b^{\dagger}_{\beta}b_{\gamma}b_{\delta}a_{\alpha}\,|\,ph\rangle \tag{6.115}$$

which imply the following constraints for configurations,

$$(p=\alpha),(h=\delta\gamma\neq\beta),(h'=\beta\neq\delta\gamma),(p'\neq\alpha) \tag{6.116}$$

and the following terms factorized from $\mathsf{e}^{[UM]}$,

$$\left(x_{p\alpha}x_{h\delta}x_{h\gamma}x'_{h\beta}\right) . \tag{6.117}$$

Then, using $CAP$ the following final expression is obtained

$$(c|c^{+}) \approx \sum_{(1)}\sum_{(S,t)} \mathsf{e}^{[UM]} \sum_{(\alpha\beta\delta\gamma)} |V_{\alpha\beta\gamma\delta}|^2 d(p-1\neq\alpha, h-2+1\neq\beta\delta\gamma, S, t) , \tag{6.118}$$

or more explicitly,

$$(c|c^{+}) \approx \sum_{(1)}\sum_{(S,t)} \mathsf{e}^{[UM]} \sum_{(\alpha\beta\delta\gamma)} |V_{\alpha\beta\gamma\delta}|^2 d(p-\alpha, h-\delta-\gamma+\beta, S, t) , \tag{6.119}$$

where the interemediary energy and momentum are

$$U' = S = (U - \epsilon_\alpha - \epsilon_\delta - \epsilon_\gamma + \epsilon_\beta) \qquad \text{and} \qquad M' = t = (M - \mathtt{m}_\alpha - \mathtt{m}_\delta - \mathtt{m}_\gamma + \mathtt{m}_\beta) , \tag{6.120}$$

and one can show that (6.120) can be rewritten as the following convolution (see Appendix V)

$$(c|c^{+}) \approx \sum_{\binom{1}{S,t}} \mathsf{e}^{[UM]} \sum_{(\alpha\beta\delta\gamma)} |V_{\alpha\beta\gamma\delta}|^2 [d(p-\alpha, h-\delta-\gamma, S_1, t) - d(p-\alpha, h-\beta-\gamma-\delta, S_2, t)] \tag{6.121}$$

$$= \sum_{\binom{1}{S,t}} \mathtt{e}^{[UM]} [\sum_{(\mathtt{m_x})} \int d\ eV_x(1,2,x,\mathtt{m_x})\,\omega(p-1,h-2,U-x,M-\mathtt{m_x})\ ,$$

$$-\sum_{(\mathtt{m})} \int d\epsilon V(1,3,\epsilon,\mathtt{m})\,\omega(p-1,h-3,U-\epsilon,M-\mathtt{m})]\ , \tag{6.122}$$

where

$$S_1 = (U-\epsilon_\alpha-\epsilon_\gamma-\epsilon_\delta) \qquad \text{and} \qquad t_1 = (M-\mathtt{m}_\alpha-\mathtt{m}_\gamma-\mathtt{m}_\delta)\ , \tag{6.123}$$

$$S_2 = (U-\epsilon_\alpha-\epsilon_\beta-\epsilon_\gamma-\epsilon_\delta) \qquad \text{and} \qquad t_2 = (M-\mathtt{m}_\alpha-\mathtt{m}_\beta-\mathtt{m}_\gamma-\mathtt{m}_\delta)\ , \tag{6.124}$$

$x{=}(\epsilon_1{+}\epsilon_3{+}\epsilon_4)$, $y{=}(\epsilon_3{+}\epsilon_4)$, $\mathtt{m_x}{=}(\mathtt{m}_\alpha{+}\mathtt{m}_\gamma{+}\mathtt{m}_\delta)$ and $\mathtt{m_y}{=}(\mathtt{m}_\gamma{+}\mathtt{m}_\delta)$,

$$V_x(1,2,x,\mathtt{m_x}){=}\sum_{(\mathtt{m_y},\mathtt{m}_\delta)} \int dy\, d\epsilon_4\, V_{(\alpha\gamma\delta)}\,\omega(1,0,x-y,\mathtt{m_x}-\mathtt{m_y})\,\omega(0,1,y-\epsilon_4,\mathtt{m_y}-\mathtt{m}_\delta)\,\omega(0,1,\epsilon_4,\mathtt{m}_\delta), \tag{6.125}$$

$$V_{(\alpha\gamma\delta)} = \int d\epsilon_2 \sum_{(\mathtt{m}_\beta)} |V_{\alpha\beta\gamma\delta}|^2\,\omega(0,1,\epsilon_2,\mathtt{m}_\beta) = \sum_{(\beta)} |V_{\alpha\beta\gamma\delta}|^2\ , \tag{6.126}$$

and

$$V(1,3,\epsilon,\mathtt{m}) = \sum_{(\mathtt{m_x},\mathtt{m_y},\mathtt{m}_\delta)} \int dx\, dy\, d\epsilon_4\, |V_{\alpha\beta\gamma\delta}|^2\,\omega(1,0,\epsilon-x,\mathtt{m}-\mathtt{m_x})$$

$$\times\,\omega(0,1,x-y,\mathtt{m_x}-\mathtt{m_y})\,\omega(0,1,y-\epsilon_4,\mathtt{m_y}-\mathtt{m}_\delta)\,\omega(0,1,\epsilon_4,\mathtt{m}_\delta)\ , \tag{6.127}$$

with $\epsilon{=}(\epsilon_1{+}\epsilon_2{+}\epsilon_3{+}\epsilon_4)$, $\mathtt{m}{=}(\mathtt{m}_\alpha{+}\mathtt{m}_\beta{+}\mathtt{m}_\gamma{+}\mathtt{m}_\delta)$, $x{=}(\epsilon_2{+}\epsilon_3{+}\epsilon_4)$, $\mathtt{m_x}{=}(\mathtt{m}_\beta{+}\mathtt{m}_\gamma{+}\mathtt{m}_\alpha)$, $y{=}(\epsilon_3{+}\epsilon_4)$ and $\mathtt{m_y}{=}(\mathtt{m}_\gamma{+}\mathtt{m}_\alpha)$.

The terms corresponding to $(c|c^\dagger)_{\,\beta\neq\delta}$ with off-shell transitions for $(\beta)$ and $(\delta)$ in $c$ and $c^\dagger$ are

$$(c|c^\dagger)_{\,\beta\neq\delta} = \sum_{\binom{12}{UM}} \sum_{(\alpha\beta\delta\gamma)} \langle ph\,|\,a_\alpha^+ b_\beta^+ b_\gamma^+ b_\beta\,|\,p'h'\rangle\langle p'h'\,|\,b_\delta^+ b_\gamma b_\delta a_\alpha\,|\,ph\rangle \tag{6.128}$$

which imply the following constraints,

$$(p=\alpha), (h=\delta\gamma\beta), (h'=\beta\delta{\neq}\gamma), (p'{\neq}\alpha) \tag{6.129}$$

and the terms factorized from $\mathsf{e}^{[U,M]}$ are

$$\left(x_{p\alpha}x_{h\delta}x_{h\gamma}x_{h\beta}x'_{h\delta}x'_{h\beta}\right) , \tag{6.130}$$

giving the following expression in terms of the degeneracies of the intermediary states,

$$\sum_{(1)}\sum_{(S,t)}\mathsf{e}^{[UM]}\sum_{(\alpha\beta\delta\gamma)}|V_{\alpha\beta\gamma\delta}|^2 d(p-1{\neq}\alpha, h-3+2{\neq}\gamma, S, t) , \tag{6.131}$$

or

$$\sum_{(1)}\sum_{(S,t)}\mathsf{e}^{[UM]}\sum_{(\alpha\beta\delta\gamma)}|V_{\alpha\beta\gamma\delta}|^2 d(p-\alpha, h-\beta-\delta-\gamma+\beta+\delta, S, t) , \tag{6.132}$$

where

$$U' = S = (U-\epsilon_\alpha-\epsilon_\gamma) \qquad \text{and} \qquad M' = t = (M-\epsilon_\alpha-\epsilon_\gamma) . \tag{6.133}$$

Finally, equation (6.133) can be rewritten as,

$$(c|c^\dagger)_{\beta\neq\delta} \approx \sum_{((1),S,t)}\mathsf{e}^{[UM]}\int d\epsilon\sum_{(\mathtt{m})}V_\epsilon(1,3,\epsilon,\mathtt{m})\omega(p-1,h-3,U-\epsilon,M-\mathtt{m}) \tag{6.134}$$

where

$$V_\epsilon(1,3,\epsilon,\mathtt{m}) = \int dxdyd\epsilon_4\sum_{\binom{\mathtt{m_x m_y}}{\mathtt{m}_\delta}}|V_{\alpha\beta\gamma\delta}|^2\omega(1,0,\epsilon-x,\mathtt{m}-\mathtt{m_x})$$

$$\times\omega(0,1,x-y,\mathtt{m_x}-\mathtt{m_y})\,\omega(0,1,y-\epsilon_4,\mathtt{m_y}-\mathtt{m}_\delta)\,\omega(0,1,\epsilon_4,\mathtt{m}_\delta) , \tag{6.135}$$

$$\epsilon = \epsilon_\alpha+\epsilon_\beta+\epsilon_\gamma+\epsilon_\delta \qquad \text{and} \qquad \mathtt{m} = \mathtt{m}_\alpha+\mathtt{m}_\beta+\mathtt{m}_\gamma+\mathtt{m}_\delta , \tag{6.136}$$

$$x = \epsilon_\beta+\epsilon_\gamma+\epsilon_\delta \qquad \text{and} \qquad \mathtt{m_x} = \mathtt{m}_\beta+\mathtt{m}_\gamma+\mathtt{m}_\delta \tag{6.137}$$

and

$$y = \epsilon_\gamma+\epsilon_\delta \qquad \text{and} \qquad \mathtt{m_y} = \mathtt{m}_\gamma+\mathtt{m}_\delta . \tag{6.138}$$

# 7. Results and conclusions

We have shown in Appendix IV (see also related notes in App.I.1) that if one relies on the Laplace transform as the main formalism for the transition strengths (TS), some microscopic transitions are not correctly described, or not well defined. Therefore, a direct microscopic formalism is preferred.

The formalism presented in this paper is a consequence of the microscopic definitions of the first and second momenta of the nuclear Hamiltonian with a residual interaction term. It provides a direct and intuitive understanding of the TS (for given number of excitons and energy of nuclear state) and its dependence on the strengths of less complex states, which can be expressed in terms of *convolutions* of these functions, in a similar way to the traditional relations obtained for nuclear densities. These expressions have been used for the evaluation of the TS and the densities and compared with the corresponding parameters of the exciton model (EXM).

The Model Space was defined by arbitrarily fixing the Fermi level, $E_F$, as greater than or equal to the minimum energy necessary to accomodate all protons and neutrons of the system in the fundamental state, and by considering the levels of the basis above and below $E_F$, to describe the fields of holes and particles respectively, up to a maximum energy per exciton, $E_{\mathtt{max}}$. The latter was then fixed arbitrarily as a model parameter.

Various "sizes" of the Model Space were considered by varying $E_{\mathtt{max}}$ and $E_F$. For sake of simplicity a not too large nuclide was used, $^{40}$Ca, to permit to test the model with relatively small runtime computation, but even so the number of configurations for increasing $n$ increased very fast making it unpractical to perform direct calculations for $n$ greater than 4. Calculations with larger number of excitons, would probably need an approximate statistical approach and should be attempted only for larger nuclei, as the consideration of large $n$ in small nuclei is usually of little physical interest, or otherwise if a specific practical situation demands it.

For example, for $^{40}$Ca, "$n$=3" and using a single "personal computer" (PC) the number of configurations is of the order of $10^5$-$10^6$ with computation time of a few minutes up to a couple of hours, depending on the considered transition and the size of the Model Space. For "$n$=4" this number approximately doubles and the maximum runtime reaches many hours. For "$n$=5" the number of configurations can be 10 times greater and the estimated runtime would reach many days. In addition, the calculated strengths increase by two or more orders of magnitude for an increase of $\Delta n$=+1 making it more difficult the direct computation of the resulting large numbers while keeping the same numerical precision. Nonetheless, the results seem clear and meaningful enough even with 4 excitons or less and no more than that will be considered in the present analysis.

The idea of convolution numerically results from the fact that all strengths, for all numbers of excitons, energies and momenta can be obtained from a *core* of elements of the transition matrix with no constraint over the microscopic initial state. This permits to obtain a basic set of TS for the simplest configurations for which the transition is possible and the TS for states of higher excitations by doing the convolution of the basic TS with the adequate nuclear densities.

The core of matrix elements can be calculated independently for given maximum excitation $U_{\mathtt{max}}$, given $E_F$ and the assumptions about the single particle level density, then its components are used to obtain the transition strengths systematically for different number of excitons, $n$, as a function of $U$. In addition, the initial configuration is constrained by the necessary presence of the sp-states that will be "destroyed" during the collision process and the final by the exclusion principle for the newly created sp-states.

For example, in Eq.(6.122) it was shown that the sum of TS designated by $(c|c^{+})$ can be written as

$$(c|c^{+}) = \sum_{\binom{12}{UM}} \sum_{(\alpha\beta\delta\gamma)} \langle ph\,|a_{\alpha}^{\dagger} b_{\delta}^{\dagger} b_{\gamma}^{\dagger} b_{\beta}|p'h'\rangle\langle p'h'\,|b_{\beta}^{\dagger} b_{\gamma} b_{\delta} a_{\alpha}|ph\rangle \ ,$$

$$\approx \sum_{\binom{1}{S,t}} \mathtt{e}^{[UM]}[\sum_{(\mathtt{m_x})} \int dx V_x(1,2,x,\mathtt{m_x})\, \omega(p-1,h-2,U-x,M-\mathtt{m_x}) \ ,$$

$$-\sum_{(\mathtt{m})} \int d\epsilon V(1,3,\epsilon,\mathtt{m})\, \omega(p-1,h-3,U-\epsilon,M-\mathtt{m})] \ , \qquad (7.1)$$

where $V_x(1{,}2{,}x{,}\mathtt{m_x})$ and $V(1{,}3{,}\epsilon{,}\mathtt{m})$ are the TS corresponding to initial configurations with 3 excitons, ($p$=1,$h$=2), and 4 excitons, ($p$=1,$h$=3), respectively, which are the simplest possible configurations in this case. The strengths $V_x(1{,}2{,}x{,}\mathtt{m_x})$ and $V(1{,}3{,}\epsilon{,}\mathtt{m})$ are defined as sums over the elements of the core of transition matrix elements that are independent of $p$, $h$, $U$ and $M$, therefore, they are functions of the Model Space but not of the level of complexity of the excited nuclear state.

The densities that enter in the definition of the transition strengths are calculated independently either by combinatorial analysis, for the initial nuclear configurations, or by direct counting of the available states in connection with each valid transition.

Using this prescription, a computer code called TRANSNU[8] was created to calculate the TS. We also assumed a "billiard balls" model for the microscopic description of the system of excitons, i. e., a quasi-free movement of the excitons in the nuclear matter and interaction by a macroscopic-type "collision", i. e., a distance independent satationary redefinition of the local phase space of the interacting sp-states with conservation of energy and angular momentum.

Therefore, the collision was described by the simplest possible meachanism, which is the free propagation of the of the initial and final exciton states in the Fock space, followed by the destruction of part of them and the creation of the final ones constrained by the conservation laws and the exclusion principle. No assumption regarding the dependence of the interaction on the relative distance of the excitons during the collision was made, therefore, one may also regard this as a type of "black box" description for the microscopic two-body interaction.[15]

The Greeen function for the pair of "colliding" excitons was the propagator of the non interacting two-body Hamiltonian[16]

$$\frac{\mathrm{e}^{ik|\vec{r_1}-\vec{r_2}|}}{\vec{r_1}-\vec{r_2}} = k\sum_{(l=0)}^{\infty}(2l+1)\mathrm{j}_l(kr_<)h_l^{(+)}(kr_>)P_l(\cos\alpha) \tag{7.2}$$

and the sp-state basis was the set of Harmonic Oscillator (HO) wave-functions, following the prescription of Brussaard and Glaudemans[5] and as wave number for the stationary excitons in (7.2) we used the inverse of the characteristic length of the HO[5]

$$k = A^{(-1/6)}\ . \tag{7.3}$$

The transition matrix elements were then obtained by direct integration over the entire space of relative positions of the colliding pair of excitons.

## 7.1 Numerical results

Inasmuch as the HO basis can be considered as a non phenomenological assumption, the above definition for the propagation of the colliding excitons can be considered as non phenomenological. With this prescription, the integrals of the transition matrix often produced very large numbers due to the non normalization of the two-body propagator.

Then, taking into account the intrinsically hypothetical nature of the HO description, an *ad hoc* model parameter was introduced to obtain always finite matrix elements that could be compared among themselves as functions of the nuclear excitation, as well as with the analogous parameters of other models. In this sense the results we present now have only qualitative validity, but they can be made more physically meaningful by fitting of the adquate experimental data.

The various transitions of Eq.(6.10) were classified according to their $\Delta n$ and those with negative $\Delta n$ were neglected, in agreement with the usual procedure of semi-classical analyses. The transitions in which the number of excitons increase correspond to the terms $(b)$ and $(c)$ of (6.10), which have the same TS as their Hermitean conjugates, terms (g) and ($h$) respectively. In the numerical analysis these terms have been divided into the following subprocesses

| case | representation | description |
|---|---|---|
| 1 | (2100++1000) | proton-h-propagates and proton-ph-pair is created |
| 2 | (1011++1000) | proton-h-propagates and neutron-ph-pair is created |
| 3 | (1110++0010) | neutron-h-propagates and proton-ph-pair is created |
| 4 | (0021++0010) | neutron-h-propagates and neutron-ph-pair is created |
| 5 | (1200++0100) | proton-p-propagates and proton-ph-pair is created |
| 6 | (0111++0100) | proton-p-propagates and neutron-ph-pair is created |
| 7 | (1101++0001) | neutron-p-propagates and proton-ph-pair is created |
| 8 | (0012++0001) | neutron-p-propagates and neutron-ph-pair is created |

and the terms which keep the number of excitons constant in Eq.(6.10), i. e. ($d$), (e) and ($f$), were divided into the following subprocesses:

| case | representation | description |
|---|---|---|
| 9 | (1100++0011) | neutron-ph-pair is destroyed and proton-ph-pair is created |
| 10 | (2000++2000) | scattering of two proton-h excitons |
| 11 | (1010++1010) | scattering of a neutron-h and a proton-h |
| 12 | (0020++0020) | scattering of two neutron-h |
| 13 | (0200++0200) | scattering of two proton-p |
| 14 | (0101++0101) | scattering of a neutron-p and a proton-p |
| 15 | (0002++0002) | scattering of two neutron-p |
| 16 | (1100++1100) | scattering of a proton-ph pair |
| 17 | (1001++1001) | scattering of a proton-h-neutron-p pair |
| 18 | (0110++0110) | scattering of a neutron-h-proton-p pair |
| 19 | (0011++0011) | scattering of a neutron-ph pair |

In the following, the results of cases 1 to 8 are designated by VPL1 results. The 9th case also corresponds to a pair creation process and is designated isolately by VPL2. At last, the cases from 10 to 19 are collectively designated by V00. In the graphs shown below the contributions of VPL2 and V00 were added and compared with VPL1.

One may question the physical meaningfulness of the comparison of the microscopic calculations with the EXM, having in sight the important criticisms of the model made by Pompeia and Carlson.[17] In fact, many of the features of the EXM result from statistical approximations that can be more or less physically precise, specially regarding the hypothesis of the attainment of equal occupation probabilities for all configurations at each stage of the increasing complexity chain, i. e. the attainment of perfect configuration mixing in the ensemble of nuclear microstates or "equilibrium hypothesis" (EH).

One of the arguments against EH is that the transition rates (TR) of the exciton model ("$\lambda_0$" for transitions that keep the number of excitons invariable and "$\lambda_+$" for transitions that increase it by 2 units) indicate that "$\lambda_0$", as predicted by the EXM, is *not large enough*, in comparison with "$\lambda_+$", to warrant the achievement of equilibrium between any two steps of the chain. Using a Monte Carlo calculation it was determined that $\lambda_0$ should be at least three orders of magnitude greater than $\lambda_+$ to validate EH.[17]

On the other hand, the EXM and the hybrid model are two largely used tools in nuclear data evaluation due to their simplicity in terms of easiness of implementation and allowance for improvement in comparison with the observed data. Then, one may analyse the important parameters of the microscopic formalism by direct comparison with well established EXM results, like the TNG model code[12], to have an idea of the quality of the estimates of the present formulation.

The microscopic calculations using TRANSNU, for different exciton numbers and assumptions about the energy level structure of the sp-states (SPL), clearly show that for all exciton numbers and SPL densities the TS corresponding to the "$\lambda_+$" of the EXM, "$t_+$", can in fact be *more than 3 orders of magnitude* greater than "$t_0$" (the analogous of "$\lambda_0$" of the EXM) in the region of not very small and not very large excitations. Then, one of the basic arguments *against* the EH would be invalidated by the present microscopic calculations.

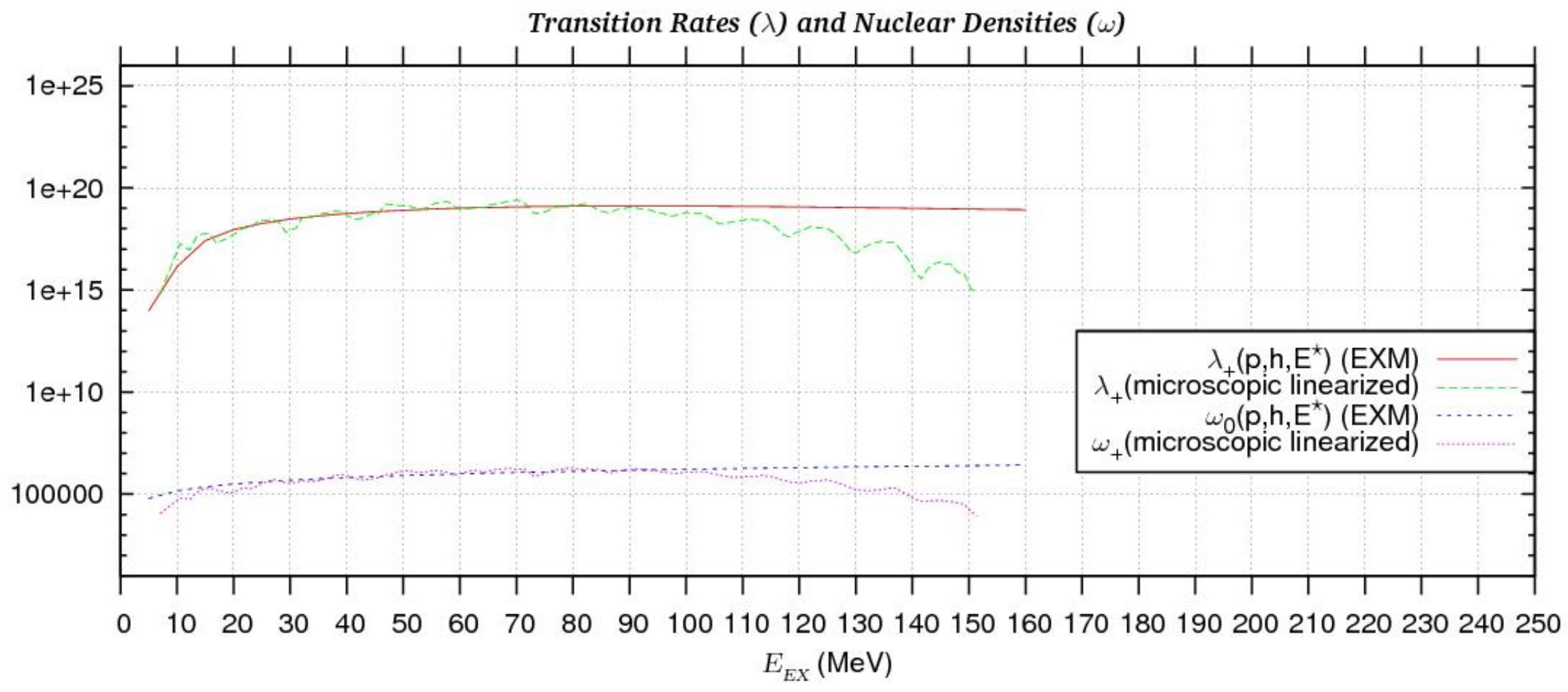

**Figure 1**. Transition strengths and nuclear densities as functions of the excitation energy for transitions that keep the number of excitons constant, for an initial configuration with $n$=3.

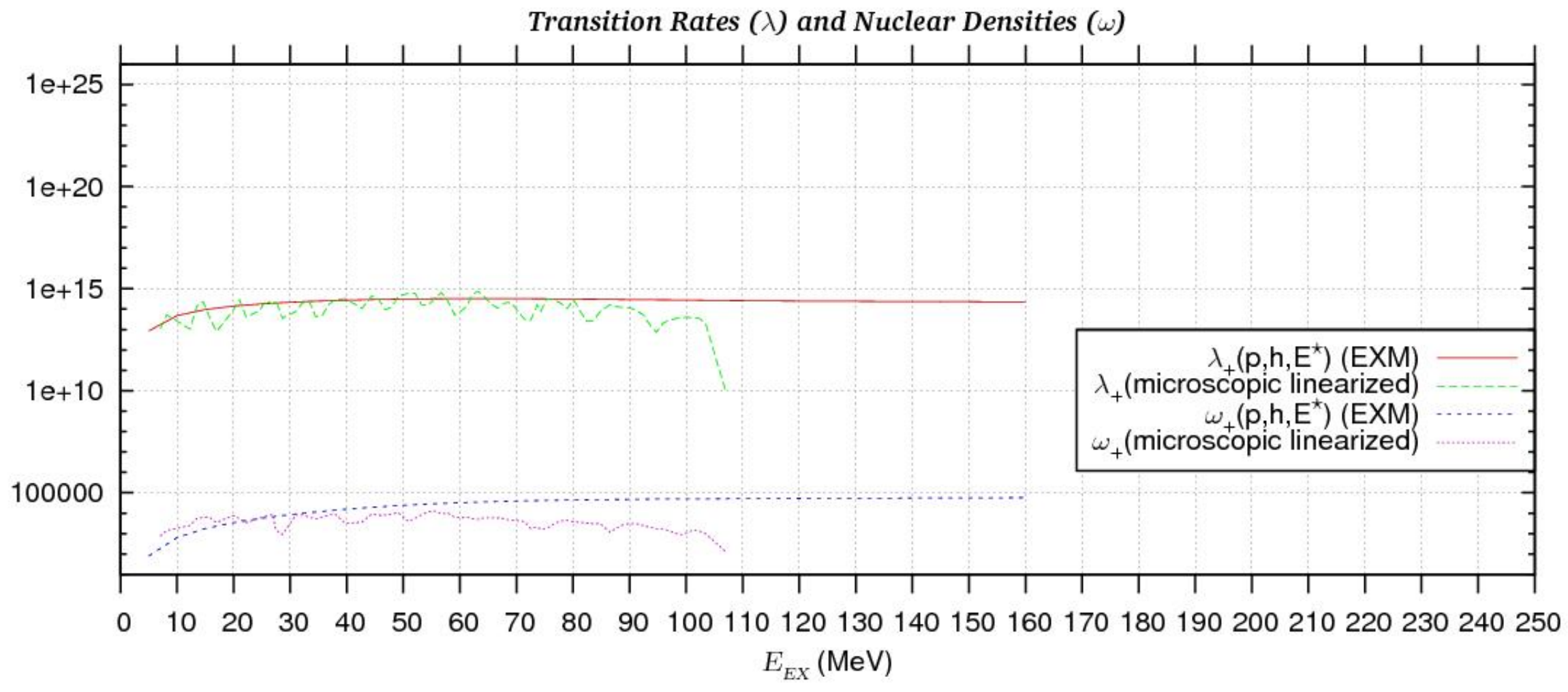

**Figure 2**. Transition strengths and nuclear densities as functions of the excitation energy for transitions that increase the number of excitons by 2, for an initial configuration with 1 exciton.

A typical example of this relation is shown in figures 1 and 2, for $t_0$ and $t_+$ respectively, for configurations with $n$=3.

In this case $t_0$ and $t_+$ have variable ratios as a function of the nuclear excitation, $E_{ex}$, but in the region of maxima $t_0$ is more than 4 orders of magnitude greater than $t_+$, which according to Ref.[17] would be enough to validate EH.

Note that the maximum of the densities in Fig. 2 is only 2 orders of magnitude smaller than the maximum of Fig. 1, indicating that the details of the collision process are the main causes of the differences between $t_0$ and $t_+$. For $n$=4, $t_0$ and $t_+$ are shown in figures 3 and 4, respectively, and one notices that the ratio between $t_0$ and $t_+$ is smaller but still greater than 3 orders of magnitude in the region of maxima.

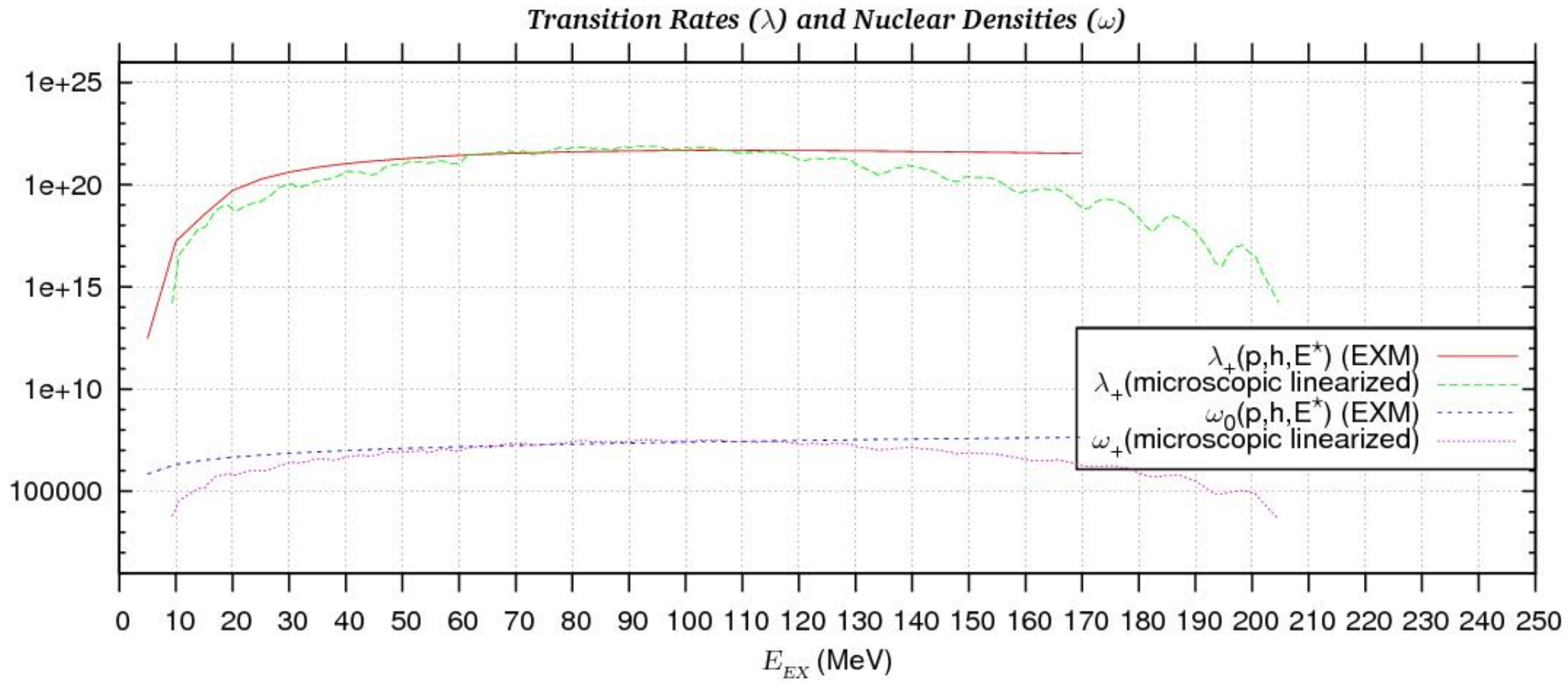

**Figure 3**. Transition strengths and nuclear densities as functions of the excitation energy for transitions that keep the number of excitons constant, for an initial configuration with 4 excitons.

Notice that the agreement with EXM is restricted to a region of not too high or too low excitations and becomes worse for higher $n$, suggesting that the EXM should be revised in these cases. Our calculations indicate that the points at which the TS fastly decrease to zero depend on the level structure of the Model Space and $n$.

The details of the factors involving the angular momentum sums have been given in a previous work Ref.[8] and the present calculations show that, although the ratio between $t_0$ and $t_+$ is smaller in the region of maxima, when these factors are fully considered, it is still greater than 3 orders of magnitude for all exciton numbers and sizes of the Model Space considered in this study. This indicates that these factors do not play an important role in the definition of the TR and could be replaced, with good approximation, by fixed phenomenological parameters.

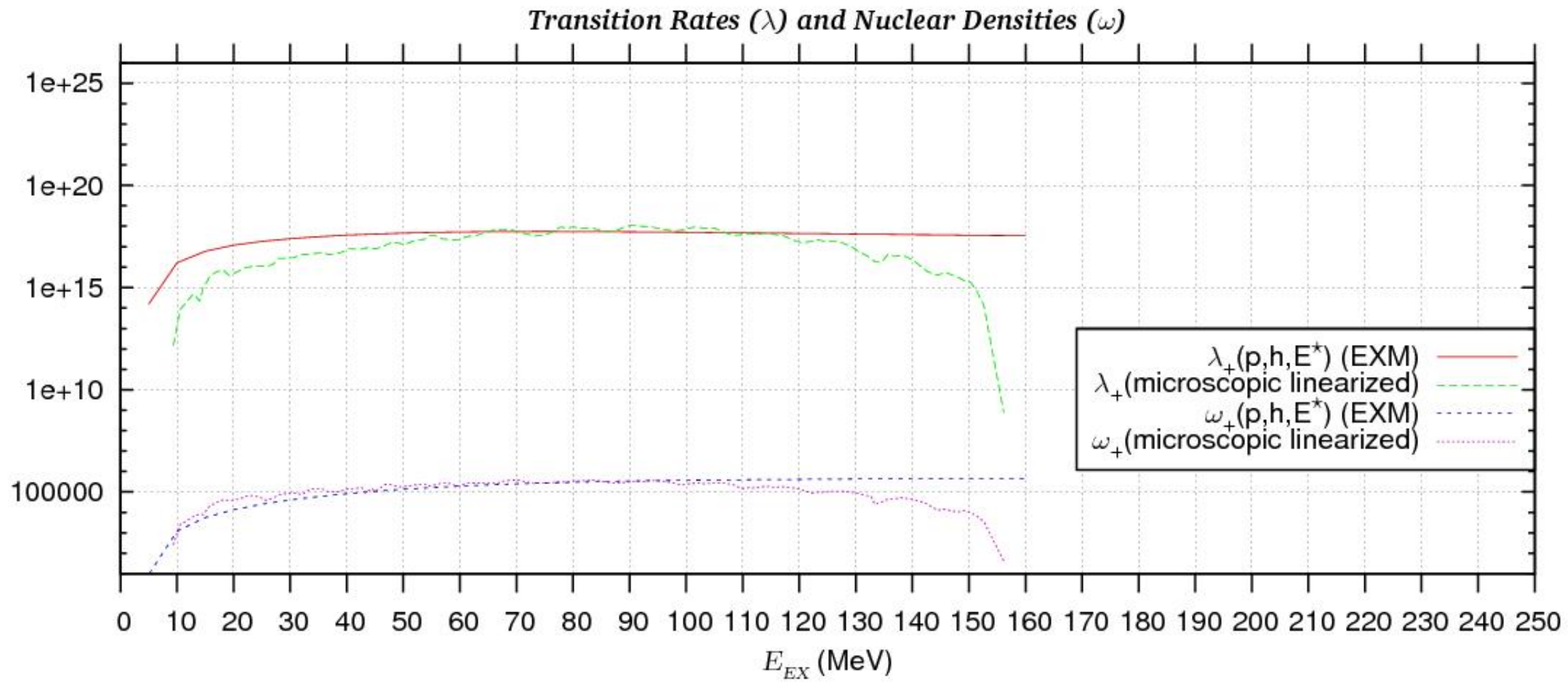

**Figure 4**. Transition strengths and nuclear densities as functions of the excitation energy for transitions that increase the number of excitons by two, for an initial configuration with 2 excitons.

In the microscopic results both the transition strengths and the densities show *strong oscillations* from one energy to the next, reflecting the existence of regions with very different degeneracies even for relatively close excitations. These differences have essentially a combinatorial nature and the physically meaningful degeneracies are those associated with the local maxima of the degeneracy curve, corresponding to the most probable configurations.

For example, Fig. 5 shows these oscillations for the same transition elements shown in Fig. 3. The results of Fig. 3 were obtained by fitting a smooth curve over the rapidly oscillating microscopic results, by neglecting the local minima and constraining the area below both curves to be the same. Therefore, the smoothed curve in Fig. 5, although approximate, follows closely the local maxima of the rapidly oscillating one and is also a physically meaningful description.

In summary, the main results and conclusions of this work are, first, to show a precise direct microscopic formalism (DMF) connecting the definition of the momenta of the PE Hamiltonian and the usual TR parameters of the semi-classical models (EXM and hybrid models), revealing the connection with the traditional approach based on the Laplace transform and to notify about some specific cases in which the latter gives physically incorrect results.

Second, the TS calculated with DMF show good comparison with the corresponding semi-classical parameters in a certain interval of not very low and not very large excitations, becoming smaller than the semi-classical estimate outside this interval. For increasing $n$ the relative decrease of the TS in comparison with the semi-classical TR is faster. We interpret these results as indicating that the EXM tends to describe better PE states with low excitation because of the competition, for higher excitations, with other processes like PE emission and the possibility of transition to the compound stage.

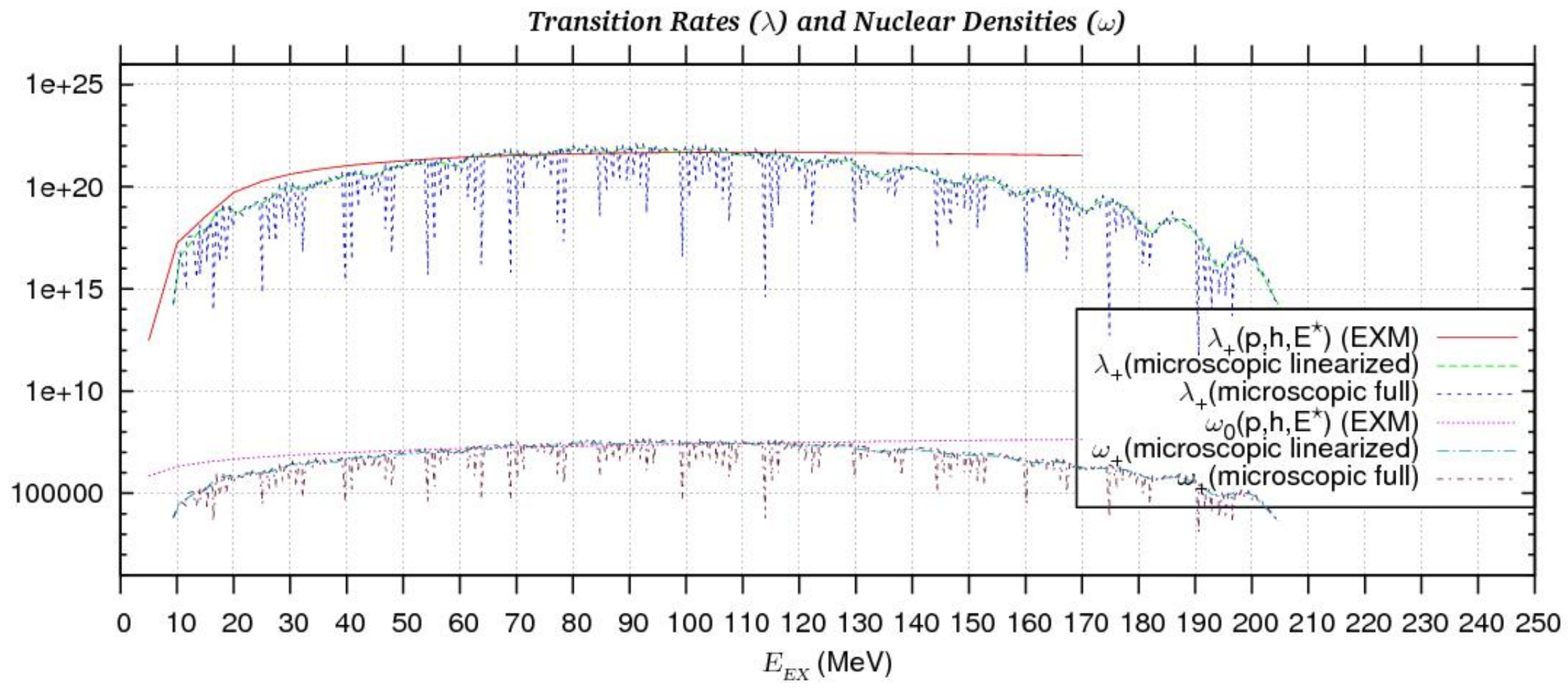

**Figure 5**. The same as Fig. 3, showing also the rapidly oscillating results of the microscopic calculations and the smoother curve that represents them.

A third conclusion is that in the region of good similarity with EXM the ratio between $t_0$ and $t_+$ is usually greater than the minimum that permits to validade the "equilibrium hypotesis", as analyzed in Ref.[17], as one of the central aspects of the EXM. This indicates that the estimate for the interaction parameter $|M|^2$ of semi-classical rates of transitions should be revised.

Fourth, the details of the angular momentum coupling play no major role in the above results and yield closely the same strengths and ratios between $t_+$ and $t_0$ as those obtained with an adequately defined *constant*.

At last, the results shown in the previous figures correspond to a modified HO basis in which an arbitrary constant spacing between sp-levels has been assumed. Without this assumption the comparison with EXM becomes *not so good* as we see in Fig.6 for transitions with $\Delta n$=0 and Fig.7 for transitions with $\Delta n$=+2.

Nonetheless, despite the remarkable difference with usual EXM estimates graphs 6 and 7 still show a ratio between $t_0$ and $t_+$ that is still compatible with EH in the region of maxima.

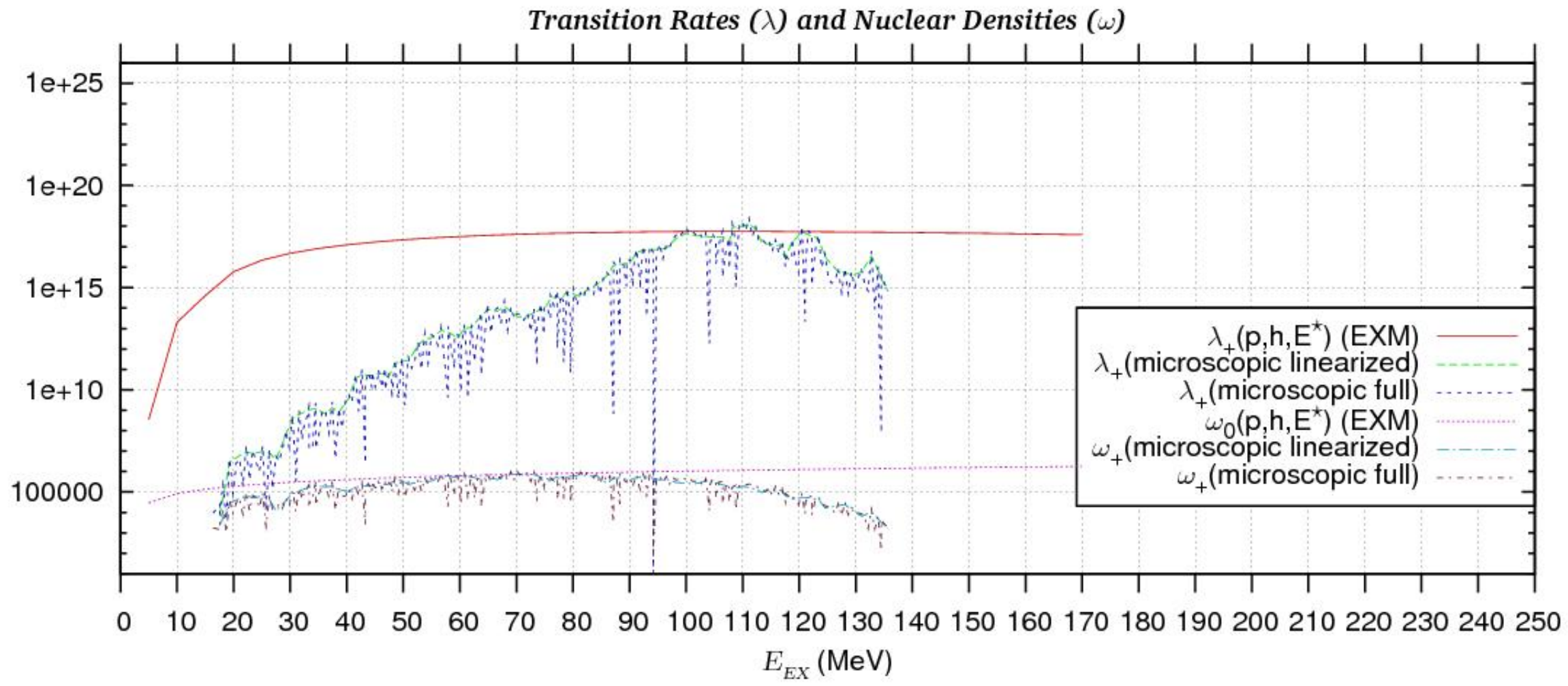

**Figure 6**. Transition strengths and nuclear densities as functions of the excitation energy for transitions that keep the number of excitons constant. It includes the rapidly oscillating results of the microscopic calculations and the smoother "linearized" curve that represents them, using the HO sp-states levels for $n$=4.

.

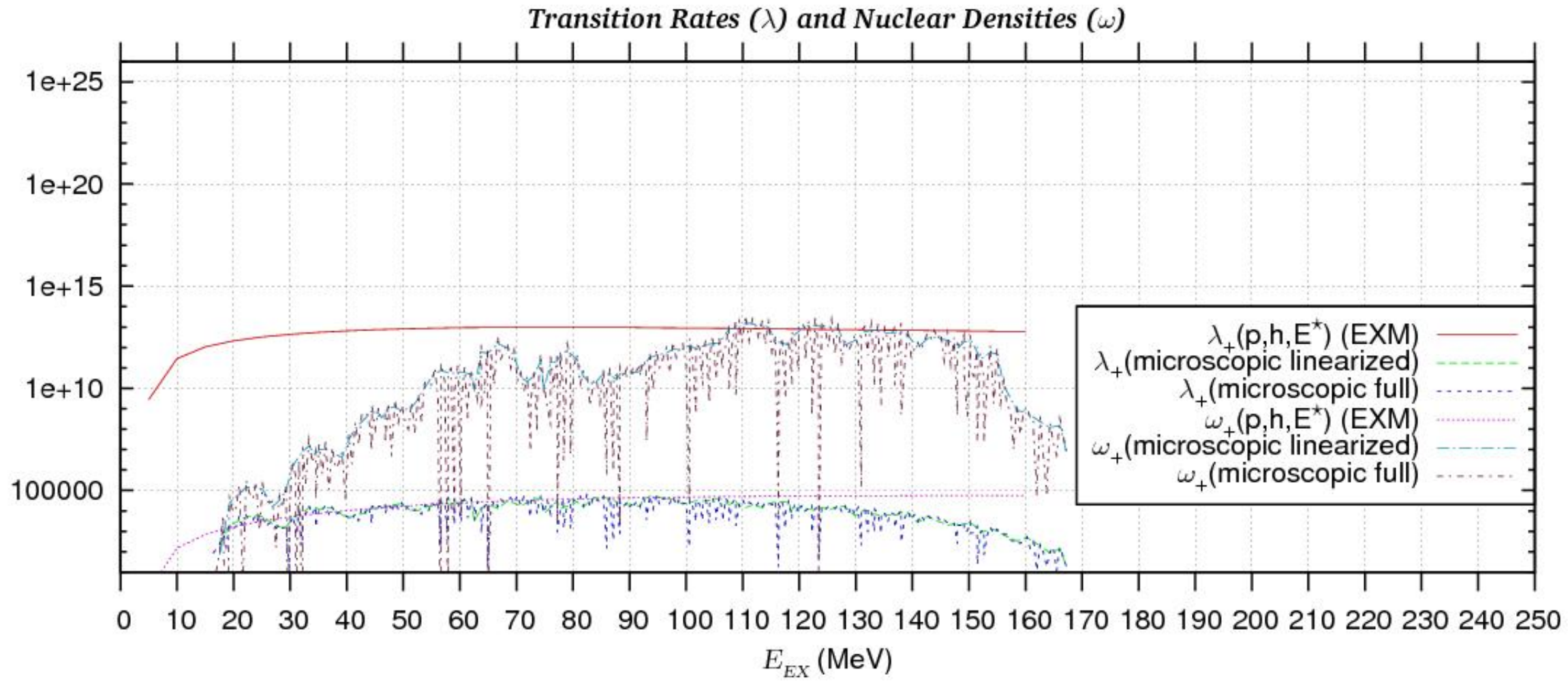

**Figure 7**. The same as Fig. 6 for transitions that increase the number of excitons by two.

## 8. Acknowledgments

The author is pleased to acknowledge the friendship and support of the colleagues IEAv/DCTA during the realization of this work.

# Appendix I - A few comments on the Darwin-Fowler method

Using the Darwin-Fowler formalism the level density can be defined as the pole of the grand canonical generating function $f(x,y)$ divided by $x^{A+1}y^{E+1}$ [3, 1] and, consequently, it can also can be interpreted as the inverse Laplace transform of a linear combination of nuclear densities for given number of excitons ($A$) and excitation energy($U$). Here we present a detailed description of the connection between these two definitions and the direct microscopic approach with the purpose of serving as a brief referential analysis of this very important method.

The general expression of the inverse Laplace transform can be written as[19]

$$F(E) = \mathcal{L}^{-1}(f(\beta)) = \frac{1}{2\pi i}\int_{\beta_r - i\infty}^{\beta_r - i\infty} \mathsf{e}^{\beta E} f(\beta) d\beta \ , \qquad \text{where } \beta_r = \text{real constant} \ , \tag{I.1}$$

while the part of the expression for the nuclear densities associated with the total nuclear energy in (2.11), using the Cauchy theorem and the grand canonical generator function, has the general form

$$\frac{1}{2\pi i}\oint \frac{f\left(y^N\right)}{y^{N+1}} dy \ , \tag{I.2}$$

where $N$ is an integer connected with the nuclear energy through $E{=}N\epsilon$, defining $E$ as discrete with values proportional to a small energy unit, $\epsilon$. Then, one performs the following transformation of variables

$$y = \mathsf{e}^{-\beta\epsilon} \tag{I.3}$$

and assumes $y$ has constant absolute value (constant nuclear temperature) and write

$$\mathsf{e}^{-\beta\epsilon} = \mathsf{e}^{-\beta_r\epsilon - i\beta_i\epsilon} = y = |y|\mathsf{e}^{i\theta_y} \ , \tag{I.4}$$

and, by hypothesis, one has

$$\beta_r\epsilon = \log\left(\frac{1}{|y|}\right) = \text{fixed} \qquad \text{and} \qquad \beta_i\epsilon = -\theta_y = -\text{arctg}\left(\frac{y_i}{y_r}\right) \ , \tag{I.5}$$

then,

$$-\epsilon d\beta = -\epsilon i d\beta_i = i d\theta_y \ . \tag{I.6}$$

If interval of integration for the argument of $y$ is conveniently taken as

$$\theta_y \in [-\pi, +\pi] \ , \tag{I.7}$$

then the integral over the closed path in the $y$-plane becomes

$$\oint (\cdots) dy = \int_{\theta_y=-\pi}^{\pi} (\cdots)|y|\mathsf{e}^{i\theta_y} i d\theta_y = -\int_{\beta_i=\pi/\epsilon}^{-\pi/\epsilon} (\cdots)|y|\mathsf{e}^{i\theta_y} \epsilon d\beta \tag{I.8}$$

and using (I.4) results

$$\oint (\cdots) dy = \int_{\beta=\beta_r-\pi/\epsilon}^{\beta_r+\pi/\epsilon} (\cdots)\mathsf{e}^{-\beta\epsilon} \epsilon d\beta \ . \tag{I.9}$$

Now, if $\epsilon$ is taken as infinitesimally small the interval of the imaginary integration becomes arbitrarily large,

$$\beta_i = (-\theta_y/\epsilon) \in [-\pi/\epsilon, +\pi/\epsilon] \approx (-\infty, +\infty), \tag{I.10}$$

and, $\beta_r$ can be considered as constant (for very small $\epsilon$ this would correspond in (I.5) to $|y|$ very close to "1"). Now, if one considers the function in (I.2) the expression becomes

$$\frac{1}{2\pi i} \oint \frac{f\left(y^N\right)}{y^{N+1}} dy = \frac{1}{2\pi i} \int_{\beta=\beta_r-\pi/\epsilon}^{\beta_r+\pi/\epsilon} \frac{f\left(\mathsf{e}^{-\beta N\epsilon}\right)}{\mathsf{e}^{-\beta(N+1)\epsilon}} \mathsf{e}^{-\beta\epsilon} \epsilon d\beta \tag{I.11}$$

$$= \frac{1}{2\pi i} \int_{\beta=\beta_r-\pi/\epsilon}^{\beta_r+\pi/\epsilon} f(\mathsf{e}^{-\beta E})\mathsf{e}^{\beta E} \epsilon d\beta \ , \tag{I.12}$$

where we have used the relation $E{=}N\epsilon$. Then, by definition, when $\epsilon$ becomes infinitesimally small and the total expression is divided by $\epsilon$, Eq. (I.12) becomes formally equal to $\mathcal{L}^{-1}(f(\mathsf{e}^{-\beta E}))$. For (I.12) to be a true inverse Laplace transform one must still assume that $E \in (0{,}\infty)$ so that the term proportional to $y^{\mathcal{N}}$ in (2.17) will also be the Laplace transform of the nuclear density for given $A$ and $E$.

For a given microstate of the grand canonical ensemble, with given finite total energy $E$ and total number of excitons $n$, the above argumentation would imply that $N$ becomes very large when $\epsilon$ becomes infinitesimal. This idea corresponds precisely with the idea of the continuum approximation (*CAP*), and there is no conflict with the direct microscopic description.

A possible problem in the above deduction could be that $\beta_r{=}\ (-1/\epsilon)\log(|y|)$, then if $\epsilon$ becomes infinitesimally small, $\beta_r$ would become very large unless $\log(|y|)$ becomes infinitesimal too. Now, if one interprets $\beta_r$ as $1/\kappa T$, then $\beta_r{\rightarrow}\infty$ would mean $T{\rightarrow}\ 0$, which is a physically non necessary condition for *CAP*.

On the other hand, one may also think that this is irrelevant for our present discussion because the variable of integration is $\beta$ instead of $\beta\epsilon$, therefore, only what affects $\beta$ directly could affect the general argumentation. If log$|y|$ is assumed to be proportional to $\epsilon$ with finite and constant $\beta_r$ this would mean that the correspondint points in the $y$-plane belong to a circle infinitesimally close to "1", without essential contradiction.

The whole reasoning supporting this "interpretation" is therefore coherent with the direct microscopic description.

## 1.1 The Saddle Point problem

We saw in equations (2.10) or (2.12) that in formulation of the Shell Model based on the Darwin-Fowler statistics the grand canonical generating functional $f(x, y)$ can be written as a sum of terms of the type $(xy^{\nu_i})$, where $\nu_i$ is an integer. In references [3] and [20] one finds that these terms are supposed to decrease in modulus when $x$ or $y$ vary over complex circles around the origin in comparison with their values when they belong to the positive real axis and, therefore, they should have a maximum in this direction. This assumption corresponds to the idea of a *saddle point* located on the positive real axes of the variables $x$ and $y$, in the original analysis of Darwin and Fowler.[20]

The formalism of Darwin and Fowler[20] considers the existence of a saddle point for the integrand of (2.11) in connection with the *steepest descent method*. It assumes the existence of a point of minimum along the positive real axis and, using a qualitative analysis of the generating function of the ensemble, concludes that this point is also a "strong maximum" along the direction of the path of integration, taken to be the circle centered at the origin with radius equal to the abscissa on the positive real axis where the minimum occurs.

Reference [20] then concludes that if the integrand has a saddle point with these characteristics, it would be sufficient to obtain approximate equations of state in the usual form, for given $A$ and $E$, in the case of systems obeying the Bose-Einstein statistics ("Planck vibrators").

Here we briefly analyze the assertion of the existence of this "strong maximum" in the case of systems of many fermions, corresponding to the $f(x,y)$ given by (2.10) or (2.12), and show that although the saddle point may exist the maximum cannot be characterized as "strong", in the sense of having much larger magnitude than the other points along the path of integration. The fact that the approximated equation of state obtained with the Darwin Fowler method works very well in various applications should then be the object of an independent analysis, which is beyond the scope of the present paper.[18]

A saddle point of a function is one that is stationary but not a local extremum. For functions of two variables, it is a maximum for the variation of one variable and a minimum for the variation of the other. More precisely, it is a point $(x_\star,y_\star) \in \mathcal{R}^{n+\mathtt{m}}$ that satisfies[21]

$$L(x_\star, y) \leq L(x_\star, y_\star) \leq L(x, y_\star) \ , \quad \forall x \in \mathcal{R}^n \text{ and } \forall y \in \mathcal{R}^{\mathtt{m}} \ , \tag{I.13}$$

or, equivalently,

$$\min_x \max_y L(x,y) = L(x_\star, y_\star) = \max_y \min_x L(x,y) \ , \tag{I.14}$$

and the definition for complex variables would correspond to the dimensions $n$=$\mathtt{m}$=2.

In a first stage of the formal application of the Darwin-Fowler method the variables $x$ and $y$ can be thought as possessing no direct physical meaning and are created only to keep track of the counting of the number of particles and the energy of the nuclear levels. This tracking is formally performed using the Residue's Theorem in the definition of the "density" associated with $y$ for a given $A$ (number of particles, excitons, etc.). Therefore, expression (2.12) for the nuclear density *would be exact* if $x$ and $y$ could be considered as continuous variables, but we have seen that this is true only in the $CAP$ limit.[20]

In addition, the Darwin-Fowler method go one step ahead and attempts to obtain a simpler, more practical algebraic expression by performing a qualitative analysis of $f(x,y)$ and using the "steepest descent" method, which we will show now can only be considered as approximately correct.

In the following we perform a quick numerical analysis of $f(x,y)$ to show that in at least one particular case the hypothesis of the existence of a saddle point with strong maximum along the path of integration at the point located on the positive axis is incorrect and, therefore, it cannot be considered as having the general validity suggested by Ref.[20]. One can analyze the general expression of $f(x,y)$ by considering a simplified version of it, for example, a product of the type

$$r(x,y) = \frac{f(x,y)}{xy} = \prod_{k=1}^{\mathtt{k}_{max}} \frac{(1+xy^k)}{xy} \tag{I.15}$$

which would correspond to a simplified "ensemble" with maximum number of excitons per configuration equal to "$\mathtt{k_{max}}$" and exciton energies $\epsilon_k$=$k\epsilon$, with fixed $\epsilon$. The division by "$xy$" corresponds to the description of a system with no excitons, i. e., the fundamental state. We then consider the variables $x$ and $y$ as belonging to circumferences around the origin in their respective complex planes and variable radii for these circumferences to have the analysis extended to their corresponding circles.

This function can be easily calculated numerically by fixing one variable, for example fixed $|x|$ and $\arg(x)$, and varying $|y|$ with values smaller that one (in accordance to its physical interpretation as $y$=$\mathtt{e}^{-\beta\epsilon}$, $\beta$ and $\epsilon$ real) and variable arg($y$) in a reasonable range, for example between 0 and $2\pi$, with steps of $2\pi/16$.

In figure 8 we show a typical result for $k_{\mathtt{max}}$=20 and arg($x$)=0, in which there are indeed points of minimum along the real axes of $x$ and $y$, i. e., for arguments of $x$ and $y$ equal to 0, $2\pi$, etc.. In addition, the points of the curve along the positive axis are also the maxima with respect to the domain of points situated along the circumferences centered at the origin to which they belong. Therefore, these are saddle points in the respective complex planes. On the other hand, one can see that these maxima are not "strong" in the sense that one cannot assume that the entire path integral along these circles can be reduced to the contribution of a small region in the vicinity of these points. Therefore the reasoning of Darwin and Fowler is only approximately valid in this case.

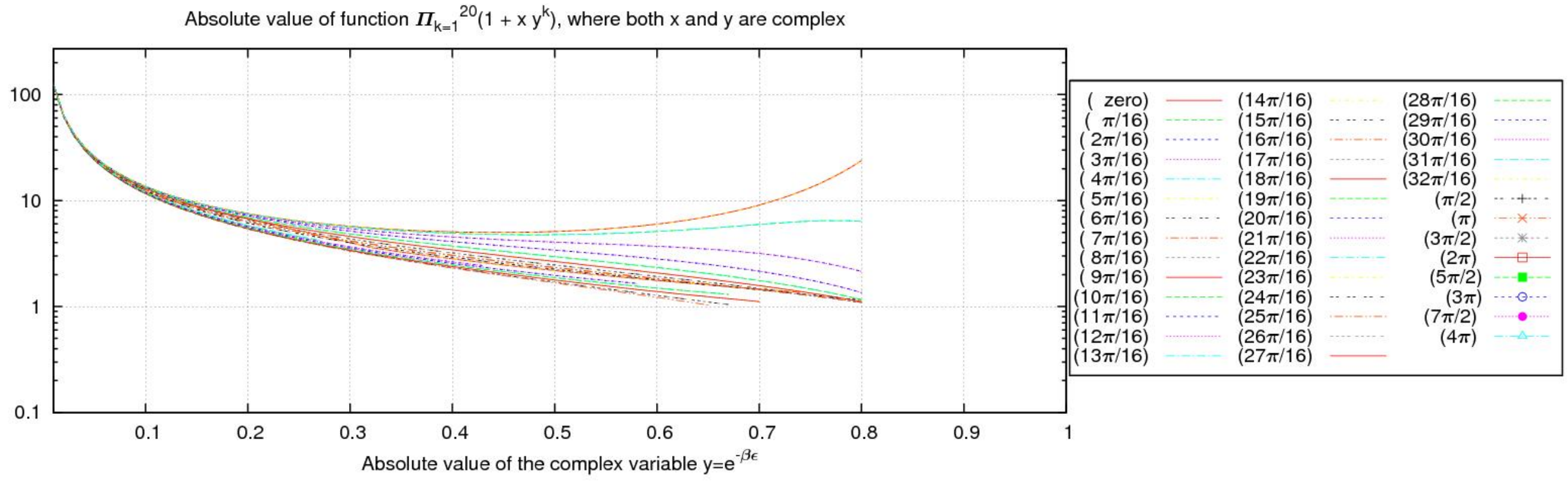

**Figure 8**. Simplified generating function, $r(x,y)$, for variable $y$ and fixed $x$, with arg($x$)=0, in the case of 20 sp-states. The lines in the box on the right side of the graph are the values of arg($y$).

These conclusions are more clearly shown in figures 9 and 11 below. Figure 9 is the 3 dimensional correspondent of the previous graph.

In figure 8 the curves for arg($y$) equal zero, $2\pi$, etc. have a minimum for some point between zero and 1 and if $x$ is real, this point is also the maximum along the circumference around the origin to which it belongs. Therefore, this minimum is a saddle point.

Notice that the maxima along the real axis in Fig. 8 have not much larger magnitude than the other points along the corresponding circumferences, less than 10 times greater for $|y| \leq 0.7$, in the given example. The relative magnitude increase for $|y|$ close to 1, corresponding to nuclear temperatures tending to $\infty$, but these points are not saddle points with respect to $y$.

Figures 10 and 11 are the analogous of the previous two figures, now with arg($x$)=$\pi$. Notice that the maxima along the real axis in comparison with other points of the corresponding circumferences have completely disappeared and they have become *minima* instead. In addition, there are no points of minimum along the real axis anymore, or in fact in any other direction of the $y$ plane. Therefore, for arg($x$)=$\pi$ the reasoning of Darwin-Fowler method becomes *totally incorrect*.

Therefore, in general, although the maxima on the real axis of $x$ and $y$ may be saddle points of $r(x, y)$ they cannot be considered as strong maxima and, if $x$ is considered a complex variable, the saddle points can only be defined in certain directions of its complex plane.

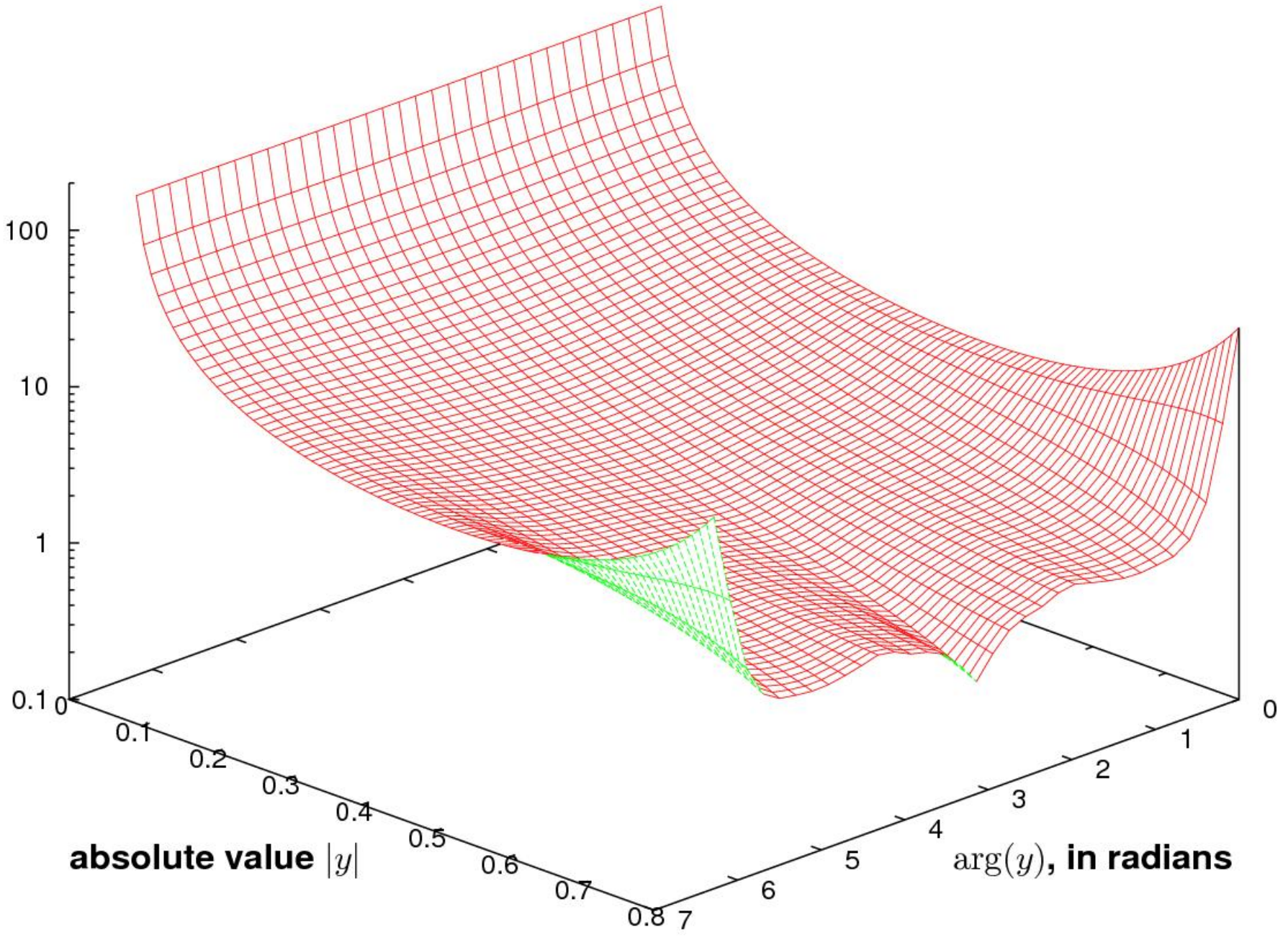

**Figure 9**. The three dimensional correspondent of Fig. 8 for 20 sp-states with arg($x$)=0. Notice that the real axis is clearly the loci of local maxima in comparison with the other points along a circumference with fixed absolute value of $y$

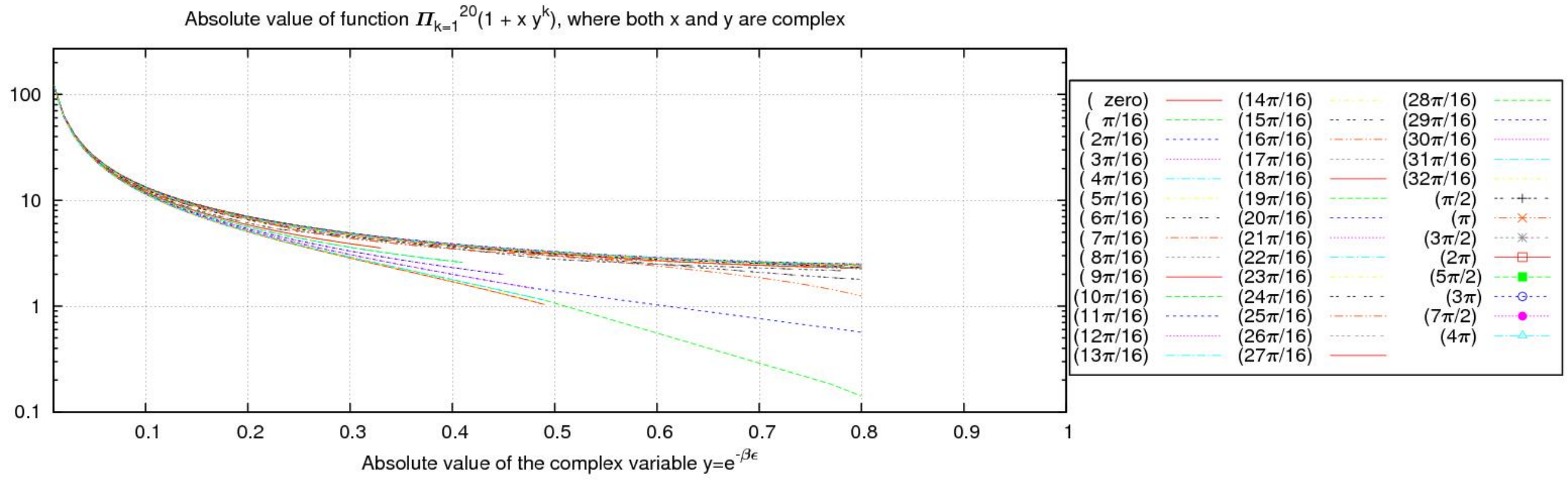

**Figure 10**. Function $r(x,y)$ in the case of 20 sp-states, with $\arg(x)=\pi$. The lines in the box on the right side of the graph correspond to the values of $\arg(y)$.

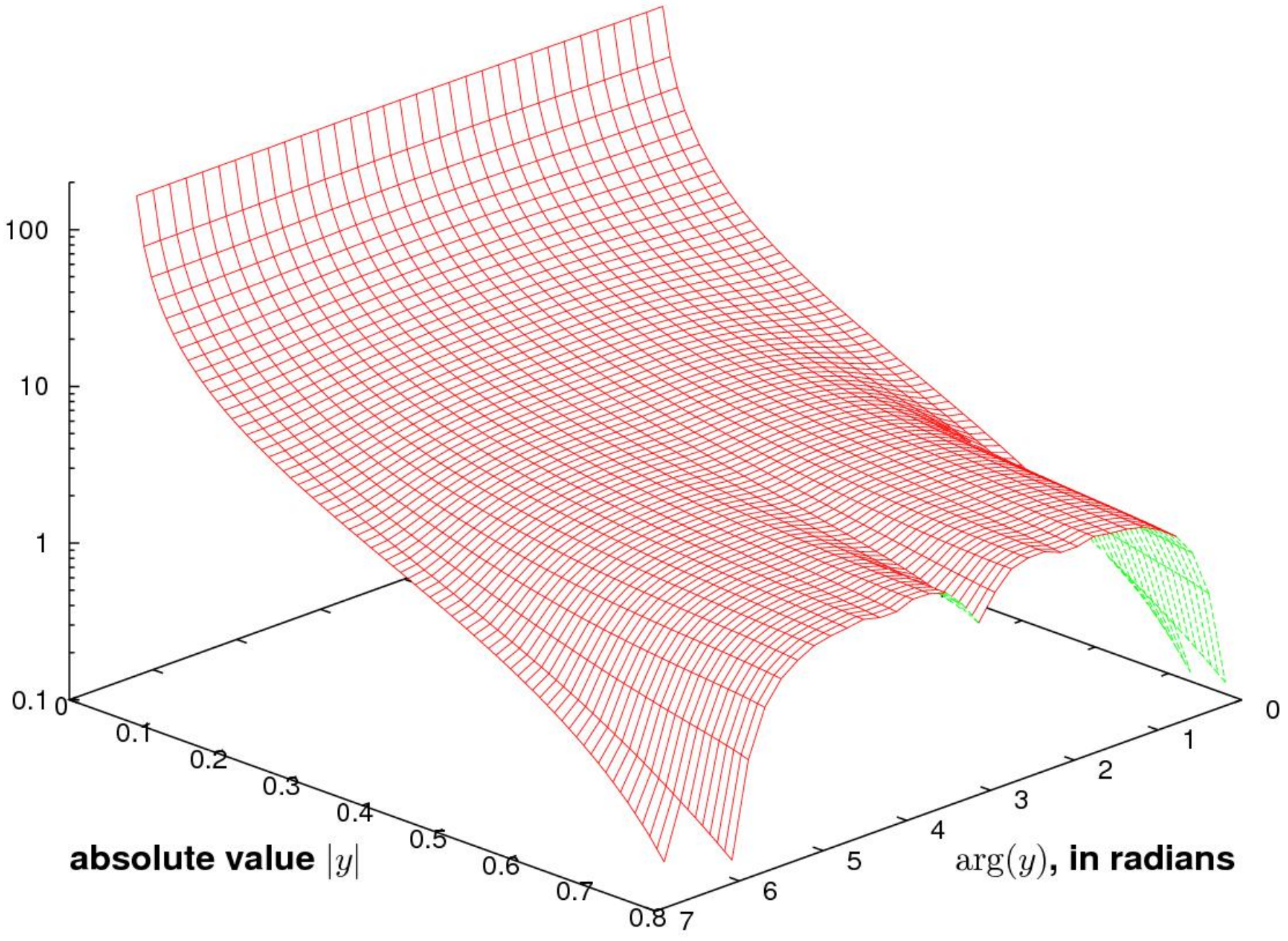

**Figure 11**. Three dimensional correspondent of Fig. 10 for 20 sp-states with $\arg(x)=\pi$. Notice that the points of the real axis are not anymore the loci of local maxima along the circumferences with fixed absolute value of $y$.

This simple analysis indicates that the entire idea of using the Cauchy theorem in the analysis of the generating function of the grand canonical ensemble, and the consequent connection with the Laplace transform, can be misleading. In this case, the use of more direct algebraic approaches as the one presented in this paper is naturally more appropriate for a microscopic description of the Shell Model, even in the $CAP$ limit.

# Appendix II - The basic convolution relation

The expressions for the degneracies in this section result from the combinatorial analysis of the corresponding sets of configurations. They are based on the assumption that the maximum number of sp-states for particles and holes in the Model Space is arbitrarily defined and the actual number of excited particles and holes used to describe the nuclear system corresponds to the simple combinatorial idea of placement of "balls" (effectively excited particle or hole) into "boxes" (available sp-states for excitation).

From the operators defined in (5.4) it was shown that $(a|a^\dagger)$ is given by Eq.(5.33)

$$(a|a^\dagger) = \sum_{(12)(UM)} \sum_{(\alpha\beta)} \langle ph\,|a^+_\alpha a_\beta|p'h'\rangle\langle p'h'\,|a^+_\beta a_\alpha|ph\rangle \tag{II.1}$$

$$\approx \sum_{(1)} \sum_{(S,t)} \mathtt{e}^{[UM]} \sum_{(\alpha\beta)} O_{\alpha\beta} d(p-1+1{\neq}\alpha{=}\beta, h, S, t)\ , \tag{II.2}$$

with $S$=($U_\alpha$-($\epsilon_\alpha$-$\epsilon_\beta$)) and $t$=($M_\alpha$-($\mathtt{m}_\alpha$-$\mathtt{m}_\beta$)). This expression can be rewritten more explicitly using the identity

$$d(p-\alpha+\beta, h, U_{(\alpha,\neq\beta)}-\epsilon_\alpha+\epsilon_\beta, M_{(\alpha,\neq\beta)}-\mathtt{m}_\alpha+\mathtt{m}_\beta) =$$

$$d(p-\alpha{\neq}\beta, h, U_{(\alpha,\neq\beta)}-\epsilon_\alpha, M_{(\alpha,\neq\beta)}-\mathtt{m}_\alpha) =$$

$$d(p-\alpha, h, U_\alpha-\epsilon_\alpha, M_\alpha-\mathtt{m}_\alpha) - d(p-\alpha=\beta, h, U_{\alpha\beta}-\epsilon_\alpha, M_{\alpha\beta}-\mathtt{m}_\alpha)\ , \tag{II.3}$$

where ($U_{(\alpha,\neq\beta)}$,$M_{(\alpha,\neq\beta)}$) are the energies and angular momenta of the configurations in which ($\alpha$) is present and ($\beta$) is not and ($U_{\alpha\beta}$,$M_{\alpha\beta}$) are the parameters of the configurations in which *both* ($\alpha$) and ($\beta$) are present, etc. If one considers the set of all configurations in which ($\alpha$) is present they have degeneracy, $d(p{=}\alpha,h,U_\alpha,M_\alpha)$, equal to the first term of the last expression of (II.3)

$$d(p=\alpha, h, U_\alpha, M_\alpha) = d(p-\alpha, h, U_\alpha-\epsilon_\alpha, M_\alpha-\mathtt{m}_\alpha) \tag{II.4}$$

and it includes also configurations in which ($\beta$) is possibly present. The degeneracy of the configurations in which both ($\alpha$) and ($\beta$) are simultaneously present is

$$d(p=\alpha\beta, h, U_{\alpha\beta}, M_{\alpha\beta}) = d(p-\alpha=\beta, h, U_{\alpha\beta}-\epsilon_\alpha, M_{\alpha\beta}-\mathtt{m}_\alpha) \tag{II.5}$$

therefore, the subtraction of (II.5) from (II.4) produces the degeneracies of the configurations in which ($\beta$) is *not* present, which is (II.3). Note that one can also write (II.3) as

$$d(p=\alpha{\neq}\beta,h,U_{\alpha\neq\beta},M_{\alpha\neq\beta}) = d(p=\alpha,h,U_\alpha,M_\alpha) - d(p=\alpha\beta,h,U_{\alpha\beta},M_{\alpha\beta}) \;, \tag{II.6}$$

then (II.2) can be rewritten as

$$\sum_{\binom{(1)}{(S,t)}} \mathtt{e}^{[UM]}\sum_{(\alpha\beta)} O_{\alpha\beta}[d(p-\alpha,h,U_\alpha-\epsilon_\alpha,M_\alpha-\mathtt{m}_\alpha) - d(p-\alpha=\beta,h,U_{\alpha\beta}-\epsilon_\alpha,M_{\alpha\beta}-\mathtt{m}_\alpha)] \;, \tag{II.7}$$

and, if $O_{\alpha\beta}$ is given for all ($\alpha$,$\beta$), (II.7) can be determined from the combinatorial results for $d(p\text{-}1,h,U\text{-}\epsilon_1,M\text{-}\mathtt{m}_1)$ and $d(p\text{-}2,h,U\text{-}\epsilon_1\text{-}\epsilon_2,M\text{-}\mathtt{m}_1\text{-}\mathtt{m}_2)$

Similarly, the corresponding term for holes, (5.39), can be expressed as

$$(\mathtt{e}|\mathtt{e}^+) = \sum_{(12)(UM)}\sum_{(\alpha\beta)} \langle ph\,|\,b^+_\beta b_\alpha\,|\,p'h'\rangle\langle p'h'\,|\,b^+_\alpha b_\beta\,|\,ph\rangle \tag{II.8}$$

$$\approx \sum_{\binom{(1)}{(S,t)}} \mathtt{e}^{[UM]}\sum_{(\alpha\beta)} O_{\alpha\beta}\, d(p,h-1+1{\neq}\beta{=}\alpha,S,t) \;, \tag{II.9}$$

with $S$=($U_\beta$-$\epsilon_\beta$+$\epsilon_\alpha$) and $t$=($M_\beta$-$\mathtt{m}_\beta$+$\mathtt{m}_\alpha$), which can be rewritten as

$$\sum_{\binom{(1)}{(S,t)}} \mathtt{e}^{[UM]}\sum_{(\alpha\beta)} O_{\alpha\beta}[d(p,h-\beta,U_\beta-\epsilon_\beta,M_\beta-\mathtt{m}_\beta) - d(p,h-\beta=\alpha,U_{\alpha\beta}-\epsilon_\beta,M_{\alpha\beta}-\mathtt{m}_\beta)] \;, \tag{II.10}$$

The above expressions can be rewritten as convolutions with the various sp-state densities, for example, if one consider the following "interpretation" of the sum over one sp-state,

$$\sum_{(\alpha)} f(\epsilon_\alpha,\mathtt{m}_\alpha) = \int d\epsilon \sum_{(\mathtt{m}_\alpha)}\sum_{(\epsilon_\alpha)} \delta(\epsilon-\epsilon_\alpha) f(\epsilon,\mathtt{m}) \tag{II.11}$$

and

$$\sum_{(\epsilon_\alpha \mathtt{m}_\alpha)} \delta_{(\mathtt{m},\mathtt{m}_\alpha)}\delta(\epsilon-\epsilon_\alpha) = \omega(1,0,\epsilon,\mathtt{m}) \;, \tag{II.12}$$

where usually the sp-state "$\alpha$" is degenerate with respect to $\epsilon_\alpha$ but not with respect to the pair ($\epsilon_\alpha$,$\mathtt{m}_\alpha$). Then,

$$\sum_{(\alpha)} f(\epsilon_\alpha,\mathtt{m}_\alpha) = \int d\epsilon \sum_{(\mathtt{m}_\alpha)} f(\epsilon,\mathtt{m}_\alpha)\omega(1,0,\epsilon,\mathtt{m}_\alpha) \;, \tag{II.13}$$

with the following straightforward inclusion of a second sp-state,

$$\sum_{(\alpha\beta)} f(\epsilon_\alpha,\epsilon_\beta,\mathtt{m}_\alpha,\mathtt{m}_\beta) = \int d\epsilon_1 d\epsilon_2 \sum_{(\mathtt{m}_\alpha \mathtt{m}_\beta)} f(\epsilon_1,\epsilon_2,\mathtt{m}_\alpha,\mathtt{m}_\beta)\omega(1,0,\epsilon_1,\mathtt{m}_\alpha)\omega(1,0,\epsilon_2,\mathtt{m}_\beta) \;, \tag{II.14}$$

where all sp-energies are positive and the sp-densities are functionals that restrict the contributions of the degeneracies to points where they are well defined.

Then, one can rewrite the sum over ($\alpha$,$\beta$) in the first term of (II.7) as

$$\sum_{(\alpha\beta)} O_{\alpha\beta} d(p-\alpha, h, U_\alpha - \epsilon_\alpha, M_\alpha - \mathtt{m}_\alpha) \ , \tag{II.15}$$

$$= \int d\epsilon_1 d\epsilon_2 \sum_{(\mathtt{m}_\alpha \mathtt{m}_\beta)} O_{\alpha\beta} \omega(p-1, h, U-\epsilon_1, M-\mathtt{m}_\alpha)\omega(1,0,\epsilon_1,\mathtt{m}_\alpha)\omega(1,0,\epsilon_2,\mathtt{m}_\beta) \ , \tag{II.16}$$

$$= \int_0^U d\epsilon_1 \sum_{(\mathtt{m}_\alpha)} \left( \int d\epsilon_2 \sum_{(\mathtt{m}_\beta)} \omega(1,0,\epsilon_2,\mathtt{m}_\beta) O_{\alpha\beta} \right) \omega(p-1, h, U-\epsilon_1, M-\mathtt{m}_\alpha)\omega(1,0,\epsilon_1,\mathtt{m}_\alpha) \ , \tag{II.17}$$

$$= \int_0^U d\epsilon_1 \sum_{(\mathtt{m}_\alpha)} O_\alpha \omega(p-1, h, U-\epsilon_1, M-\mathtt{m}_\alpha)\omega(1,0,\epsilon_1,\mathtt{m}_\alpha) \ , \tag{II.18}$$

where

$$O_\alpha = \left( \int d\epsilon_2 \sum_{(\mathtt{m}_\beta)} \omega(1,0,\epsilon_2,\mathtt{m}_\beta) O_{\alpha\beta} \right) \ , \tag{II.19}$$

and the sum over ($\alpha$,$\beta$) in the second term of (II.7) as

$$\int d\epsilon_1 d\epsilon_2 \sum_{(\mathtt{m}_\alpha \mathtt{m}_\beta)} O_{\alpha\beta} \omega(p-2, h, U-\epsilon_1-\epsilon_2, M-\mathtt{m}_\alpha-\mathtt{m}_\beta)\omega(1,0,\epsilon_1,\mathtt{m}_\alpha)\omega(1,0,\epsilon_2,\mathtt{m}_\beta) \tag{II.20}$$

$$= \int_0^U d\epsilon \int_0^\epsilon d\epsilon_2 \sum_{(\mathtt{m},\mathtt{m}_\beta)} O_{\alpha\beta} \omega(p-2, h, U-\epsilon, M-\mathtt{m})\omega(1,0,\epsilon-\epsilon_2,\mathtt{m}-\mathtt{m}_\beta)\omega(1,0,\epsilon_2,\mathtt{m}_\beta) \tag{II.21}$$

$$= \int_0^U d\epsilon \sum_{(\mathtt{m})} O_\epsilon(2,0,\epsilon,\mathtt{m}) \omega(p-2, h, U-\epsilon, M-\mathtt{m}) \ , \tag{II.22}$$

where $\epsilon$=($\epsilon_1$+$\epsilon_2$) and $\mathtt{m}$=($\mathtt{m}_\alpha$+$\mathtt{m}_\beta$) and

$$O_\epsilon(2,0,\epsilon,\mathtt{m}) = \int_0^\epsilon d\epsilon_2 \sum_{(\mathtt{m}_\beta)} O_{\alpha\beta} \omega(1,0,\epsilon-\epsilon_2,\mathtt{m}-\mathtt{m}_\beta)\omega(1,0,\epsilon_2,\mathtt{m}_\beta) \ , \tag{II.23}$$

Therefore, both terms can be expressed as convolutions if adequate definitions are given for the nuclear degeneracies.

In Sec.4 a different approach is used for the nuclear density, without Dirac deltas, which is equivalent to the above one. For each ($U$,$M$), the density was defined by the following approximated expression,

$$\omega(A,U,M) \approx d(A,U,M)/\delta U = constant,$$

which implies that the integral of $\omega(A,U,M)$ over $[U_{prev},U]$ is equal to the degeneracy at $U$,

$$d(A,U,M) = \sum_{(i=1)}^{d(A,U,M)} (1) \approx \omega(A,U,M)\delta U = \int_{U_{prev}}^{U} \omega(A,U,M)dU. \tag{II.24}$$

then,

$$\sum_{(U=E_{min})}^{E_{max}} \sum_{(i=1)}^{d(A,U,M)} (...) \approx \int_{U=E_{min}}^{E_{max}} \omega(A,U,M)(...)dU. \tag{II.25}$$

In the case of sp-states a similar definition can be used in which the degeneracy in (II.23), at each point ($\epsilon_\alpha$,$\mathtt{m}_\alpha$), is usually equal one, then

$$1 \approx \int_{\epsilon_{\alpha,prev}}^{\epsilon_\alpha} \omega(1,0,\epsilon,\mathtt{m}_\alpha)d\epsilon. \tag{II.26}$$

where $\omega(1,0,\epsilon,\mathtt{m}_\alpha)=1/\delta\epsilon$, and $\delta\epsilon=(\epsilon_\alpha\text{-}\epsilon_{\alpha,prev})$ and

$$\sum_{(\alpha)}(...) = \sum_{(\epsilon_\alpha=\epsilon_{min})}^{\epsilon_{max}} \sum_{(\mathtt{m}_\alpha)}(...) \approx \int_{\epsilon_{min}}^{\epsilon_{max}} \sum_{(\mathtt{m}_\alpha)} \omega(1,0,\epsilon,\mathtt{m}_\alpha)(...)d\epsilon\ . \tag{II.27}$$

Then, replacing (...) by "$O_\alpha d(p\text{-}\alpha,h,U_\alpha\text{-}\epsilon_\alpha,M_\alpha\text{-}\mathtt{m}_\alpha)$", with $O_\alpha$ given by (II.19), and using (II.24) yields

$$\sum_{(\alpha)} O_\alpha d(p-\alpha,h,U_\alpha-\epsilon_\alpha,M_\alpha-\mathtt{m}_\alpha) = \sum_{(\epsilon_\alpha=\epsilon_{min})}^{\epsilon_{max}} \sum_{(\mathtt{m}_\alpha)} O_\alpha d(p-\alpha,h,U_\alpha-\epsilon_\alpha,M_\alpha-\mathtt{m}_\alpha) \tag{II.28}$$

$$\approx \int_{\epsilon_{min}}^{\epsilon_{max}} \sum_{(\mathtt{m}_\alpha)} \omega(1,0,\epsilon,\mathtt{m}_\alpha) O_\alpha \int_{U_{\alpha,prev}}^{U_\alpha} \omega(p-\alpha,h,U-\epsilon,M_\alpha-\mathtt{m}_\alpha)dUd\epsilon\ , \tag{II.29}$$

and if one assumes that the approximate densities vanish outside the energy interval used in their definition, one can rewrite (II.29) as

$$\int_{U_{\alpha,prev}}^{U_\alpha} dU \sum_{(\mathtt{m}_\alpha)} O_\alpha \int_0^U \omega(1,0,\epsilon,\mathtt{m}_\alpha)\omega(p-1,h,U-\epsilon,M_\alpha-\mathtt{m}_\alpha)d\epsilon\ , \tag{II.30}$$

and if one assumes that the convolution integral is approximately constant in the interval $(U_{\alpha,prev},U_\alpha]$ results

$$\sum_{(\alpha)} O_\alpha d_\alpha(p-\alpha,h,U_\alpha-\epsilon_\alpha,M_\alpha-\mathtt{m}_\alpha) \tag{II.31}$$

$$\approx \delta U \sum_{(\mathtt{m}_\alpha)} O_\alpha \int_0^U \omega(p-1, h, U-\epsilon, M-\mathtt{m})\omega(1, 0, \epsilon, \mathtt{m}) d\epsilon \tag{II.32}$$

where $\delta U$=($U_\alpha$-$U_{\alpha,prev}$), then the sum of "$\sum_{(\alpha)} O_\alpha d_\alpha$" is approximatetly proportional to the convolution of nuclear densities. Now, under *CAP* one may assume that $\delta U$ is approximately constant and small,

$$\delta U \approx \epsilon_{nuc} = \text{ infinitesimal nuclear energy spacing} \tag{II.33}$$

then, taking $\delta U \approx dU$ results

$$\sum_{(\alpha)} O_\alpha d_\alpha \overset{CAP}{\approx} \left( \sum_{(\mathtt{m}_\alpha)} O_\alpha \int_0^U \omega(p-1, h, U-\epsilon, M-\mathtt{m}_\alpha)\omega(1, 0, \epsilon, \mathtt{m}_\alpha) d\epsilon \right) dU \tag{II.34}$$

which can be interpreted as the elementary variation of the inverse Laplace transform of the summed transition strengths.

# Appendix III - Analysis of the Model Space description

In this section the expression of the one body operator

$$O = \sum_{\binom{\alpha \in p}{\beta \in p}} O_{\alpha\beta} a_\alpha^\dagger a_\beta \tag{III.1}$$

is analyzed and its Model Space interpretation, based on the combinatorial analysis of the occupation of sp-states, is compared with the physical interpretation of the operators.

To obtain an expression involving also the hole operators so that transitions in which "new particle" and "new hole" states can be excited, as opposed to transitions corresponding to the simple scattering of the already excited "particle" states, one introduces the operators associated with the "$h$" states and replaces

$$a_\beta \;\text{ by }\; (a_\beta + b_\beta^\dagger) \qquad \text{and} \qquad a_\alpha^\dagger \;\text{ by }\; (a_\alpha^\dagger + b_\alpha) \tag{III.2}$$

then,

$$O = \sum_{\binom{\alpha \in p}{\beta \in p}} O_{\alpha\beta} a_\alpha^\dagger a_\beta + \sum_{\binom{\alpha \in p}{\beta \in h}} O_{\alpha\beta} a_\alpha^\dagger b_\beta^\dagger + \sum_{\binom{\alpha \in h}{\beta \in p}} O_{\alpha\beta} b_\alpha a_\beta + \sum_{\binom{\alpha \in h}{\beta \in h}} O_{\alpha\beta} b_\alpha b_\beta^\dagger \,. \tag{III.3}$$

When compared with (III.1) the last expression needs some clarification because, on the one hand, the statement ($\alpha \in p, \beta \in p$) in (III.1) means that a "$p$" was destroyed in ($\beta$) and another "$p$" was re-created in ($\alpha$), therefore both ($\beta$) and ($\alpha$) are sp-states of the "$p$" type in the Model Space. On the other hand, if ($\alpha$) is not equal to ($\beta$), does the possible creation of "$p$" in ($\alpha$) necessarily means that ($\alpha$) is "empty" in the sense that that it is a "$h$" sp-state?

To answer this question one may think initially within a combinatorial point of view and consider the "$p$"-type states defined by the Model Space as "empty boxes" in which the various "$p$" *excitons* will be placed when the $a_\alpha^\dagger$ operator acts and, similarly, the action of the operator $a_\alpha$ will decrease the number "$p$" excitons but will not alter the number of "$p$"-type states originally defined in the Model Space, which is also the *maximum number of "p" excitons* in the model.

The interpretation of $b_\alpha$ and $b_\alpha^\dagger$ in the Model Space is similar to the "$p$" operators with the caveat that one is dealing now with the placement or withdrawal of "holes" into and out of "empty boxes", which may not be a physically intuitive idea.

In a physical point of view a "$p$" sp-state can only be created on states occupied by "$h$" sp-state *and* vice versa. Then, when $b_\alpha^\dagger$ operates it can act exclusively on states occupied by "$p$" although in combinatorial terms the action of $b_\alpha^\dagger$ means only that a new "$h$" state ($\alpha$) has been "placed" into one of the available "empty boxes" for holes of the Model Space. Then, which one is correct? To say that the domain of sp-states on which $b^\dagger_\alpha$ acts is "$\alpha \in p$" or "$\alpha \in h$"?

In the Model Space description "particles" and "holes" are considered independent fields, then the creation of "$h$" can happen wherever there are "empty boxes" and the consideration of its "previous occupation by a particle" becomes meaningless. Then, one can use the fermion relation

$$b^\dagger_\alpha b_\beta + b_\beta b^\dagger_\alpha = \delta_{\alpha,\beta} \;, \tag{III.4}$$

where both ($\alpha$) and ($\beta$) are "$h$-boxes" of the Model Space ("empty" or "filled"), with no direct relation with the "$p$" sp-states. The case $\alpha{=}\beta$ corresponds either to the action of "filling a box that has just be emptied" *or* "emptying a box that has just been filled" and the case $\alpha{\neq}\beta$ gives the "zero operator" meaning that the both processes of creating a "$h$" ($\alpha$) followed or preceded by the destruction of a *different* "$h$" ($\beta$) have zero probability.

On the other hand, if one thinks more physically that $b^\dagger_\alpha$ is acting on states occupied by "$p$" and $b_\alpha$ corresponds in fact to the creation of a "$p$" on states previously occupied by holes, then the first term of (III.4) is acting on the "$h$" sp-states that have already been excited, while the second term is acting on the "$p$" sp-states that also have already been excited and neither of them are acting over the entire set of "$h$-boxes" of the Model Space.

In combinatorial terms the action of $a_\alpha$ on $|ph\rangle$ selects configurations ($p$) that contain ($\alpha$) while $b^\dagger_\alpha$ acting on $|ph\rangle$ selects the configurations ($h$) that do *not* contain ($\alpha$) . Then, ($a_\alpha{+}b^\dagger_\alpha$) acting on $|ph\rangle$ select the configurations ($ph$) in which ($\alpha$) is present in ($p$) *or* not present in ($h$) and the number of such configurations is

$$\binom{\mathsf{g}-1}{p-1}\left(\binom{b}{h}-\binom{b}{h-1}\right)=\binom{\mathsf{g}-1}{p-1}\left(\frac{b-h}{h}\right)\binom{b-1}{h-1}=\binom{\mathsf{g}-1}{p-1}\binom{b-1}{h}\;, \tag{III.5}$$

where the maximum number of "$p$" excitons in the Model Space is being called $\mathsf{g}$ and the maximum number of "$h$" excitons is being called $b$ and these letters will be used exclusively to designate these numbers in this paper. These maxima are related with the physical limits for the possible energies of excitons beyond which the excited "$p$" is supposed to be emitted and the nuclear system will be essentially altered in the sense that the Model Space must be modified too.

Because the sp-operators $a_\alpha$ and $b^\dagger_\alpha$ act on different fields and the fact that the action of $a_\alpha$ decreases the excitation energy, $U$, while $b^\dagger_\alpha$ increases $U$, one concludes that the sp-states designated by ($\alpha$) in both operators are in fact independent, therefore the use of the same index ($\alpha$) to designate both sp-states can be misleading.

The dynamics of the whole many-body system is defined independently of the Model Space itself. For example, the idea that "$p$"s and "$h$"s can only be created or destroyed in $ph$-pairs is described in the model strictly by additional constraints to correctly describe the conservation laws and the change of the number of excitons, with the corresponding change of $U$ and angular momenta, by the "correct definition" of the residual interaction.

Then, the $(\alpha)$ on $a_\alpha$ acts on the "$p$" sp-states which have already been excited in the available nuclear configurations $(ph)$ while the $(\alpha)$ of $b^\dagger_\alpha$ acts on the "$h$" sp-states of the Model Space that have *not yet* been occupied by a "$h$". Similarly, in the case of $(a^\dagger_\alpha{+}b_\alpha)$, $a^\dagger_\alpha$ acts on the "$p$" sp-states that have not been excited yet while $b_\alpha$ destroys the already existing "$h$".

Now, when one puts both operators together

$$(a^\dagger_\alpha + b_\alpha)(a_\alpha + b^\dagger_\alpha) \; , \tag{III.6}$$

the interpretation must follow logically from the above one. In particular, the term $b_\alpha b^\dagger_\alpha$ corresponds to the creation of a "$h$" on an unoccupied state $(\alpha)$ of the Model Space followed by the destruction of the same sp-state. Therefore, it counts the number of unoccupied ("available") sp-states of the type "$h$" while $b^\dagger_\alpha b_\alpha$ counts the number of occupied sp-states of the type "$h$". The sum of these operators, $(b^\dagger_\alpha b_\alpha{+}b_\alpha b^\dagger_\alpha)$, counts the total number of sp-states of the type "$h$" in the Model Space and, using the notation of (III.5), this number is $b$.

The fact that the excitation of "$p$"s and "$h$"s is *physically not indenpendent* of each other complicates the description a little because $b$ is usually greater than $\mathtt{g}$ (because one cannot define more "$p$" sp-states than the actual number of nucleons in the nuclear system, while the number of "$h$" sp-states has no constraints except for the maximum nuclear excitation energy) and, on the other hand, for a given Model Space, the maximum excitation of "$h$" sp-states cannot in fact go beyond $\mathtt{g}$. Then,

$$\sum_{(\alpha\beta\,\in\,\mathtt{filled}-h)} \langle ph\,|\,b^\dagger_\alpha b_\beta|\,ph\rangle + \sum_{(\alpha\beta\,\in\,-h)}\langle ph\,|\,b_\beta b^\dagger_\alpha|\,ph\rangle = \sum_{(\alpha\beta\,\in\,\mathtt{all}-h)} \langle ph\,|\,\delta_{\alpha,\beta}|\,ph\rangle = b \; , \tag{III.7}$$

and the normal ordered operator $O$ in Eq. (III.1) can be rewritten as

$$O = \underbrace{\sum_{\left(\substack{\alpha\,\in\,-p \\ \beta\,\in\,\mathtt{filled}-p}\right)} O_{\alpha\beta}a^\dagger_\alpha a_\beta}_{\text{(a)}} + \underbrace{\sum_{(\alpha\,\in\,\mathtt{filled}-p)} O_\alpha a^\dagger_\alpha a_\alpha}_{\text{(b)}} + \underbrace{\sum_{\left(\substack{\alpha\,\in\,-p \\ \beta\,\in\,-h}\right)} O_{\alpha\beta}a^\dagger_\alpha b^\dagger_\beta}_{\text{(c)}} + \underbrace{\sum_{\left(\substack{\alpha\,\in\,\mathtt{filled}-h \\ \beta\,\in\,\mathtt{filled}-p}\right)} O_{\alpha\beta}b_\alpha a_\beta}_{\text{(d)}}$$

$$- \underbrace{\sum_{\left(\substack{\alpha\,\in\,-h \\ \beta\,\in\,-h}\right)} O_{\alpha\beta}b^\dagger_\beta b_\alpha}_{\text{(e)}} - \underbrace{\sum_{(\alpha\,\in\,\mathtt{filled}-h)} b^\dagger_\alpha b_\alpha O_{\alpha\alpha}}_{\text{(f)}} + \underbrace{\sum_{(\alpha\,\in\,\mathtt{all}-h)} O_{\alpha\alpha}}_{\text{(g)}} \tag{III.8}$$

In reference to (III.8), the last term is

$$\sum_{(\alpha \in \mathtt{all}-h)} O_{\alpha\alpha} = \sum_{(\alpha,\beta \in \mathtt{all}-h)} \langle \alpha | O_{\alpha\beta} \delta_{\alpha,\beta} | \beta \rangle ,$$

and if one considers, for example, the number of particles operator (III.8) gives

$$N = \sum_{(\alpha \in \mathtt{filled}-p)} a^{\dagger}_{\alpha} a_{\alpha} - \sum_{(\alpha \in \mathtt{filled}-h)} b^{\dagger}_{\alpha} b_{\alpha} + \sum_{(\alpha \in \mathtt{all}-h)} (1) , \tag{III.9}$$

then

$$\langle N \rangle = \langle ph | N | ph \rangle = p - h + b , \tag{III.10}$$

which agrees with the idea expressed in (III.5) and is equal to the total number of already excited "$p$" sp-states ("filled" $p$ states) plus the total number of "$h$" sp-states that have not yet been excited ("empty" $h$ states). This description equates filled "$p$" sp-states with empty "$h$" sp-states, in which "$h$" sp-states can be created. This would be physically correct if "$h$" sp-states could only be created by the destruction of a "$p$" in the same sp-state, but this is not necessarily true in the Model Space description.

In other words, the total number of actually excited "$p$"s, which is the expected value of $a^{\dagger}_{\alpha} a_{\alpha}$, has been increased by the total number of "potential holes", which coincides in the Model Space with the total number of "empty-$h$" sp-states, but this reasoning does not take into account the constraint due to pair creation and annihilation. If this condition is taken into account, the actual number of potential holes ("empty-$h$" states) cannot be greater than "$\mathsf{g}$-$p$" instead of "$b$-$h$", corresponding to the interpretation of holes as empty particle states, then (III.10) becomes $\langle N \rangle = \mathsf{g}$, which is the physically correct result.

In conclusion, the apparent simplicity of the basic formulation of the Model Space that leads from (III.1) to (III.3), may produce meaningless results if it is not epistemologically enhanced by the correct physical interpretation of each step. This happens due to the inherent incompleteness of the Model Space description.

# Appendix IV - A problem with the Laplace transform approach

In the following discussion we drop the dependence of the degeneracies and densities on the total angular momentum, $M$, to have a clearer and easier to follow presentation, but it could be included in the argument in a straightforward way. The following analysis was inspired by the results for the evaluation of the moments of the Hamiltonian, in Sec.6, in which a given sp-state is destroyed and created in the same transition, in particular by Eqs. (6.17), (6.18) and (6.19).

In general one may write the generating function of the grand canonical ensemble as

$$f(x,y) = \prod_{\mu,\nu}(1+x_{p\mu})(1+x_{h\nu}) = \sum_{(p,h=0)}^{\infty} x^p y^h \sum_{(\epsilon_1<\cdots<\epsilon_{(p+h)})} d(p,h,U)\mathsf{e}^{-\beta U} \ , \tag{IV.1}$$

where $U{=}\epsilon_1{+}\cdots{+}\epsilon_{(p+h)}$, and for a given "particle" sp-state ($\alpha$) one may write

$$f_\alpha(x,y) = \frac{\prod_{\mu,\nu}(1+x_{p\mu})(1+x_{h\nu})}{1+x_{p\alpha}} = \sum_{(p,h=0)}^{\infty} x^p y^h \sum_{(\epsilon_i\neq\epsilon_\alpha)} d(p{\neq}\alpha,h,U_{\neq\alpha})\mathsf{e}^{-\beta U_{\neq\alpha}} \ , \tag{IV.2}$$

where $U$ is the energy of $p$-particles and $h$-holes with all sp-states considered, while $U_{\neq\alpha}$ is the energy of $p$-particles and $h$-holes in which all $p$-particles are different from $\alpha$. Then, one can rewrite (IV.1) as

$$f(x,y) = (1+x_{p\alpha})f_\alpha(x,y) = f_\alpha(x,y) + x_{p\alpha}f_\alpha(x,y) \tag{IV.3}$$

and the last term corresponds to the terms summed on the RHS of (6.17).

Notice that for given ($p$,$h$) the energy grids of $U_{\neq\alpha}$ and $U_\alpha$ are "complementary" with respect to the grid of $U$ in the sense that if $\mathsf{g}$ and $b$ are the maximum values of $p$ and $h$ in the Model Space, then one can show by combinatorial analysis of the corresponding sets of configurations that the number of configurations associated with the domains of $U$, $U_\alpha$ and $U_{\neq\alpha}$, which will be called $NOC(U)$, $NOC(U_\alpha)$ and $NOC(U_{\neq\alpha})$, respectively, are given by

$$NOC(U) = \binom{\mathsf{g}}{p}\binom{b}{h} \ , \quad NOC(U_\alpha) = \binom{\mathsf{g}{-}1}{p{-}1}\binom{b}{h} \quad \text{and} \quad NOC(U_{\neq\alpha}) = \binom{\mathsf{g}-1}{p}\binom{b}{h} \ , \tag{IV.4}$$

where the conventional notation for binomial coefficients has been used. Therefore,

$$NOC(U) = NOC(U_\alpha) + NOC(U_{\neq\alpha}) \ , \tag{IV.5}$$

and $NOC(U)$ is $\mathsf{g}$ times greater than $NOC(U_{\neq\alpha})$, corresponding to *very different grids* usually.

From (IV.2) one derives an approximate connection with the Laplace transform

$$x\mathsf{e}^{-\beta\epsilon_\alpha} f_\alpha(x,y) = \sum_{(p,h=0)}^{\infty} x^{p+1} y^h \sum_{(\epsilon_i\neq\epsilon_\alpha)} d(p{\neq}\alpha,h,U_{\neq\alpha})\mathsf{e}^{-\beta(U_{\neq\alpha}+\epsilon_\alpha)} \tag{IV.6}$$

$$\overset{CAP}{\approx} \sum_{(p,h=0)}^{\infty} x^{p+1}y^h \int_0^{\infty} \omega(p{\neq}\alpha, h, R)\mathsf{e}^{-\beta(R+\epsilon_\alpha)}dR \tag{IV.7}$$

$$= \sum_{(p,h=0)}^{\infty} x^{p+1}y^h \int_{\epsilon_\alpha}^{\infty} \omega(p{\neq}\alpha, h, s-\epsilon_\alpha)\mathsf{e}^{-\beta s}ds \tag{IV.8}$$

$$= \sum_{(p,h=0)}^{\infty} x^{p+1}y^h \mathcal{L}\{u(s-\epsilon_\alpha)\omega(p{\neq}\alpha, h, s-\epsilon_\alpha)\}, \tag{IV.9}$$

where "$u(s\text{-}\epsilon_\alpha)$" is the Heaviside step function and "$\omega(p{\neq}\alpha,h,R)$" is the continuous nuclear density corresponding to the discrete degeneracy "$d(p{\neq}\alpha,h,U_{\neq\alpha})$" (and "$R$" is the continuous counterpart of $U_{\neq\alpha}$, as defined in Sec.4). On the other hand, from (6.18) one may write

$$x_{p\alpha}f_\alpha(x,y) = x\mathsf{e}^{-\beta\epsilon_\alpha}f_\alpha(x,y) \overset{CAP}{\approx} \sum_{(1)}\sum_{(St)}\mathsf{e}^{[St]}\mathsf{e}^{-\beta\epsilon_\alpha}d(p-1{\neq}\alpha, h, S, t)\ , \tag{IV.10}$$

$$= \sum_{(1)}\mathcal{L}\Big\{\mathsf{e}^{-\beta\epsilon_\alpha}\omega(p-1{\neq}\alpha, h, S, t)\Big\}\ , \tag{IV.11}$$

where $S{=}(U_\alpha\text{-}\epsilon_\alpha)$. This expression would again lead to a Laplace transform involving the Heaviside step function, but one could also use (6.19) to obtain

$$x_{p\alpha}f_\alpha(x,y) = x\mathsf{e}^{-\beta\epsilon_\alpha}f_\alpha(x,y) \overset{CAP}{\approx} \sum_{(1)}\mathcal{L}\{\omega(p=\alpha, h, U_\alpha, M_\alpha)\}\ , \tag{IV.12}$$

which, due to the identity (6.16), is incompatible with (IV.11) unless $\epsilon_\alpha$=0.

Equation (IV.12) is the correct one in this case because it is in agreement with (IV.5) while (IV.11) is not, as it will be demonstrated next.

Initially notice that in (IV.12) the nuclear density can also be written as $\omega(p\text{-}\alpha\text{+}\alpha,h,U_\alpha,M_\alpha)$ because the selection of configurations is determined by the destruction of ($\alpha$) and not altered by the following re-creation of ($\alpha$).

On the other hand, the degeneracy "$d(p{\neq}\alpha,h,U_{\neq\alpha})$" is *not* the same as "$d(p\text{-}1{\neq}\alpha,h,U_\alpha\text{-}\epsilon_\alpha)$" in (6.18) because the energy grids of the arguments of both functions are different (in fact, complementary in the sense of (IV.4) and (IV.5)), but it is easy to show by combinatorial analysis that "$d(p{\neq}\alpha,h,U_{\neq\alpha})$" is equal to "$d(p\text{+}1\text{-}\alpha,h,U_{\neq\alpha})$", corresponding to the grid of energies obtained by the creation of a new particle in a given configuration, increasing "$p$" to "$p$+1", followed by the "selection" of configurations containing ($\alpha$) by the destruction of ($\alpha$).

From (IV.9) one has for given ($\alpha$) and ($p$,$h$) the following inverse Laplace transform of (IV.4)

$$\mathcal{L}^{-1}\Big\{x\mathsf{e}^{-\beta\epsilon_\alpha}f_\alpha(x,y)\Big\} = \mathcal{L}^{-1}\Big\{\mathsf{e}^{-\beta\epsilon_\alpha}\mathcal{L}\{\omega(p{\neq}\alpha, h, U_{\neq\alpha}\}\Big\} = u(U_{\neq\alpha})\omega(p{\neq}\alpha, h, U_{\neq\alpha})\ , \tag{IV.13}$$

and using $\mathcal{L}^{-1}\{f_\alpha(x,y)\}$=$\omega(p{\neq}\alpha,h,U_{\neq\alpha})$, $\mathcal{L}^{-1}\{f(x,y)\}$=$\omega(p,h,U)$ and (IV.5) one would conclude that

$$\omega(p,h,U) = \omega(p{\neq}\alpha,h,U_{\neq\alpha}) + u(U_{\neq\alpha})\omega(p{\neq}\alpha,h,U_{\neq\alpha}) \; , \tag{IV.14}$$

which is obviously a *false* relation because $(\alpha)$ is present in the domain of the configurations described by the LHS but is not present on the RHS. On the other hand, according to (IV.5) and using (IV.12), the correct relation would be

$$\omega(p,h,U) \approx \omega(p-1{\neq}\alpha,h,U_\alpha-\epsilon_\alpha) + \omega(p{\neq}\alpha,h,U_{\neq\alpha}) \; , \tag{IV.15}$$

or

$$\omega(p,h,U) \approx \omega(p=\alpha,h,U_\alpha) + \omega(p{\neq}\alpha,h,U_{\neq\alpha}) \; , \tag{IV.16}$$

which, in this case, is an obviously true relation.

This result means that the exponential term involving the sp-energy, in the term $x_{p\alpha}f_\alpha(x,y)$ of (IV.7), should be "absorbed" into the integral defining the Laplace transform instead of being considered as a multiplicative term that produces an inverse Laplace transform in which the Heaviside step function is present. This result corresponds to the identity between Eqs.(6.18) and (6.19).

Therefore, the direct use of the Laplace transform in the analysis of $\langle a\rangle$ in (6.12) may lead to wrong conclusions.

Sometimes the creation of sp-states brings problems of interpretation under the Laplace transform approach, which can be easily overcome within the direct microscopic formalism.

# Appendix V - Transition strengths expressed as convolutions

This Appendix is directly related to Sec.6 and presents some details of the algebraic deductions of a few transition strengths (TS) of the PE Hamiltonian, in particular, of how they can be expressed as convolutions with nuclear densities of less excited states. Each TS is preceded by a title identical to the corresponding subsection of Sec.6.

**Terms that increases the number of $p$ and $h$ by 2.**

The expression of $(a|a^{+})$ in (6.61) is

$$(a|a^{+}) = \sum_{\binom{12}{UM}} \sum_{(\alpha\beta\delta\gamma)} \langle ph\,|\,a^{+}_{\alpha} a^{+}_{\beta} b^{+}_{\delta} b^{+}_{\gamma}\,|\,p'h'\rangle\langle p'h'\,|\,b_{\gamma} b_{\delta} a_{\beta} a_{\alpha}\,|\,ph\rangle \tag{V.1}$$

$$= \sum_{(1)} \sum_{(St)} \mathtt{e}^{[UM]} \sum_{(\alpha\beta\delta\gamma)} |V_{\alpha\beta\gamma\delta}|^2 d(p-2{\neq}\alpha\beta, h-2{\neq}\delta\gamma, S, t)\ , \tag{V.2}$$

which are analogous of the first moment expressions.

This expression can be rewritten, using (**??**), in terms of the convolution over the energies of the selected sp-states by rewriting the sum over $(\alpha\beta\gamma\delta)$ as,

$$\sum_{(\alpha\beta\delta\gamma)} |V_{\alpha\beta\gamma\delta}|^2 d(p-2{\neq}\alpha\beta, h-2{\neq}\delta\gamma, S, t) \tag{V.3}$$

$$= \sum_{(\alpha\beta\gamma\delta)} |V_{\alpha\beta\gamma\delta}|^2 d(p-\alpha-\beta, h-\delta-\gamma, U_{(\alpha,\beta,\gamma,\delta)}-\epsilon, M_{(\alpha,\beta,\gamma,\delta)}-\mathtt{m})\ , \tag{V.4}$$

$$\approx \int d\epsilon_1 d\epsilon_2 d\epsilon_3 d\epsilon_4 \sum_{(\mathtt{m}_\alpha \mathtt{m}_\beta \mathtt{m}_\gamma \mathtt{m}_\delta)} |V_{\alpha\beta\gamma\delta}|^2 d(p-2, h-2, U-\epsilon, M-\mathtt{m})\ ,$$

$$\times\omega(1,0,\epsilon_1,\mathtt{m}_\alpha)\omega(1,0,\epsilon_2,\mathtt{m}_\beta)\omega(0,1,\epsilon_3,\mathtt{m}_\gamma)\omega(0,1,\epsilon_4,\mathtt{m}_\delta)\ , \tag{V.5}$$

$$= \sum_{(\mathtt{m})} \int d\epsilon V(2,2,\epsilon,\mathtt{m})\,\omega(p-2, h-2, U-\epsilon, M-\mathtt{m}) \tag{V.6}$$

where $\epsilon{=}\epsilon_\alpha{+}\epsilon_\beta{+}\epsilon_\gamma{+}\epsilon_\delta$, $\mathtt{m}{=}\mathtt{m}_\alpha{+}\mathtt{m}_\beta{+}\mathtt{m}_\gamma{+}\mathtt{m}_\delta$ and

$$V(2,2,\epsilon,\mathtt{m}) = \sum_{(\mathtt{m_x},\mathtt{m_y},\mathtt{m}_\delta)} \int dx\,dy\,d\epsilon_4\,|V_{\alpha\beta\gamma\delta}|^2\,\omega(1,0,\epsilon-x,\mathtt{m}-\mathtt{m_x})$$

$$\times\,\omega(1,0,x-y,\mathtt{m_x}-\mathtt{m_y})\,\omega(0,1,y-\epsilon_4,\mathtt{m_y}-\mathtt{m}_\delta)\,\omega(0,1,\epsilon_4,\mathtt{m}_\delta) \tag{V.7}$$

and $y{=}\epsilon_3{+}\epsilon_4$, $x{=}\epsilon_2{+}y$, $\mathtt{m_x}{=}\mathtt{m}_\beta{+}\mathtt{m_y}$ and $\mathtt{m_y}{=}\mathtt{m}_\gamma{+}\mathtt{m}_\alpha$

Equation (V.6) cannot be further simplified unless explicit assumptions are made about the dependence of $|V_{\alpha\beta\gamma\delta}|^2$ on the energy and angular momentum of the sp-states.

**Terms that increases the number of $p$ and $h$ by one, $(b|b^+)$.**

From (6.76) one has

$$(b|b^\dagger) = \sum_{\binom{12}{UM}} \sum_{(\alpha\beta\delta\gamma)} |V_{\alpha\beta\gamma\delta}|^2 \langle ph\,|\,a^+_\alpha a^+_\beta b^+_\delta a_\gamma|\,p'h'\rangle\langle p'h'\,|\,a^+_\gamma b_\delta a_\beta a_\alpha|\,ph\rangle \tag{V.8}$$

$$= \sum_{(1)} \sum_{(S,t)} \mathsf{e}^{[UM]} \sum_{(\alpha\beta\delta\gamma)} |V_{\alpha\beta\gamma\delta}|^2 d(p-\alpha-\beta+\gamma, h-\delta, S, t)\ . \tag{V.9}$$

and one can rewrite the sum over $(\alpha\beta\gamma\delta)$ as (see (III.3))

$$\sum_{(\alpha\beta\delta\gamma)} |V_{\alpha\beta\gamma\delta}|^2 (d(p-\alpha-\beta, h-\delta, S, t) - d(p-\alpha-\beta-\gamma, h-\delta, S, t))\ . \tag{V.10}$$

The first term is (see (**??**))

$$\sum_{(\alpha\beta\delta\gamma)} |V_{\alpha\beta\gamma\delta}|^2 d(p-\alpha-\beta, h-\delta, S, t)\ ,$$

$$\approx \int d\epsilon_1 d\epsilon_2 d\epsilon_3 d\epsilon_4 \sum_{(\mathtt{m}_\alpha \mathtt{m}_\beta \mathtt{m}_\gamma \mathtt{m}_\delta)} |V_{\alpha\beta\gamma\delta}|^2 d(p-2, h-1, U-\epsilon, M-\mathtt{m})\ ,$$

$$\times\omega(1,0,\epsilon_1,\mathtt{m}_\alpha)\omega(1,0,\epsilon_2,\mathtt{m}_\beta)\omega(0,1,\epsilon_3,\mathtt{m}_\gamma)\omega(0,1,\epsilon_4,\mathtt{m}_\delta) \tag{V.11}$$

$$= \sum_{(\mathtt{m})} \int d\epsilon V_x(2,1,x,\mathtt{m_x})\, \omega(p-2, h-1, U-x, M-\mathtt{m_x})\ , \tag{V.12}$$

where $y{=}(\epsilon_2{+}\epsilon_4)$, $x{=}(\epsilon_1{+}y)$, $\mathtt{m_x}{=}(\mathtt{m}_\alpha{+}\mathtt{m_y})$ and $\mathtt{m_y}{=}(\mathtt{m}_\beta{+}\mathtt{m}_\delta)$,

$$V_x(2,1,x,\mathtt{m_x}) = \sum_{(\mathtt{m_y},\mathtt{m}_\delta)} \int dy\, d\epsilon_4\, V_{(\alpha\beta\delta)}\, \omega(1,0,x-y,\mathtt{m_x}-\mathtt{m_y})\, \omega(0,1,y-\epsilon_4,\mathtt{m_y}-\mathtt{m}_\delta)\, \omega(0,1,\epsilon_4,\mathtt{m}_\delta) \tag{V.13}$$

and

$$V_{(\alpha\beta\delta)} = \int d\epsilon_3 \sum_{(\mathtt{m}_\gamma)} |V_{\alpha\beta\gamma\delta}|^2\, \omega(1,0,\epsilon_3,\mathtt{m}_\gamma)\ , \tag{V.14}$$

and the second term of (V.10) is

$$\sum_{(\alpha\beta\delta\gamma)} |V_{\alpha\beta\gamma\delta}|^2 d(p-\alpha-\beta-\gamma, h-\delta, U-\epsilon, M-\mathtt{m}) \ ,$$

$$\approx \int d\epsilon_1 d\epsilon_2 d\epsilon_3 d\epsilon_4 \sum_{(\mathtt{m}_\alpha \mathtt{m}_\beta \mathtt{m}_\gamma \mathtt{m}_\delta)} |V_{\alpha\beta\gamma\delta}|^2 d(p-3, h-1, U-\epsilon, M-\mathtt{m}) \ ,$$

$$\times \omega(1,0,\epsilon_1,\mathtt{m}_\alpha)\omega(1,0,\epsilon_2,\mathtt{m}_\beta)\omega(0,1,\epsilon_3,\mathtt{m}_\gamma)\omega(0,1,\epsilon_4,\mathtt{m}_\delta) \ , \tag{V.15}$$

$$= \sum_{(\mathtt{m})} \int d\epsilon V(2,2,\epsilon,\mathtt{m})\, d(p-2, h-2, U-\epsilon, M-\mathtt{m}) \tag{V.16}$$

where $\epsilon{=}(\epsilon_1{+}\epsilon_2{+}\epsilon_3{+}\epsilon_4)$, $\mathtt{m}{=}(\mathtt{m}_\alpha{+}\mathtt{m}_\beta{+}\mathtt{m}_\gamma{+}\mathtt{m}_\delta)$, $y{=}(\epsilon_3{+}\epsilon_4)$, $x{=}(\epsilon_2{+}y)$, $\mathtt{m_x}{=}(\mathtt{m}_\beta{+}\mathtt{m_y})$ and $\mathtt{m_y}{=}(\mathtt{m}_\gamma{+}\mathtt{m}_\alpha)$ and

$$V(2,2,\epsilon,\mathtt{m}) = \sum_{(\mathtt{m_x},\mathtt{m_y},\mathtt{m}_\delta)} \int dx\, dy\, d\epsilon_4\, |V_{\alpha\beta\gamma\delta}|^2\, \omega(1,0,\epsilon-x,\mathtt{m}-\mathtt{m_x})$$

$$\times\, \omega(1,0,x-y,\mathtt{m_x}-\mathtt{m_y})\, \omega(0,1,y-\epsilon_4,\mathtt{m_y}-\mathtt{m}_\delta)\, \omega(0,1,\epsilon_4,\mathtt{m}_\delta) \tag{V.17}$$

**Terms that increase ($p$,$h$) by one: off-shell processes.**

In the cases in which Hermitean conjugates of a given sp-operator, $a_{p\alpha}$ and $a^\dagger_{p\alpha}$, act on the same nuclear configuration the corresponding sp-energies are not present in the energy conservation constraint for the total transition. Then, the processes of creation and destruction of these sp-states are off-shell with respect to the transition process.

In the expression of $(b|b^+)$ in (6.65) with $\gamma$ replaced by $\alpha$ in $(b)$ and $(b+)$ one has

$$(b_{\gamma\to\alpha}|b^+_{\gamma\to\alpha}) = \sum_{\binom{12}{UM}} \sum_{(\alpha\beta\delta)} \langle ph\,|\,a^+_\alpha a^+_\beta b^+_\delta a_\alpha\,|\,p'h'\rangle\langle p'h'\,|\,a^+_\alpha b_\delta a_\beta a_\alpha\,|\,ph\rangle$$

$$\approx \sum_{(1)} \sum_{(S,t)} \mathsf{e}^{[UM]} \sum_{(\alpha\beta\delta)} |V_{\alpha\beta\delta}|^2 d(p-\alpha-\beta+\alpha, h-\delta, S, t) \ , \tag{V.18}$$

as it was shown in (6.96), where

$$S = (U_\alpha - \epsilon_\beta - \epsilon_\delta) \qquad \text{and} \qquad t = (M_\alpha - \mathtt{m}_\beta - \mathtt{m}_\delta) \ , \tag{V.19}$$

$$d(p-\alpha-\beta+\alpha, h-\delta, S, t) = d(p-\alpha-\beta, h-\delta, U_{(\alpha\beta\delta)} - x, M_{(\alpha\beta\delta)} - \mathtt{m_x}) \ , \tag{V.20}$$

and

$$x = \epsilon_\alpha + \epsilon_\beta + \epsilon_\delta \qquad \text{and} \qquad \mathtt{m_x} = \mathtt{m}_\alpha + \mathtt{m}_\beta + \mathtt{m}_\delta \ . \tag{V.21}$$

The indices of $U$ and $M$ correspond to the sp-states that are *necessarily present* in the nuclear configurations associated with these parameters, for the transition to be possible. Although the configurations after the transition have 2 excitons less than the initial one, the degeneracy describing the transition strength corresponds to configurations with 3 excitons less because $(\alpha)$ was selected by an additional off-shell process.

Then,

$$(b_{\gamma\to\alpha}|b^+_{\gamma\to\alpha}) \approx \int dx \sum_{(\mathtt{m_x})} V_x(2,1,x,\mathtt{m_x})\omega(p-2,h-1,U-x,M-\mathtt{m_x}) \tag{V.22}$$

where

$$V_x(2,1,x,\mathtt{m_x}) = \int dy d\epsilon_3 \sum_{(\mathtt{m_y}\mathtt{m}_\delta)} |V_{\alpha\beta\gamma\delta}|^2 \omega(1,0,x-y,\mathtt{m_x}-\mathtt{m_y})$$

$$\times\omega(1,0,y-\epsilon_3,\mathtt{m_y}-\mathtt{m}_\delta)\,\omega(0,1,\epsilon_3,\mathtt{m}_\delta) \ , \tag{V.23}$$

which is expected to be valid in the $CAP$ limit.

Similarly, another off-shell process can be obtained from (6.85) with $\alpha$ replaced by a different index $\gamma$ in $(b+)$, i.e., $\gamma{\neq}\alpha$, then

$$(b_{\gamma\to\alpha}|b^+_{\alpha\to\gamma})_{\gamma\neq\alpha} = \sum_{\binom{12}{UM}} \sum_{(\alpha\beta\delta\gamma)} \langle ph\,|\,a^\dagger_\alpha a^\dagger_\beta b^\dagger_\delta a_\alpha|p'h'\rangle\langle p'h'\,|\,a^\dagger_\gamma b_\delta a_\beta a_\gamma|ph\rangle. \tag{V.24}$$

$$\approx \sum_{(1)} \sum_{(S,t)} \mathsf{e}^{[UM]} \sum_{(\alpha\beta\delta\gamma)} d(p-\alpha-\gamma-\beta+\alpha+\gamma, h-\delta, S, t) \ , \tag{V.25}$$

where $S$ and $t$ are still given by (V.19) and now both $\alpha$ and $\gamma$ transitions are described as off-shell. The degeneracy is

$$d(p-\alpha-\gamma-\beta+\alpha+\gamma, h-\delta, S, t) = d(p-\alpha-\beta-\gamma, h-\delta, U-\epsilon, M-\mathtt{m}) \ ,$$

$$= d((p=\alpha\gamma)-\beta, h-\delta, U_{(\alpha\gamma)}-\epsilon_\beta-\epsilon_\delta, M_{(\alpha\gamma)}-\mathtt{m}_\beta-\mathtt{m}_\delta) \ , \tag{V.26}$$

and

$$x = \epsilon_\alpha + \epsilon_\beta + \epsilon_\delta \qquad \text{and} \qquad \mathtt{m_x} = \mathtt{m}_\alpha + \mathtt{m}_\beta + \mathtt{m}_\delta \ . \tag{V.27}$$

Although the configurations after the transition have 2 excitons less than the initial one, the degeneracy describing the transition strength corresponds to configurations with 4 excitons less because $(\alpha)$ and $(\gamma)$ were selected by an additional off-shell processes.

Then,

$$(b_{\gamma\to\alpha}|b^+_{\alpha\to\gamma})_{\gamma\neq\alpha} \approx \int d\epsilon \sum_{(\mathtt{m})} V_\epsilon(3,1,\epsilon,\mathtt{m})\omega(p-3,h-1,U-\epsilon,M-\mathtt{m}) \tag{V.28}$$

where

$$V_\epsilon(3,1,\epsilon,\mathtt{m}) = \int dxdyd\epsilon_4 \sum_{\binom{\mathtt{m_x m_y}}{\mathtt{m}_\delta}} |V_{\alpha\beta\gamma\delta}|^2 \omega(1,0,\epsilon-x,\mathtt{m}-\mathtt{m_x})$$

$$\times\omega(1,0,x-y,\mathtt{m_x}-\mathtt{m_y})\,\omega(1,0,y-\epsilon_4,\mathtt{m_y}-\mathtt{m}_\delta)\,\omega(0,1,\epsilon_4,\mathtt{m}_\delta) \ , \tag{V.29}$$

which is valid in the $CAP$ limit.

**Other terms with on-shell transitions.**

The term corresponding to $(c|c+)$ in (6.10) can be written analogously

$$(c|c^+) = \sum_{\binom{12}{UM}} \sum_{(\alpha\beta\delta\gamma)} |V_{\alpha\beta\gamma\delta}|^2 \langle ph\,|\,a^\dagger_\alpha b^\dagger_\delta b^\dagger_\gamma b_\beta\,|\,p'h'\rangle\langle p'h'\,|\,b^\dagger_\beta b_\gamma b_\delta a_\alpha\,|\,ph\rangle \tag{V.30}$$

$$\approx \sum_{(1)} \sum_{(S,t)} \mathsf{e}^{[UM]} \sum_{(\alpha\beta\delta\gamma)} |V_{\alpha\beta\gamma\delta}|^2 d(p-\alpha, h-\delta-\gamma+\beta, S, t) \ ,$$

$$= \sum_{(1)} \sum_{(S,t)} \mathsf{e}^{[UM]} \sum_{(\alpha\beta\delta\gamma)} \big(|V_{\alpha\beta\gamma\delta}|^2 d\big(p-\alpha, h-\delta-\gamma, U_{(\alpha\delta\gamma)}-x, M_{(\alpha\delta\gamma)}-\mathtt{m_x}\big)$$

$$-|V_{\alpha\beta\gamma\delta}|^2 d\big(p-\alpha, h-\delta-\gamma-\beta, U_{(\alpha\beta\gamma\delta)}-\epsilon, M_{(\alpha\beta\gamma\delta)}-\mathtt{m}\big)\big) \ , \tag{V.31}$$

where

$$S = (U-\epsilon_\alpha-\epsilon_\gamma-\epsilon_\delta+\epsilon_\beta) \qquad \text{and} \qquad t = (M-\mathtt{m}_\alpha-\mathtt{m}_\gamma-\mathtt{m}_\delta+\mathtt{m}_\beta) \ , \tag{V.32}$$

$$\epsilon = (\epsilon_\alpha+\epsilon_\beta+\epsilon_\gamma+\epsilon_\delta) \qquad \text{and} \qquad \mathtt{m} = (\mathtt{m}_\alpha+\mathtt{m}_\beta+\mathtt{m}_\gamma+\mathtt{m}_\delta) \ , \tag{V.33}$$

$$x = (\epsilon_\alpha+\epsilon_\gamma+\epsilon_\delta) \qquad \text{and} \qquad \mathtt{m_x} = (\mathtt{m}_\alpha+\mathtt{m}_\gamma+\mathtt{m}_\delta) \ , \tag{V.34}$$

and the indices of $U$ and $M$ indicate the sp-states which are supposed to be present in the nuclear configurations associated with these parameters.

The first term is

$$\sum_{(\alpha\beta\delta\gamma)} |V_{\alpha\beta\gamma\delta}|^2 d(p-\alpha, h-\delta-\gamma, S, t) \ ,$$

$$\approx \int d\epsilon_1 d\epsilon_2 d\epsilon_3 d\epsilon_4 \sum_{(\mathtt{m}_\alpha \mathtt{m}_\beta \mathtt{m}_\gamma \mathtt{m}_\delta)} |V_{\alpha\beta\gamma\delta}|^2 \omega(p-1, h-2, U-x, M-\mathtt{m}_\mathtt{x}) \ ,$$

$$\times\omega(1,0,\epsilon_1,\mathtt{m}_\alpha)\omega(0,1,\epsilon_2,\mathtt{m}_\gamma)\omega(0,1,\epsilon_3,\mathtt{m}_\delta)\omega(1,0,\epsilon_4,\mathtt{m}_\beta) \ , \tag{V.35}$$

$$= \sum_{(\mathtt{m}_\mathtt{x})} \int dx V_x(1,2,x,\mathtt{m}_\mathtt{x})\, \omega(p-1, h-2, U-x, M-\mathtt{m}_\mathtt{x}) \tag{V.36}$$

where

$$V_x(1,2,x,\mathtt{m}_\mathtt{x}) = \sum_{(\mathtt{m}_\mathtt{y}\mathtt{m}_\delta)} \int dy\, d\epsilon_3\, V_{(\alpha\gamma\delta)}\, \omega(1,0,x-y,\mathtt{m}_\mathtt{x}-\mathtt{m}_\mathtt{y})\, \omega(0,1,y-\epsilon_3,\mathtt{m}_\mathtt{y}-\mathtt{m}_\delta)\, \omega(0,1,\epsilon_3,\mathtt{m}_\delta), \tag{V.37}$$

$$y = (\epsilon_\gamma + \epsilon_\delta) \qquad \text{and} \qquad \mathtt{m}_\mathtt{y} = (\mathtt{m}_\gamma + \mathtt{m}_\delta) \ , \tag{V.38}$$

and

$$V_{(\alpha\gamma\delta)} = \int d\epsilon_4 \sum_{(\mathtt{m}_\beta)} |V_{\alpha\beta\gamma\delta}|^2\, \omega(1,0,\epsilon_4,\mathtt{m}_\beta) = \sum_{(\beta)} |V_{\alpha\beta\gamma\delta}|^2$$

and the second term is

$$\sum_{(\alpha\beta\delta\gamma)} |V_{\alpha\beta\gamma\delta}|^2 d(p-\alpha, h-\beta-\delta-\gamma, S, t) \ ,$$

$$\approx \int d\epsilon_1 d\epsilon_2 d\epsilon_3 d\epsilon_4 \sum_{\binom{\mathtt{m}_\alpha \mathtt{m}_\beta}{\mathtt{m}_\gamma \mathtt{m}_\delta}} |V_{\alpha\beta\gamma\delta}|^2 \omega(p-1, h-3, U-\epsilon, M-\mathtt{m}) \ ,$$

$$\times\, \omega(1,0,\epsilon_1,\mathtt{m}_\alpha)\, \omega(0,1,\epsilon_4,\mathtt{m}_\beta)\, \omega(0,1,\epsilon_2,\mathtt{m}_\gamma)\, \omega(0,1,\epsilon_3,\mathtt{m}_\delta) \ , \tag{V.39}$$

$$= \sum_{(\mathtt{m})} \int d\epsilon V_\epsilon(1,3,\epsilon,\mathtt{m})\, \omega(p-1, h-3, U-\epsilon, M-\mathtt{m}) \tag{V.40}$$

where

$$V_\epsilon(1,3,\epsilon,\mathtt{m}) = \sum_{\binom{\mathtt{m_x m_y}}{\mathtt{m}_\delta}} \int dx\, dy\, d\epsilon_3\, |V_{\alpha\beta\gamma\delta}|^2 \omega(1,0,\epsilon - x, \mathtt{m} - \mathtt{m_x})$$

$$\times\, \omega(1,0,x-y,\mathtt{m_x}-\mathtt{m_y})\, \omega(0,1,y-\epsilon_3,\mathtt{m_y}-\mathtt{m}_\delta)\, \omega(0,1,\epsilon_3,\mathtt{m}_\delta)$$

with $(x,\mathtt{m_x})$ redefined as

$$x = (\epsilon_\gamma + \epsilon_\delta + \epsilon_\beta) \qquad \text{and} \qquad \mathtt{m_x} = (\mathtt{m}_\gamma + \mathtt{m}_\delta + \mathtt{m}_\beta)\ , \tag{V.41}$$

The terms corresponding to $(c|c^\dagger)$ with $\beta\neq\delta$ are

$$(c|c^\dagger)_{\,\beta\neq\delta} = \sum_{\binom{12}{UM}} \sum_{(\alpha\beta\delta\gamma)} \langle ph\,|\,a_\alpha^+ b_\beta^+ b_\gamma^+ b_\beta\,|\,p'h'\rangle\langle p'h'\,|\,b_\delta^+ b_\gamma b_\delta a_\alpha\,|\,ph\rangle$$

$$\approx \sum_{((1),S,t)} \mathtt{e}^{[UM]} \sum_{(\alpha\beta\delta\gamma)} d(p-\alpha, h-\delta-\gamma-\beta+\beta+\delta, U-\epsilon_\alpha-\epsilon_\gamma, M-\mathtt{m}_\alpha-\mathtt{m}_\gamma)\ ,$$

$$= \sum_{((1),S,t)} \mathtt{e}^{[UM]} \sum_{(\alpha\beta\delta\gamma)} (|V_{\alpha\beta\gamma\delta}|^2 d\big(p-\alpha, h-\delta-\gamma-\beta, U_{(\alpha\beta\gamma\delta)}-\epsilon, M_{(\alpha\beta\gamma\delta)}-\mathtt{m}\big)$$

$$-|V_{\alpha\beta\gamma\delta}|^2 d(p-\alpha, h-\delta-\gamma-\beta, U_{\alpha\beta\gamma\delta}-\epsilon, M_{\alpha\beta\gamma\delta}-\mathtt{m}))\ , \tag{V.42}$$

$$\approx \sum_{((1),S,t)} \mathtt{e}^{[UM]} \int d\epsilon \sum_{(\mathtt{m})} V_\epsilon(1,3,\epsilon,\mathtt{m})\omega(p-1,h-3,U-\epsilon,M-\mathtt{m}) \tag{V.43}$$

where

$$V_\epsilon(1,3,\epsilon,\mathtt{m}) = \int dx dy d\epsilon_4 \sum_{\binom{\mathtt{m_x m_y}}{\mathtt{m}_\delta}} |V_{\alpha\beta\gamma\delta}|^2 \omega(1,0,\epsilon-x,\mathtt{m}-\mathtt{m_x})$$

$$\times\omega(0,1,x-y,\mathtt{m_x}-\mathtt{m_y})\, \omega(0,1,y-\epsilon_4,\mathtt{m_y}-\mathtt{m}_\delta)\, \omega(0,1,\epsilon_4,\mathtt{m}_\delta)\ , \tag{V.44}$$

$$\epsilon = \epsilon_\alpha + \epsilon_\beta + \epsilon_\gamma + \epsilon_\delta \qquad \text{and} \qquad \mathtt{m} = \mathtt{m}_\alpha + \mathtt{m}_\beta + \mathtt{m}_\gamma + \mathtt{m}_\delta\ , \tag{V.45}$$

$$x = \epsilon_\beta + \epsilon_\gamma + \epsilon_\delta \qquad \text{and} \qquad \mathtt{m_x} = \mathtt{m}_\beta + \mathtt{m}_\gamma + \mathtt{m}_\delta \tag{V.46}$$

and

$$y = \epsilon_\gamma + \epsilon_\delta \qquad \text{and} \qquad \mathtt{m_y} = \mathtt{m}_\gamma + \mathtt{m}_\delta\ . \tag{V.47}$$